\documentclass{aa}  

\usepackage{graphicx}
\usepackage{txfonts}
\usepackage{natbib}
\bibpunct{(}{)}{;}{a}{}{,} 
\usepackage{graphicx}	
\usepackage{amsmath}	
\usepackage{epsfig}
\usepackage{animate}
\usepackage{xcolor}
\usepackage{tabularx}
\usepackage{threeparttable}
\usepackage{hyperref}
\hypersetup{colorlinks=true, citecolor=blue, linkcolor=blue}

\newcommand{\hii}{H\,{\sc ii}}
\newcommand{\nha}{[N\,{\sc ii}]/H$\alpha$}
\newcommand{\sha}{[S\,{\sc ii}]/H$\alpha$}
\newcommand{\ohb}{[O\,{\sc iii}]/H$\beta$}
\newcommand{\no}{[N\,{\sc ii}]/[O\,{\sc ii}]}
\newcommand{\oo}{[O\,{\sc iii}]/[O\,{\sc ii}]}
\newcommand{\so}{[S\,{\sc iii}]/[O\,{\sc iii}]}

\begin{document}

   \title{The need for multicomponent dust attenuation in modeling nebular emission: Constraints from SDSS-IV MaNGA}
   \titlerunning{Multicomponent dust attenuation for nebular emission lines}

   \author{Xihan Ji
          \inst{1}\fnmsep\thanks{E-mail: xji243@uky.edu}
          , 
          Renbin Yan\inst{2}\fnmsep\thanks{E-mail: rbyan@cuhk.edu.hk},
          Kevin Bundy\inst{3},
          Médéric Boquien\inst{4},
          Adam Schaefer\inst{5},
          Francesco Belfiore\inst{6},
          Matthew A. Bershady\inst{7,8,9},
          Niv Drory\inst{10},
          Cheng Li\inst{11},
          Kyle B. Westfall\inst{3},
          {Zesen
          Lin\inst{2},}
          Dmitry Bizyaev\inst{12},
          David R. Law\inst{13},
          Rogério Riffel\inst{14,16},
          \and
          Rogemar A. Riffel\inst{15,16}
          }

   \institute{Department of Physics and Astronomy, University of Kentucky, 505 Rose Street, Lexington, KY 40506, USA\          
   \and
    Department of Physics, The Chinese University of Hong Kong, Shatin, N.T., Hong Kong S.A.R., China\
    \and
    University of California Observatories - Lick Observatory, University of California Santa Cruz, 1156 High St., Santa Cruz, CA 95064, USA\
    \and
    Centro de Astronomía, Universidad de Antofagasta, Avenida Angamos 601, Antofagasta 1270300, Chile\
    \and
    Max-Planck-Institut für Astrophysik, Karl-Schwarzschild-Str. 1, D-85748 Garching, Germany\
    \and
    INAF — Osservatorio Astrofisico di Arcetri, Largo E. Fermi 5, I50125, Florence, Italy\
    \and
    University of Wisconsin - Madison, Department of Astronomy,
    475 N. Charter Street, Madison, WI 53706-1582, USA\
    \and
    South African Astronomical Observatory, PO Box 9, Observatory 7935, Cape Town, South Africa\
    \and
    Department of Astronomy, University of Cape Town, Private Bag X3, Rondebosch 7701, South Africa\
    \and
    McDonald Observatory, The University of Texas at Austin, 2515 Speedway, Stop C1402, Austin, TX 78712, USA\
    \and
    Department of Astronomy, Tsinghua University, Beijing 100084, China\
    \and
    Apache Point Observatory, P.O. Box 59, Sunspot, NM 88349, USA\
    \and
    Space Telescope Science Institute, 3700 San Martin Drive, Baltimore, MD 21218, USA\
    \and
    Instituto de Física, Universidade Federal do Rio Grande do Sul, Campus do Vale, 91501-970, Porto Alegre, RS, Brazil\
    \and
    Departamento de F\'isica, CCNE, Universidade Federal de Santa Maria, 97105-900, Santa Maria, RS, Brazil\
    \and
    Laborat\'orio Interinstitucional de e-Astronomia - LIneA, Rua Gal. Jos\'e Cristino 77, Rio de Janeiro, RJ - 20921-400, Brazil
    \\
             }

   \date{Received 27 September 2022 / Accepted 12 December 2022}

 
  \abstract{
  A fundamental assumption adopted in nearly every extragalactic study that analyzes optical emission lines is that the attenuation of different emission lines can be described by a single attenuation curve, scaled by a single reddening parameter, usually E(B$-$V). 
  Here we show this assumption fails in many cases with important implications for derived results.
  We developed a new method to measure the differential nebular attenuation among three kinds of transitions: the Balmer lines of hydrogen; high-ionization transitions ($>13.6$ eV) including [Ne\,{\sc iii}], [O\,{\sc iii}], and [S\,{\sc iii}]; and low-ionization transitions ($\lesssim 13.6$ eV) including [O\,{\sc ii}], [N\,{\sc ii}], and [S\,{\sc ii}].
  This method bins the observed data in a multidimensional space spanned by attenuation-insensitive line ratios. 
  Within each small bin, the variations in nebular parameters such as the metallicity and ionization parameter are negligible compared to the variation in the nebular attenuation.
  This allowed us to measure the nebular attenuation using both forbidden lines and Balmer lines.
  We applied this method to a sample of 2.4 million star-forming (SF) spaxels from the Mapping Nearby Galaxies at Apache Point Observatory (MaNGA) survey.
  We found that the attenuation of high ionization lines and Balmer lines can be well described by a single \cite{fitzpatrick1999} extinction curve with $R_V =3.1$.
  However, no single attenuation curve can simultaneously account for these transitions and the derived attenuation of low-ionization lines.
  This strongly suggests that different lines have different effective attenuations, likely because spectroscopy at hundreds of parsecs to kiloparsecs of resolution mixes multiple physical regions that exhibit different intrinsic line ratios and different levels of attenuation.
  As a result, the assumption that different lines follow the same attenuation curve breaks down.
  {Using a single attenuation curve determined by Balmer lines to correct attenuation-sensitive forbidden line ratios could bias the nebular parameters derived by 0.06\,--\,0.25 dex at $A_V = 1$, depending on the details of the dust attenuation model.}
  Observations of a statistically large sample of \hii\ regions with high spatial resolutions and large spectral coverage are vital for improved modeling and deriving accurate corrections for this effect.
  }

   \keywords{ISM: dust, extinction --
                ISM: \hii\ regions --
                ISM: lines and bands
               }
               
   \authorrunning{X. Ji et al.}
   \maketitle
%

\section{Introduction}

Nebular attenuation\footnote{Throughout this work, we use the word ``attenuation'' to indicate the potential presence of complex cloud geometries around the source as well as the scattering of light into and out of the line of sight. While the word ``extinction'' describes the simpler scenario where there is effectively a dust screen between the observer and the foreground source.} due to dust grains is ubiquitous in astronomical observations.
Besides the Galactic extinction, observations of emission lines from extragalactic \hii\ regions inevitably suffer from different degrees of intrinsic attenuation inside their host galaxies.
The overall attenuation of the emission is usually described by a parameterized nebular attenuation curve, whose shape is determined by the dust composition, grain size distribution, and geometry.
Understanding the attenuation curve is important for accurate measurements of fluxes of emission lines and their derived physical properties, for example, star formation rate, metallicity, nitrogen-to-oxygen ratio (N/O), ionization parameter, among others \citep[e.g.,][]{pagel1979,alloin1979,diaz1991,thurston1996,dopita2000,kennicutt2012}.
Its shape also provides information about the dust sizes and compositions around emission-line regions \citep[e.g.,][]{spitzer1978,draine2003,flaherty2007}.
In addition, by comparing the average nebular attenuation curve with the average stellar attenuation curve, we can gain information about the different dust geometries around young stellar populations and older stellar populations \citep[e.g.,][]{charlot2000,inoue2005,chevallard2013}.

Various attempts have been made to measure the nebular attenuation in both low-redshift and high-redshift emission-line galaxies \citep[e.g.,][]{reddy2004,wild2011,price2014,reddy2020,rezaee2021}.
Many recent works suggest that the average nebular attenuation curve found in external galaxies has a similar shape to the average Galactic extinction curve with some slight differences \citep{wild2011,reddy2020,rezaee2021}.
{Although considerable dispersion around the average curve depending on galaxy properties is expected \citep[e.g.,][]{zafar2015,fitzpatrick2007,salim2020}, due to the limitations of the methods and samples, this potential variation has not been investigated in depth so far.}
{Nevertheless, there is evidence that simply applying the average Galactic extinction curve to extragalactic \hii\ regions is inaccurate as the residual line ratio between the extinction-corrected data and the predictions from photoionization models show an unexpected dependence on the Balmer decrement \citep[see Figure 14 in][]{ji2022}.
Furthermore, there could be a bias in the derived attenuation due to the complicated dust geometry, which depends on the aperture size \citep{vale2020}. 
}

Most of the current studies on the optical attenuation curve rely on Balmer lines. In \hii\ regions, the ratios of the intrinsic fluxes of Balmer lines are roughly fixed and are relatively insensitive to variations in temperature and density \citep[e.g., 
$\rm flux_{H\alpha}/flux_{H\beta} \approx 2.86^{+0.18}_{-0.11}$ for an \hii\ region with $n_H\lesssim 10^{6}~cm^{-3}$ and $5,000~K \lesssim T\lesssim 20,000~K$,][]{agn3}.
Therefore, any significant deviation from the theoretical ratios for Balmer lines can be interpreted as the effect of dust attenuation. However, this method can only cover a relatively small range of wavelength in the optical (roughly 4000\,--\,6600~\AA ). 
Also, higher order Balmer lines (blueward of H$\gamma$ with upper levels at $n > 5$) are relatively weak in observations, both because they are intrinsically weak and because they are more attenuated.
In addition, absorption features in the blue part of the optical spectra add further uncertainties to the measurements of weak Balmer lines.
These make this method inapplicable for a large number of galaxies whose emission-line spectra have intermediate to low signal-to-noise ratios (S/Ns). Stacking galaxy spectra could partly circumvent this problem, but this would also miss a lot of information about the variation of the nebular attenuation curve in different galaxies or regions.

Finally, it is often implicitly assumed in extragalactic studies that all of the other nebular emission lines follow the same attenuation curve probed by Balmer lines.
There are two major caveats associated with this assumption.
First, the attenuation for emission lines not lying in the spectral range covered by Balmer lines is not well constrained.
Second, this assumption requires that all emission lines come from the same spatial location and thus experience the same level of attenuation, whose validity is rarely carefully evaluated.

Compared to high-order Balmer lines, strong optical-to-near infrared (NIR) forbidden lines cover a much larger spectral range and are easier to measure. However, the intrinsic ratios of these lines strongly depend on nebular parameters such as the metallicity and ionization parameter. 
To determine the intrinsic ratios of attenuation-sensitive forbidden lines, one has to rely on theoretical or empirical calibrations of nebular parameters, which could bring large systematic uncertainties since different calibrations generally do not agree with each other \citep[e.g.,][]{kewley2008,kewley2019,ji2022}.
The key question is whether it is possible to find an empirical way to constrain these nebular parameters without specifying their values. One potential and indirect solution is to compare the attenuated spectra from galaxies with similar physical properties including redshifts, stellar masses, star formation rates (SFRs), and so forth, but different degrees of attenuation. 
The nebular parameters of these galaxies are likely similar given the galaxy scaling relations, and thus the differences in the attenuation-sensitive strong line ratios are likely mainly driven by dust attenuation.
\cite{wild2011b}, for example, used a similar pair-matching approach to empirically measure the stellar attenuation curve in external galaxies.
Still, one needs to be cautious about the scatter in the adopted galaxy scaling relations and how they transfer into variations in the galaxy spectra. Also, the uncertainties in the controlled physical parameters (e.g., stellar masses, SFRs, etc.) require careful treatments.

In this work, we adopt a slightly different approach by taking the observed quantities rather than the derived ones as the controlled physical parameters. We start with the assumption that, {within the same observed star-forming (SF) region}, all emission lines follow the same attenuation curve scaled by the same reddening parameter [$\rm A_V$ or E(B$-$V)]. 
By using a 3D line-ratio space spanned by three attenuation-insensitive line ratios, [N\,{\sc ii}]$\lambda 6583$/H$\alpha$, [S\,{\sc ii}]$\lambda \lambda 6716,6731$/H$\alpha$, and [O\,{\sc iii}]$\lambda 5007$/H$\beta$, we constrain the nebular parameters of observed SF spatial pixels (spaxels) and thus the intrinsic ratios of forbidden lines.
{We show that small regions in this 3D line-ratio space correspond to a set of nearly constant nebular parameters, including metallicity (or O/H) and ionization parameter.}
In comparison, the variation of the dust attenuation inside these small regions can still be significant, driving most of the variations in attenuation-sensitive line ratios. This allows us to construct relative attenuation curves using a large number of extragalactic SF regions with detectable strong forbidden lines and the two strongest Balmer lines, H$\alpha$ and H$\beta$.
However, our results show that a single attenuation curve is insufficient to describe all emission lines,
{meaning this fundamental assumption of many extragalactic emission line studies breaks down.}
Based on our findings, we discuss potential explanations and suggest ways to move forward for future emission-line studies.

The layout of this paper is as follows. In Section~\ref{sec:data}, we introduce the observational data we use. In Section~\ref{sec:method}, we detail the method we use to measure the nebular attenuation curve. 
In Section~\ref{sec:results}, we show our results and compare the attenuation curve we derive with those in the literature.
In Section~\ref{sec:explain}, we examine some physical models and observational effects that could potentially explain our results.
{In Section~\ref{sec:impact}, we estimate the impact of the line-specific attenuation on nebular diagnostics.}
We discuss the robustness of our method in Section~\ref{sec:discussions}, and draw our conclusions in Section~\ref{sec:conclusions}.
We also present several tests on different linear regression algorithms and emission line measurements in Appendix~\ref{appendix} and Appendix~\ref{appendix:b}.
{In Appendix~\ref{subsec:high_resol_hii}, we apply our method to a data set with high spatial resolutions from a single galaxy, IC 342.}
For convenience, we only show the wavelength of any given forbidden line when first introduced, unless there are other emission lines with different wavelengths but produced by the same kind of ion.


\section{Data}
\label{sec:data}

\begin{figure*}
    \centering
    \includegraphics[width=0.51\textwidth]{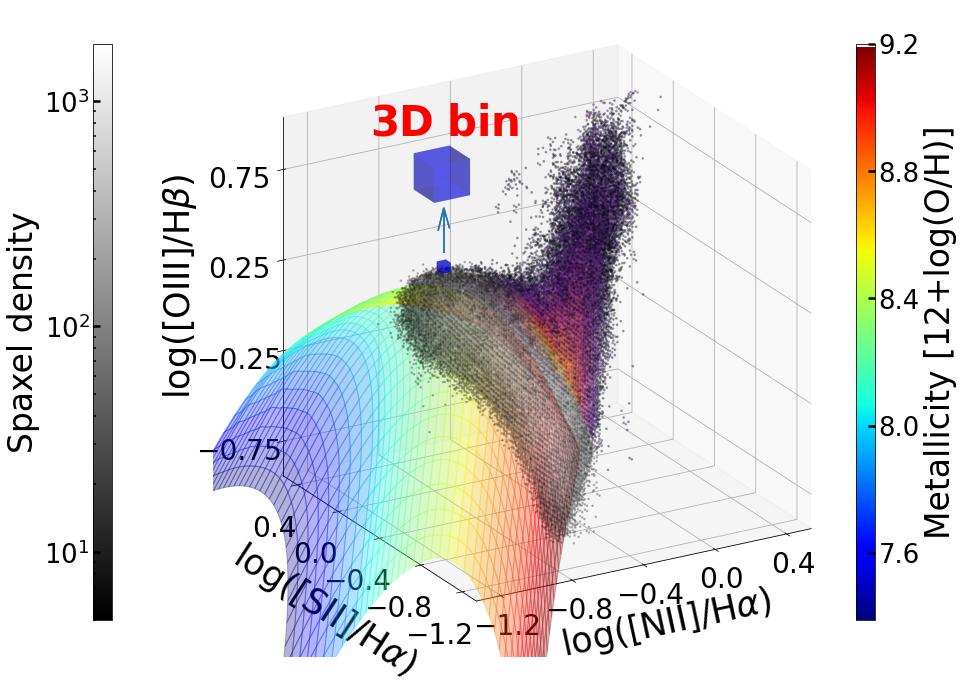}
    \includegraphics[width=0.44\textwidth]{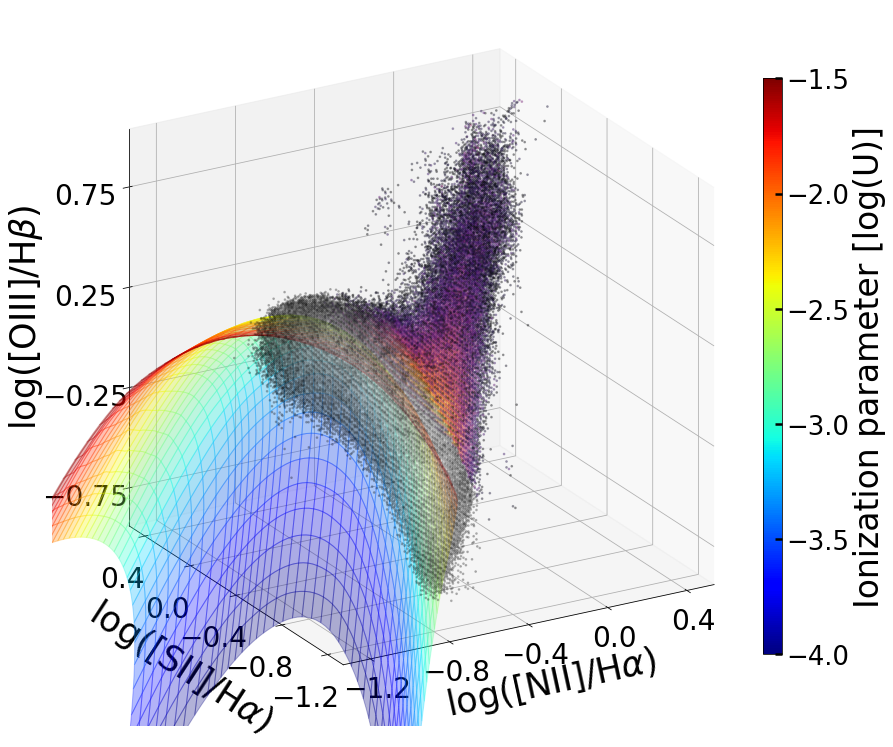}
    \includegraphics[width=0.45\textwidth]{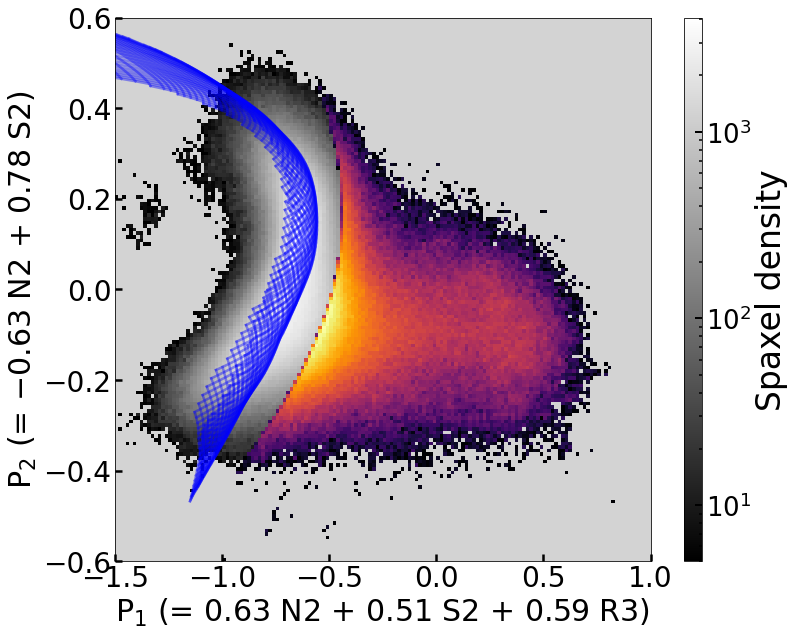}
    \caption{Distribution of the MaNGA sample viewed in a 3D line-ratio space and in one of its 2D projections. {\it Upper left}: 3D density map of the MaNGA sample, where SF spaxels are colored from black to white and other ionized regions are colored from purple to yellow. A photoionization model computed by \protect\cite{ji2020b} using {\sc cloudy} is also shown, whose color coding indicates the metallicity of the simulated \hii\ region.
    The blue cube is an exaggerated illustration of the 3D bins we used in this work.
    {\it Upper right}: the same photoionization model is shown, but color coded according to the ionization parameter of the simulated \hii\ region. {\it Bottom}: a 2D projection of the 3D space that corresponds to an ``edge-on'' view of the data (color map) and model (blue). The two axes, $P_1$ and $P_2$, are linear combinations of the original three axes, which are denoted as N2, S2, and R3. For clarity, we only show the part of the model that covers the middle 98\% of the data along the line of sight. It is clear from this view that the model surface cuts through the center of the SF spaxel distribution.
    }
    \label{fig:sample}
\end{figure*}

Our main sample is drawn from the $11\rm ^{th}$ product launch of the Mapping Nearby Galaxies at Apache Point Observatory survey \citep[MaNGA,][]{bundy2015,yan2016}, which is equivalent to the MaNGA products within the $17\rm ^{th}$ public data release of the Sloan Digital Sky Survey \citep[SDSS DR17,][]{dr17}.
MaNGA is one of the three core programs of SDSS-IV \citep{blanton2017}. It collected integral field unit (IFU) data using the 2.5 m Sloan telescope \citep{gunn2006} for over 10,000 galaxies at $0.01 < z <0.14$ with a nearly flat stellar mass distribution in $10^9 < M_{*}/M_{\odot} < 10^{11}$ \citep{wake2017}. Its IFUs are hexagonal bundles of optical fibers with numbers ranging from 19 to 127 \citep{drory2015}, and a three-point dithering scheme was adopted as the main observing strategy \citep{law2015}. Meanwhile, MaNGA used another set of small fiber bundles to target standard stars for flux calibrations, achieving an absolute calibration uncertainty less than 5\% for 89\% of the wavelength range \citep{yan2016a}. The spectra were obtained through the BOSS spectrographs with a spectral resolution of $R\sim 2,000$ and a wavelength coverage of 3,622 \AA\ $< \lambda <$ 10,354 \AA\ \citep{smee2013}. The observed spectra were first reduced by the Data Reduction Pipeline \citep[DRP,][]{law2016,law2021} and then processed by the Data Analysis Pipeline \citep[DAP,][]{westfall2019,belfiore2019} to produce science products including measurements of emission line fluxes.
Throughout this work, we used the Gaussian fitted emission line fluxes from the DAP (see Appendex~\ref{appendix:b} for the results based on summed fluxes).

We selected our sample using the optical diagnostic diagrams. Specifically, we used the 3D diagnostic diagrams introduced by \cite{ji2020b}, which combines the traditional [N\,{\sc ii}]- and [S\,{\sc ii}]-based BPT diagrams \citep{baldwin1981,veilleux1987}. The selection function is given by
\begin{equation}
    P_1 < -1.57 P_2^2 + 0.53 P_2 - 0.48,
\end{equation}
where
\begin{equation}
    P_1 = 0.63 N2 + 0.51 S2 + 0.59 R3,
\end{equation}
and
\begin{equation}
    P_2 = -0.63 N2 + 0.78 S2.
\end{equation}
Here we have used N2, S2, and R3 to denote the decimal logarithms of \nha ,\sha ,and \ohb .
Before applying this criterion, we excluded spaxels that were flagged by the DAP as problematic or have S/N in H$\alpha$, H$\beta$, [O\,{\sc iii}], [N\,{\sc ii}], [S\,{\sc ii}], [O\,{\sc ii}]$\lambda \lambda 3726,3729$, or [S\,{\sc iii}]$\lambda \lambda 9069,9531$ smaller than 3.
We included [O\,{\sc ii}] and [S\,{\sc iii}] during the sample selection as they are important for determining the blue end and red end of the attenuation curve.

{Our selection criterion corresponds to a roughly fixed fraction of star-formation contribution based on the theoretical models of \cite{ji2020b}\footnote{{We note that limited by MaNGA's spatial resolution, we cannot select individual \hii\ regions. Therefore, each spaxel in our sample showing ionization compatible with star formation possibly includes multiple \hii\ regions and some diffuse ionized regions.}}.}
It also allows us to include some SF spaxels in the outskirt of galaxies that would have been missed by the widely adopted demarcation lines \citep[e.g.,][]{kewley2001,kauffmann2003} based on the 2D diagrams.
Regardless, we note that our conclusions remain largely unchanged if we use the traditional demarcation lines for selection instead.
We also tried the empirical selection of SF spaxels based on gas kinematics \citep{law2021b}, which again produced similar results.
Our final sample consists of $\sim 2.4\times 10^6$ spaxels. Figure~\ref{fig:sample} shows the distribution of our sample in the 3D diagnostic diagram. For now we do not apply any further cut to the data. We check the effect of additional selection criteria on our results in Section~\ref{subsec:depend_obs_pro}.

Apart from our main sample, we also made use of the MaNGA observation of a nearby galaxy IC~342, which is one of the ancillary targets in MaNGA. IC~342 is a nearby \citep[3.3 Mpc,][]{saha2002} massive \citep[$M_* = 10^{9.95} M_{\odot}$,][]{zibetti2009} SF spiral galaxy.
At the distance of IC~342, MaNGA has a spatial resolution of $\sim 32$ pc, which is much higher than the typical resolution of 1\,--\,2 kpc for the MaNGA main sample.
With IC~342, we checked the attenuation law on scales close to the size of individual \hii\ regions. Although we only have one such nearby target in MaNGA, it provides important evidence on the scale-dependence of the nebular attenuation law.
We show the analyses of the IC~342 data in Appendix~\ref{subsec:high_resol_hii}.

\section{Method}
\label{sec:method}

In this section we introduce the method we used to measure the nebular attenuation.
For a given emission line with intrinsic flux $f_{\lambda 0}$, the attenuated flux $f_{\lambda }$ is given by
\begin{equation}
    f_\lambda = 10^{-0.4 A_{\lambda }}f_{\lambda 0},
\end{equation}
where $A_\lambda$ is the wavelength-dependent attenuation in magnitudes. The ratio of two emission lines can then be expressed as
\begin{equation}
    f_{\lambda _1}/f_{\lambda _2} = 10^{-0.4 (A_{\lambda _1} -A_{\lambda _2}) }f_{\lambda _1 0}/f_{\lambda _2 0}.
    \label{eq:flux_ratio}
\end{equation}
If the intrinsic flux ratio, $f_{\lambda _1 0}/f_{\lambda _2 0}$, is known, one can calculate the difference in attenuation between the two emission lines using the observed flux ratio. In SF \hii\ regions, the intrinsic flux ratios of Balmer lines are roughly constants \citep{agn3}. For example, under the case B recombination, the intrinsic ratio between the first two Balmer lines is roughly 2.86. Using the observed fluxes of Balmer lines, one can then measure the relative attenuation at their wavelengths, which is one of the widely adopted methods to determine the nebular attenuation curve.

Apart from Balmer lines, there are many strong forbidden lines from optical to NIR that can be observed in \hii\ regions, such as [O\,{\sc ii}]$\lambda \lambda 3726,3729$, [O\,{\sc iii}]$\lambda 5007$, [N\,{\sc ii}]$\lambda 6583$, [S\,{\sc ii}]$\lambda \lambda 6716,6731$, [S\,{\sc iii}]$\lambda \lambda 9069,9531$, etc. The intrinsic ratio between any two forbidden lines, however, is not a constant and varies among \hii\ regions. In principle, the intrinsic ratio is set by the physical parameters of \hii\ regions. To a first approximation, the variations in the intrinsic line ratios are driven by the variation in the metallicity and ionization parameter. Here the metallicity is defined as the overall chemical abundance in the gas phase, which is usually represented by the oxygen abundance, 12 + log(O/H). The ionization parameter, $U$, is defined as the relative strength of the ionizing radiation, or $\frac{\Phi _0}{n_{\rm H} c}$, where $\Phi _0$ is the flux of the ionizing photons, $n_H$ is the volume density of hydrogen, and $c$ is the speed of light. One way to understand this approximation is to look at the BPT diagrams, where pairs of forbidden-to-Balmer line ratios with similar wavelengths are used. The ratios between these emission lines are insensitive to attenuation and are close to their intrinsic ratios, {if one assumes that they follow the same attenuation curve}. In these diagrams, it is shown that photoioniztion model grids of varying metallicity and ionization parameter can well cover the data locus of SF regions \citep[e.g.,][]{dopita2000,kewley2001,kewley2019,dagostino2019}\footnote{However, we note that some observed line ratios, such as [O\,{\sc i}]$\lambda 6300$/H$\alpha$ and [S\,{\sc iii}]/[S\,{\sc ii}], cannot be correctly reproduced by current photoionization models \citep[see e.g.,][]{mingozzi2020,law2021b}.}.

As shown by \citet{ji2020b}, 2D diagnostic diagrams might suffer from projection effects, making it difficult to tell whether a model grid fits the data distributions in two or more diagrams in a consistent manner.
By combining more than two line ratios to form a high-dimensional diagram, one can put a much more stringent constraint on the position of the ideal model that self-consistently predicts all the line ratios.
In Figure~\ref{fig:sample}, we plotted our best-fit photoionization model grid for MaNGA SF spaxels \citep{ji2020b} by varying the metallicity and ionization parameter of a simulated \hii\ region using the photoionization code {\sc cloudy} \citep{ferland2017}. 
Comparing the area covered by the photoionization model and the spatial distribution of the SF spaxels, we see that the location of any given datum in this line-ratio space is, to a good approximation, determined by a specific combination of a given metallicity and ionization parameter. Variations in other nebular parameters (e.g., hardness of the ionizing spectrum, abundance patterns for secondary elements, etc.) manifest themselves as scatters around the model surface, and can be covered by a more complicated model grid in principle, if we allow changes in these parameters.

With this approximation, we can rewrite Equation~\ref{eq:flux_ratio} as

\begin{equation}
    \begin{aligned}
    &\log \frac{f_{\lambda _1}}{f_{\lambda _2}} = \log (\frac{f_{\lambda _1}}{f_{\lambda _2}})_0 -0.4(A_{\lambda _1} - A_{\lambda _2})\\ &=F_{\lambda _1, \lambda _2}({O/H, U, ...}) + \frac{A_{\lambda _1} - A_{\lambda _2}}{A_{\rm H\alpha} - A_{\rm H\beta}}\log \frac{f_{\rm H\alpha}/f_{\rm H\beta}}{2.86}\\&=F_{\lambda _1, \lambda _2}({O/H, U, ...})
    +m_{\lambda _1, \lambda _2}\log \frac{f_{\rm H\alpha}/f_{\rm H\beta}}{2.86},
    \end{aligned}
    \label{eq:forbidden_ratio}
\end{equation}
where $F_{\lambda _1, \lambda _2}({O/H, U, ...})$ describes the intrinsic ratios between selected lines as a function of the metallicity, ionization parameter, and other nebular parameters of \hii\ regions. Also, we have used the relation between the fluxes of the first two Balmer lines
\begin{equation}
    \log (f_{\rm H\alpha}/f_{\rm H\beta}/2.86) = -0.4(A_{H\alpha} - A_{H\beta}).
\end{equation}
If we define
\begin{equation}
    l_\lambda \equiv \frac{A_\lambda }{A_V},
\end{equation}
to describe the normalized attenuation curve, we have
\begin{equation}
    m_{\lambda _1, \lambda _2}=\frac{l_{\lambda _1}-l_{\lambda _2}}{l_{\rm H\alpha}-l_{\rm H\beta}}.
    \label{eq:slope}
\end{equation}
Here $m_{\lambda _1, \lambda _2}$ describes the relative attenuation difference between $\lambda _1$ and $\lambda _2$ with respect to the attenuation difference between H$\alpha$ and H$\beta$.
The key of our method is to constrain the variation in $F_{\lambda _1, \lambda _2}({O/H,U,...})$, making it negligible compared to the variation in attenuation. If we achieve this, we can easily derive the relative attenuation, $m_{\lambda _1, \lambda _2}$, by fitting a linear model to the ``reddening relation'' between $\log \frac{f_{\lambda _1}}{f_{\lambda _2}}$ and $\log \frac{f_{\rm H\alpha}}{f_{\rm H\beta}}$.

We summarize our method as follows. First, we selected three line ratios that are usually thought to be attenuation-insensitive, which are \nha, \sha, and \ohb. Second, we binned our sample in the 3D space spanned by log(\nha), log(\sha), and log(\ohb) (or N2, S2, and R3 for short) as illustrated in Figure~\ref{fig:sample}.
Following our argument about the variation of line ratios in the 3D diagram, there is no degeneracy between O/H and $U$ in this space.
Therefore, there exists inverse functions to describe the metallicity and ionization parameter with these line ratios, that is, $(O/H) \approx F^{-1}_{0}(N2, S2, R3)$ and $U \approx F^{-1}_{1}(N2, S2, R3)$.
Thus, any small variation in O/H or U corresponds to small variations in line ratios. For example, 
\begin{equation}
    \delta (O/H) \approx c_0 (\delta N2) + c_1 (\delta S2) + c_2 (\delta R3),
\end{equation}
where $c_0, c_1,$ and $c_2$ are coefficients also depending on N2, S2, and R3. We can then rewrite the intrinsic variations for {attenuation sensitive} line ratios as
\begin{equation}
    \begin{split}
    \delta F_{\lambda _1, \lambda _2} & \approx a_0 \delta (O/H) + a_1 \delta U + \sum_i~a_i\delta ({\rm secondary~parameters})\\
    & \approx c_0^{\prime} (\delta N2) + c_1^{\prime} (\delta S2) + c_2^{\prime} (\delta R3).
    \end{split}
\end{equation}
Again, $a_i$ and $c_i^{\prime}$ are coefficients.
It is now clear that if we make the size of each bin small enough, the variation in the metallicity, ionization parameter, and other nebular parameters would be small as well.
In addition, since we binned the data in a line-ratio space that is insensitive to attenuation, we did not constrain the variation in attenuation in each bin. As a result, the variation of the second term in the rightmost hand side of Equation~\ref{eq:forbidden_ratio} dominates over that of the first term. Finally, we performed a linear regression in each bin, treating $\log (f_{\lambda _1}/f_{\lambda _2})$ as a linear function of $\log (f_{\rm H\alpha}/f_{\rm H\beta}/2.86)$. The slope of the relation then gives the relative attenuation with respect to the Balmer lines, as shown in Equation~\ref{eq:slope}.

{We emphasize that our method does not have any dependence on photoionization models, as the selection is completely based on the data.
Our only assumption is that the three line ratios we use are able to constrain the major parameters\footnote{{{These include more than just the metallicity and ionization parameter, as the binning is done in 3D. 
Additionally, we tried using a 4D space by including the stellar mass, which correlates with the variation in the N/O versus O/H relations \citep{belfiore2017,schaefer2022}, and the results remain essentially the same. The drawback of higher-dimensional binning is having fewer data points in each bin and thus having larger statistical uncertainties.}}}
of SF spaxels, which fix their intrinsic line ratios.}

In principle, the smaller the size of the bin, the more accurate our method becomes. However, reducing the bin size also reduces the number of data points within each bin. Therefore, uncertainties associated with small number statistics would become important for bins that are too small. 
In practice, we tried different bin sizes and eventually used a size of $\rm 0.0167\times 0.0167\times 0.0167~dex^3$ for each bin. With this binning scheme, we have $\sim 3000$ bins with more than 180 data points within each bin. Meanwhile, the $\pm 2\sigma$ range of $\log (f_{\rm H\alpha}/f_{\rm H\beta}/2.86)$ is typically $\rm 0.20_{-0.02}^{+0.04}~ dex$ in each bin. 
Before fitting the reddening relations, we removed 3D bins with fewer than 180 data points.
We also tried other cuts that select bins with more than 100 data points and 300 data points.
Since the MaNGA data are more densely populated at high metallicities in the 3D line ratio space, the cut on the number of data points in each bin influences the average metallicity of our sample.
We discuss the effect of different binning scheme and bin selection criteria in detail in Section~\ref{subsec:binning_eff}.

Prior to binning data in 3D, we applied an attenuation correction to line ratios that form the 3D space, assuming a \citet{fitzpatrick1999} extinction curve with $R_V = 3.1$ (F99 hereafter), which is an initial guess as to the shape of the attenuation curve.
This step should in principle have little effect on the derived attenuation. However, as we show in Section~\ref{sec:results}, our results seem to indicate different attenuation curves for different species of lines.
If this is true, both the correction we applied and the assumption of the attenuation-insensitive lines become questionable, which we further discuss in Section~\ref{sec:explain} and Section~\ref{sec:discussions}.
There is another effect associated with our choice of the line ratios that form the 3D space. Results obtained using forbidden lines that are both involved in the 3D space could be correlated by construction. This effect is discussed in Section~\ref{sec:results} and Section~\ref{sec:discussions}.

The method we used to perform the linear regression for Equation~\ref{eq:forbidden_ratio} is a maximum likelihood method taking into account uncertainties in both variables as well as the intrinsic scatter in the dependent variable. The natural logarithm of the likelihood function for a given data point, $i$, is given by
\begin{equation}
\begin{aligned}
    \ln F_{likelihood,i} & = -0.5(y_i-mx_i-b)^2/(m^2\sigma _{x_i}^2 + \sigma _{y_i}^2 + \sigma _0^2) \\
    & = -0.5 \chi ^2_i,
\end{aligned}
\label{eq:lf_i}
\end{equation}
where $m$ and $b$ are the slope and intercept, and $\sigma _{x_i}$, $\sigma _{y_i}$, and $\sigma _0$ are the measurement uncertainties in $x_i$, $y_i$, and the intrinsic scatter in $y$ (which is the same for all data points during the fit), respectively\footnote{{The numerator of $\chi ^2$ indicates the separation is calculated along the $y$ axis in this linear model. However, we note that constructing the likelihood function from the perspective of orthogonal distance regression (ODR) using, for example, Equation 32 of \cite{hogg2010}, leads to the same final expression for the $\chi ^2$ term.}}.
We did not consider any intrinsic scatter in $x$, as the Balmer ratios are relatively insensitive to temperature and density variations \citep{agn3}. Still, we discuss the effect of including it in Section~\ref{subsec:int_scatter}.
We estimated $\sigma _{x_i}$ and $\sigma _{y_i}$ by propagating the measurement uncertainty of individual emission line.
Given the test made by \cite{belfiore2019} on emission line measurements in MaNGA, we multiplied the uncertainties reported by DAP by a factor of 1.25. We note that this adjustment does not have any significant impact on our results.
We started by adopting an initial guess for the intrinsic scatter $\sigma _0 = 0$, and minimized the total $\chi ^2$ using the {\sc minimize} function in {\sc scipy} to obtain the slope and intercept with the \cite{nelder1965} method.
We then checked whether the following condition is satisfied:
\begin{equation}
    \sum_{i=1}^{N} \chi ^2_i/(N-2) \leq 1,
    \label{eq:stop}
\end{equation}
where N is the number of data points.
If not, we changed the intrinsic scatter by an increment of $0.0001$ and recomputed the slope and intercept.
We repeated the above procedure until Equation~\ref{eq:stop} is satisfied.
The method is essentially identical to the one adopted by \citet{tremaine2002}. We note that there are other ways to estimate the slope and intercept when both heteroscedastic uncertainties and intrinsic scatter are present \citep[see e.g.,][]{kelly2007}. We have verified the robustness of this method and present the tests in Appendix~\ref{appendix}.

\section{Results}
\label{sec:results}

\begin{figure*}
    \includegraphics[width=0.31\textwidth]{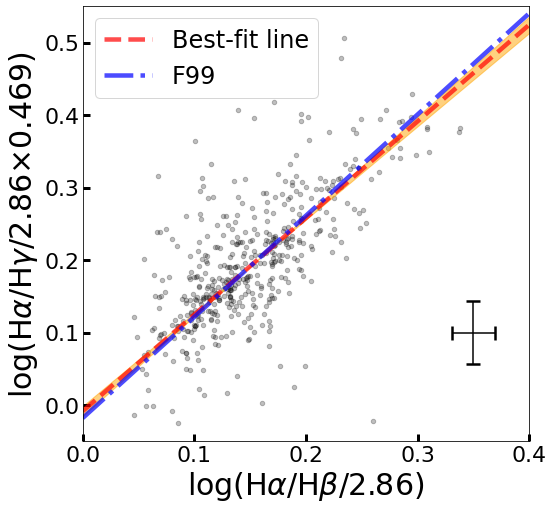}
    \includegraphics[width=0.31\textwidth]{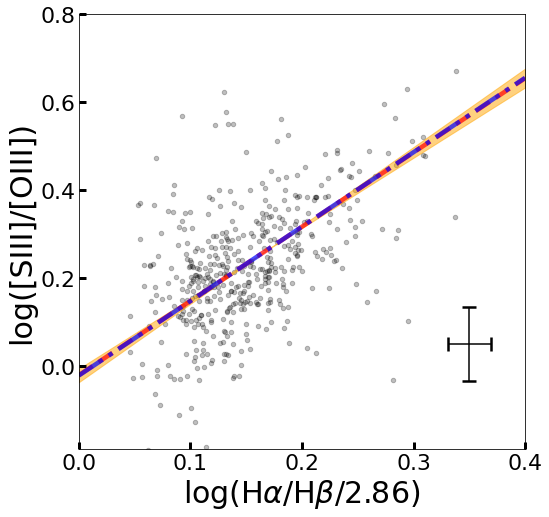}
    \includegraphics[width=0.32\textwidth]{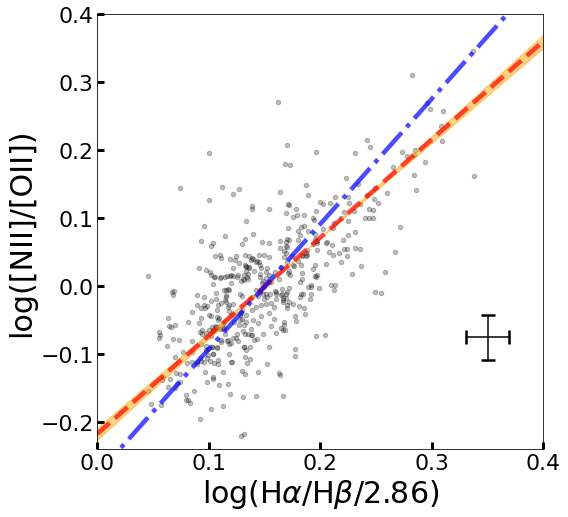}
    \caption{Fitting examples in one of the 3D bins. The dashed red lines are the best-fit lines given by the maximum-likelihood method. The orange shaded regions indicate the $1\sigma$ uncertainties of the linear models. The dash-dotted blue lines are obtained by fixing the slopes using the values from an F99 extinction curve with $R_V = 3.1$ during the fit.
    The error bars indicate the measurement uncertainties in different logarithmic line ratios.
    {\it Left}: the reddening relation between H$\alpha$/H$\beta$ and H$\alpha$/H$\gamma$.
    {\it Middle}: the reddening relation between H$\alpha$/H$\beta$ and \so.
    {\it Right}: the reddening relation between H$\alpha$/H$\beta$ and \no.
    }
    \label{fig:bin_eg}
\end{figure*}

In this section we compute the relative nebular attenuation ``seen'' by emission lines from different species of ions and atoms. We investigate three categories of lines according to their production mechanisms as well as the ionization energies (IE) of their corresponding ions: 1) Balmer lines, including $\rm H\alpha$, $\rm H\beta$, $\rm H\gamma$, and $\rm H\delta$; 2) high ionization lines ($\rm IE > 13.6~eV$), including [Ne\,{\sc iii}]$\lambda \lambda 3869,3967$, [O\,{\sc iii}]$\lambda 5007$, and [S\,{\sc iii}]$\lambda \lambda 9069,9531$; 3) low ionization lines ($\rm IE \lesssim 13.6~eV$), including [O\,{\sc ii}]$\lambda \lambda 3726,3729$, [N\,{\sc ii}]$\lambda 6583$, [S\,{\sc ii}]$\lambda \lambda 6716,6731$, and [O\,{\sc i}]$\lambda 6300$. We summarized our results in Table~\ref{tab:derived}. For the rest of the paper, we use $m_{\rm line _1, line _2}$ to represent the slope of the reddening relation, $\log (f_{\rm line _1}/f_{\rm line _2})$ versus $\log (f_{\rm H\alpha}/f_{\rm H\beta})$, and use $m^{\prime}_{\rm line _1, line _2}$ to represent the measured median value of the slope.

\subsection{Attenuation seen by Balmer lines}

\begin{figure*}
    \includegraphics[width=0.95\textwidth]{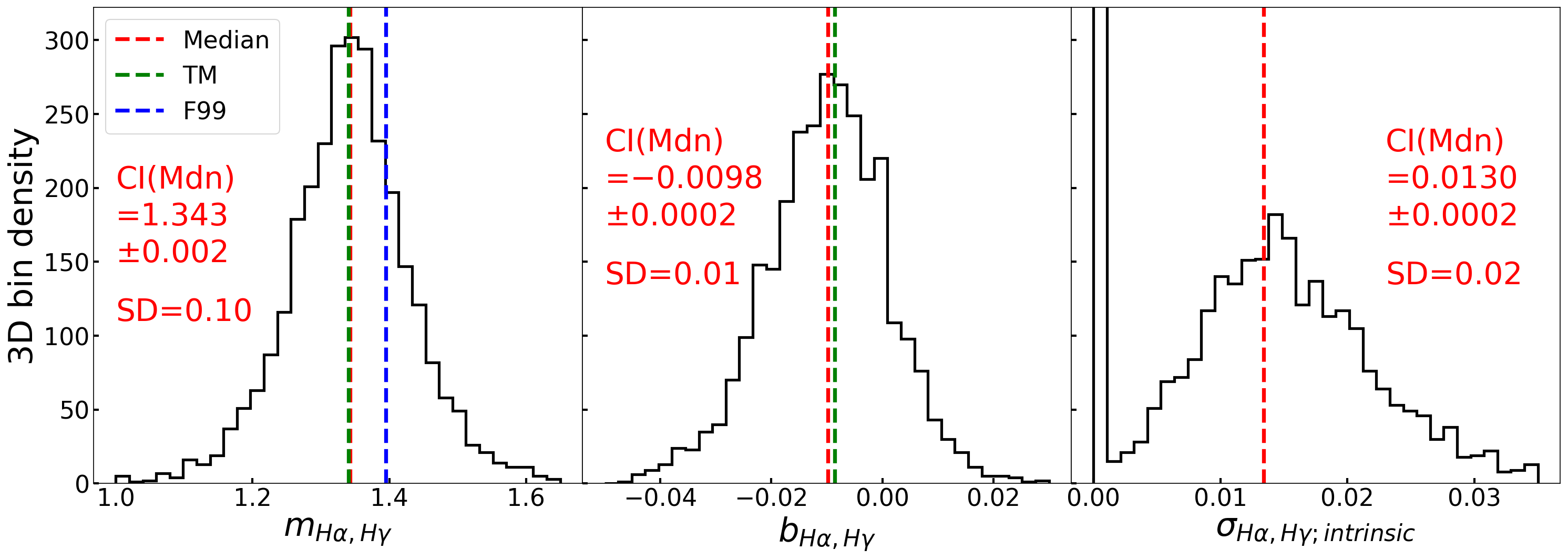}
    \includegraphics[width=0.95\textwidth]{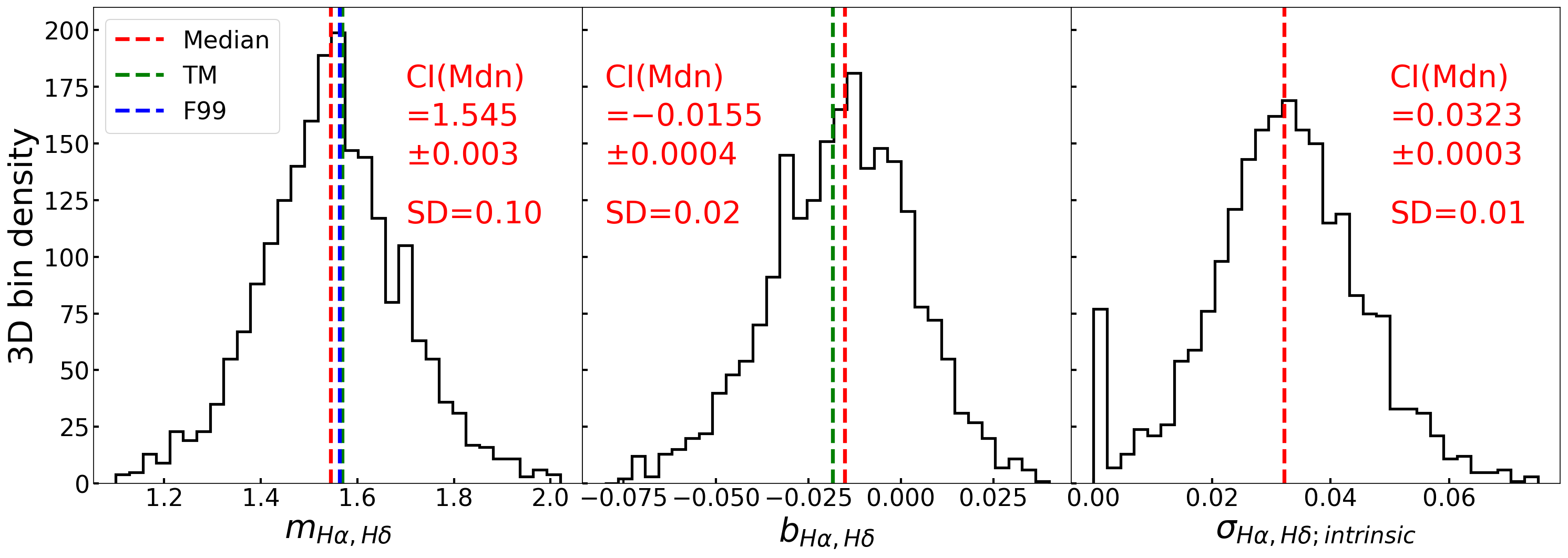}
    \caption{Relative attenuation probed by Balmer lines. The first row shows the results obtained by fitting a linear function to the log(H$\alpha$/H$\gamma$/2.86$\times 0.469$) versus log(H$\alpha$/H$\beta$/2.86) relation within each 3D bin we defined. From left to right, we plot the distributions of the slope, intercept, and the intrinsic scatter.
    {For each distribution, we show the 68\% confidence interval of the median [CI(Mdn)] and the standard deviation (SD).}
    The dashed-red lines indicate the median values among our 3D bins.
    The dashed-green lines indicate the values obtained by the traditional method where the whole sample is used for a single fit. The dashed-blue lines correspond to the values given by the F99 extinction curve. 
    The second row shows similar information, but is obtained by fitting a linear function to the log(H$\alpha$/H$\delta$/2.86$\times 0.259$) versus log(H$\alpha$/H$\beta$/2.86) relation.
    }
    \label{fig:balmer}
\end{figure*}

We calculated two slopes related to Balmer lines: 1) the slope of log(H$\alpha$/H$\gamma$/2.86$\times 0.469$) versus log(H$\alpha$/H$\beta$/2.86) relation (or $m_{\rm H\alpha ,H\gamma }$); 2) the slope of log(H$\alpha$/H$\delta$/2.86$\times 0.259$) versus log(H$\alpha$/H$\beta$/2.86) relation (or $m_{\rm H\alpha ,H\delta }$).
Here we have assumed the intrinsic Balmer ratios to follow the Case B approximation values at $n_e \sim 100~{\rm cm}^{-3}$ and $T\sim 10^4 {\rm K}$ \citep{agn3}. 
{We divided the line ratios by their intrinsic values so that the intercepts, log(line ratio$_{\rm obs.}$/line ratio$_{\rm theor.}$)|$_{\rm H\alpha/H\beta=2.86}$, ideally should be zero.} 
We note that there is a small covariance between the independent and dependent variables due to the common term of H$\alpha$, which is taken into account by introducing an extra term of $-2mCov(x,y)$ in the likelihood function (see Appendix~\ref{appendix} for more details). We also required that the S/N of all emission lines used to be greater than 3.
We calculated these slopes in two different ways. The first way is the traditional method (hereafter TM), which uses the entire sample for the calculation. The second way is our new method, where we calculated slopes in 3D bins and used the median as the representative value. For Balmer lines, these two methods should give us identical results, which serves as a sanity check for our method.

{The left panel of Figure~\ref{fig:bin_eg} shows a fitting example in one of the 3D bins, where a clear linear relation between log(H$\alpha$/H$\beta$) and log(H$\alpha$/H$\gamma$) is present. The best-fit linear model shows a small uncertainty of 0.046 on the slope, and is in good agreement with the prediction from an F99 extinction curve with $R_V = 3.1$.}
Figure~\ref{fig:balmer} shows the distributions of the slopes, intercepts, and intrinsic scatters of the two reddening relations.
For an F99 extinction curve with $R_V = 3.1$, the expected slopes are $m^{\rm F99}_{\rm H\alpha ,H\gamma} = 1.395$ and $m^{\rm F99}_{\rm H\alpha ,H\delta} = 1.563$. When $R_V = 4.05$, these values become $m^{\rm F99}_{\rm H\alpha ,H\gamma} = 1.384$ and $m^{\rm F99}_{\rm H\alpha ,H\delta} = 1.542$.
In comparison, the slopes obtained by TM are {$m_{\rm H\alpha ,H\gamma} = 1.3400\pm 0.0005$ and $m_{\rm H\alpha ,H\delta} = 1.5683\pm 0.0009$,} where the uncertainties are calculated with a Markov chain Monte Carlo (MCMC) method using the {\sc emcee} package in {\sc python} \citep{emcee}.
If we change the S/N threshold for H$\gamma$ and H$\delta$ from 3 to 10, the resulting slopes become {$m_{\rm H\alpha ,H\gamma} = 1.3293\pm 0.0005$ and $m_{\rm H\alpha ,H\delta} = 1.503\pm 0.001$.}
While the median slopes given by our new method are 
$m^{\prime}_{\rm H\alpha ,H\gamma} = 1.343\pm 0.002$
and 
{$m^{\prime}_{\rm H\alpha ,H\delta} = 1.545\pm 0.003$}, where the uncertainties correspond to the 68\% confidence intervals derived by using the biweight estimator of \citet{beers1990}.
As a sanity check, we also ran an MCMC analysis and confirmed the uncertainties given by the MCMC were in good agreement with those returned by the biweight estimator.
The $1\sigma$ widths (or the standard deviations) of the slope distributions are 0.08 and {0.16}, respectively.

The results given by TM and our new method are close to each other. Our new method inevitably produces larger uncertainties. On the one hand, the intrinsic scatter in line ratios can have a larger influence on the derived slopes for bins with relatively small number of data points.
On the other hand, there could be genuine variation in the shape of the attenuation curve depending on the nebular parameters or host galaxy properties, which are not included in the calculation of the uncertainties (see Section~\ref{subsec:depend_obs_pro}).
{The slopes derived from TM cannot be described by any single F99 extinction curve with a fixed $R_V$. Given the dependence of these slopes on the S/N cut, it is possible that the sample has intrinsic variations in $R_V$ that depends on galaxy properties \citep[see][]{salim2020}. Regardless, the corrections based on the F99 extinction curve with $R_V = 3.1$ would only introduce a bias less than {2\%} for H$\alpha$/H$\gamma$ and H$\alpha$/H$\delta$ at $A_V = 1$~mag.
}
A similar conclusion can be drawn for the median slopes derived with the new method.
These results confirm that the attenuation probed by Balmer lines in galaxies can be approximately described by an average MW extinction curve {\citep[e.g.,][]{wild2011,reddy2020, rezaee2021}}.

The intercepts obtained by two methods also show good consistency. It is noteworthy that the intercepts we derived are slightly larger than 0. This could be due to the fact that the average intrinsic Balmer ratios are different from the values we assumed. Regardless, this offset does not affect the slopes we derived. We also notice that the median intrinsic scatter in $y-$axis [i.e., log(H$\alpha$/H$\gamma$/2.86$\times 0.469$) or log(H$\alpha$/H$\delta$/2.86$\times 0.259$)] in each bin is small compared to the change in $y$. This ensures that the variation in $y$ within each 3D bin is indeed mostly driven by the variation in $x$, or log(H$\alpha$/H$\beta$/2.86).

\subsection{Attenuation seen by high ionization lines}

\begin{figure*}
    \includegraphics[width=0.95\textwidth]{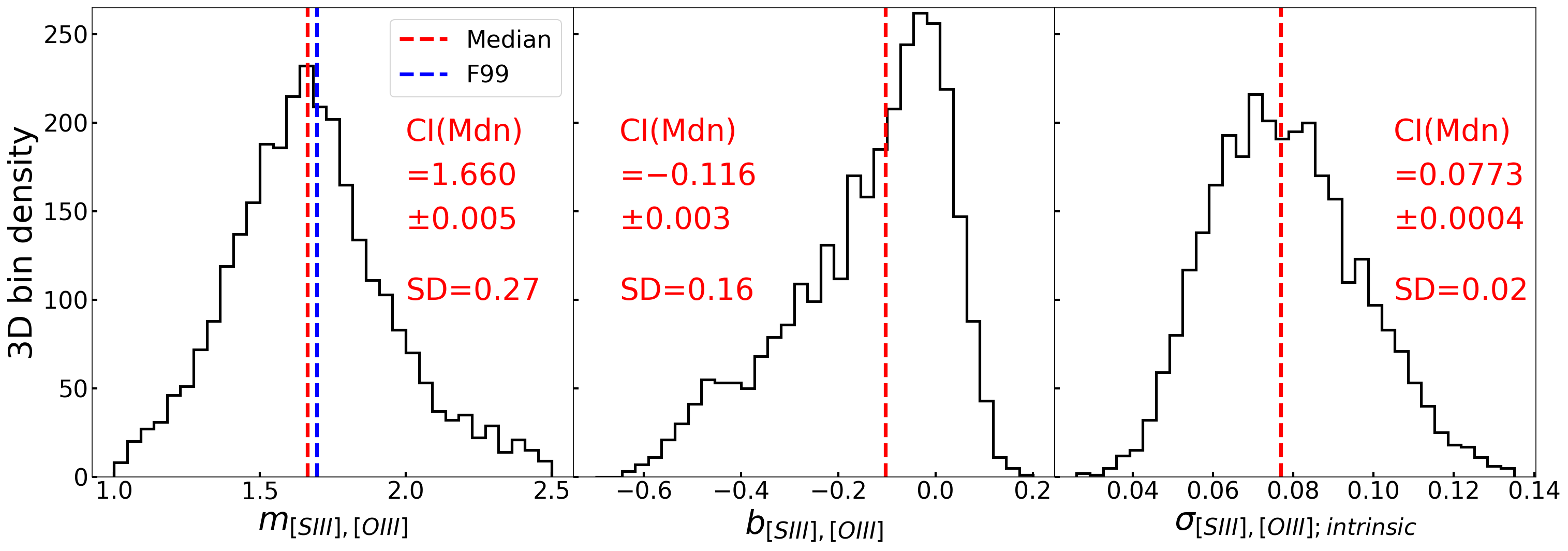}
    \includegraphics[width=0.95\textwidth]{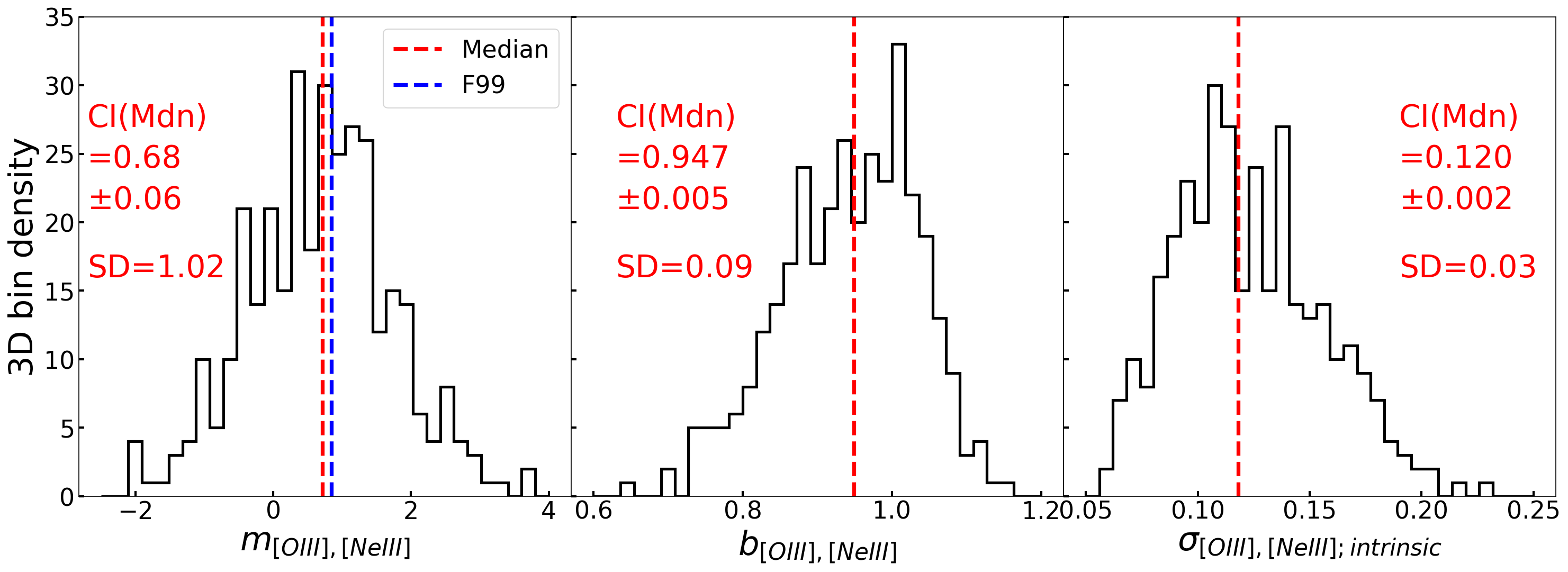}
    \caption{Relative attenuation probed by high ionization lines. The first row shows the results obtained by fitting a linear function to the log(\so) versus log(H$\alpha$/H$\beta$/2.86) relation within each 3D bin we defined. From left to right, we plot the distributions of the slope, intercept, and the intrinsic scatter.
    {For each distribution, we show the 68\% confidence interval of the median [CI(Mdn)] and the standard deviation (SD).}
    The dashed-red lines indicate the median values among our 3D bins. The dashed-blue lines correspond to the values given by the F99 extinction curve. The second row shows similar information, but is obtained by fitting a linear function to the log([Ne\,{\sc iii}]/[O\,{\sc iii}]) versus log(H$\alpha$/H$\beta$/2.86) relation.
    }
    \label{fig:high_ion}
\end{figure*}

Using our 3D binning method, we calculated slope of the log([S\,{\sc iii}]/[O\,{\sc iii}]) versus log(H$\alpha$/H$\beta$) relation as well as the slope of the log([O\,{\sc iii}]/[Ne\,{\sc iii}]) versus log(H$\alpha$/H$\beta$) relation.
Since DAP ties the attenuation-uncorrected fluxes of the [S\,{\sc iii}] doublet, making $f_{{\rm [SIII]}\lambda 9531}/f_{{\rm [SIII]}\lambda 9069}\approx 2.439$, the derived slope distributions for these two lines are identical.
Considering that the attenuation at NIR is relatively weak and is less dependent on wavelength, tying the observed fluxes of these two lines should not have a significant effect on our measured nebular attenuation. For [Ne\,{\sc iii}] doublet, DAP also sets a fixed ratio for the attenuation-uncorrected fluxes, making $f_{{\rm [NeIII]}\lambda 3869}/f_{{\rm [NeIII]}\lambda 3967}\approx 3.33$. Although the wavelengths of [Ne\,{\sc iii}] lines are much shorter compared to [S\,{\sc iii}], the wavelength separation of the [Ne\,{\sc iii}] doublet is much smaller. 
The F99 extinction curve predicts that $(m^{\rm F99}_{{\rm [OIII],[NeIII]}\lambda 3869} - m^{\rm F99}_{{\rm [OIII],[NeIII]}\lambda 3967})/m^{\rm F99}_{\rm [OIII],[NeIII]}\approx 9\%$. This is much smaller than the width of the slope distribution we derived below. Since the measurement uncertainties of [Ne\,{\sc iii}] are relatively large, an S/N cut of 3 already removes a large number of data points. To have enough 3D bins, we lowered the minimum number of spaxels required for each bin from 180 to 100. Even with this lowered cut, we only had {336} qualified 3D bins to derive the median reddening relation for log([O\,{\sc iii}]/[Ne\,{\sc iii}]).

{In the middle panel of Figure~\ref{fig:bin_eg}, we plotted the fitting result for the log(\so) versus log(H$\alpha$/H$\beta$) relation in one of the 3D bins. The overall scatter along the $y$ axis is larger compared to the case of Balmer lines, and a large intrinsic scatter of 0.06 is returned by the fitting function, indicating there is still remaining variations in the intrinsic ratio of these forbidden lines. Regardless, the best-fit linear model matches the prediction from the F99 curve well in this case.}
Figure~\ref{fig:high_ion} shows the distribution of the slopes we derived.
For the [S\,{\sc iii}] doublet, we show the result of [S\,{\sc iii}]$\lambda 9531$. For the [Ne\,{\sc iii}] doublet, we show the result of [Ne\,{\sc iii}]$\lambda 3869$. The slope distributions derived from the other [S\,{\sc iii}] and [Ne\,{\sc iii}] lines are identical to these.
We found that the 68\% confidence intervals for the central locations of the two slopes distributions are 
$m^{\prime}_{\rm [SIII],[OIII]} = 1.660\pm 0.005$
and {$m^{\prime}_{\rm [OIII],[NeIII]} = 0.68\pm 0.06$}. 
Meanwhile, the standard deviations of the two slope distributions are 0.27 and {1.04}, respectively.
Although we applied an S/N cut to [Ne\,{\sc iii}] lines, the resulting slope distribution is still very wide. 
Part of the reason could be due to our inclusion of the 3D bins having fewer data points.
In addition, the measurements of [Ne\,{\sc iii}] fluxes could be influenced by the nearby absorption features in observed spectra.
Despite the large uncertainties, our derived slopes still lie close to the values given by the F99 extinction curve with $R_V = 3.1$, which are $m^{\rm F99}_{\rm [SIII],[OIII]} = 1.695$ and $m^{\rm F99}_{\rm [OIII],[NeIII]} = 0.81$, after averaging the values for the two lines in each doublet.
{At $A_V = 1$ mag, using the F99 corrections would cause a bias of 1.3\% for \so\ and a bias of 5\% for [O\,{\sc iii}]/[Ne\,{\sc iii}].}
Therefore, the attenuation for both Balmer lines and high ionization lines in our sample galaxies can be approximately described by the F99 extinction curve. With the inclusion of the [S\,{\sc iii}] doublet, we were able to constrain the nebular attenuation curve redward of H$\alpha$.

From Figure~\ref{fig:high_ion}, one can see that the intrinsic scatter found in the reddening relations for high ionization lines is much larger compared to those for Balmer lines, which contributes to the wider slope distributions. For log([O\,{\sc iii}]/[Ne\,{\sc iii}]), the median intrinsic scatter is {0.118 dex}. 
The standard deviation of log(H$\alpha$/H$\beta$) in each 3D bin is typically 0.05 dex. Combining this value with $\rm m_{[OIII],[NeIII]} \approx 0.8$, one expects that the attenuation-driven variation in log([O\,{\sc iii}]/[Ne\,{\sc iii}]) in each 3D bin is smaller than the intrinsic scatter. This further explains why the estimation of $\rm m_{[OIII],[NeIII]}$ is more uncertain.

\subsection{Attenuation seen by low ionization lines}

\begin{figure*}
    \includegraphics[width=0.95\textwidth]{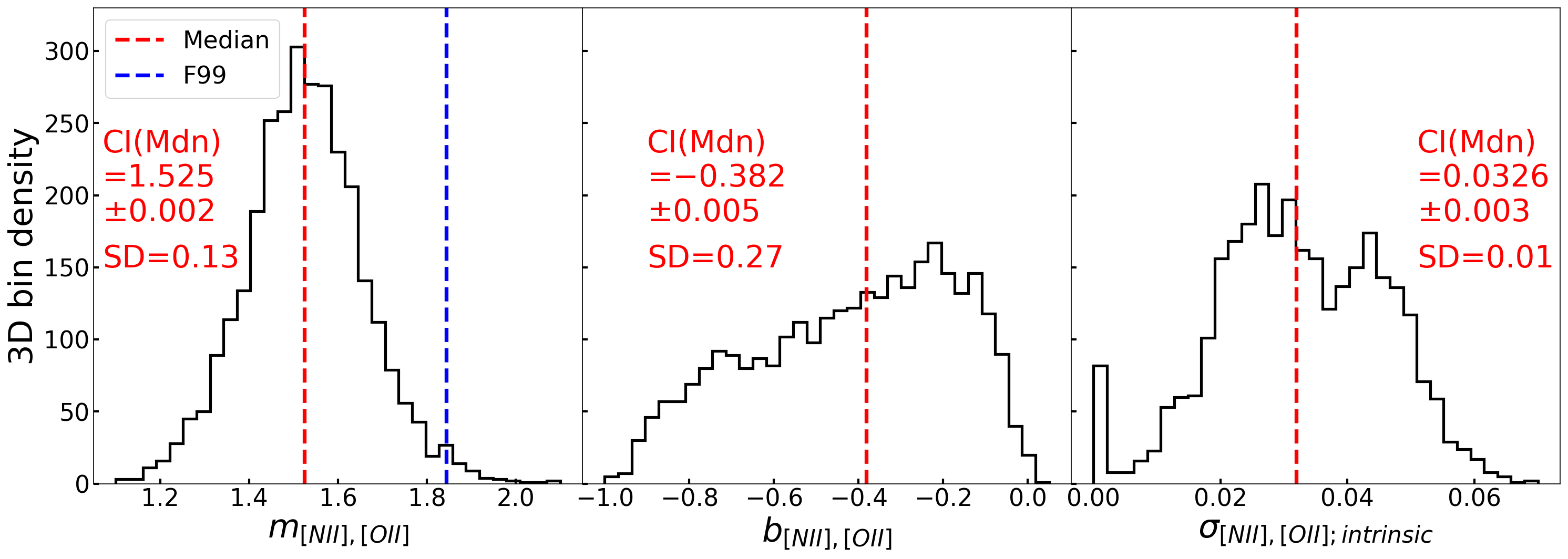}
    \caption{Relative attenuation probed by low ionization lines. We fit a linear function to the log(\no) versus log(H$\alpha$/H$\beta$/2.86) relation within each 3D bin. From left to right, we plot the distributions of the slope, intercept, and the intrinsic scatter.
    {For each distribution, we show the 68\% confidence interval of the median [CI(Mdn)] and the standard deviation (SD).}
    The dashed-red lines indicate the median values among our 3D bins. The dashed-blue lines correspond to the values given by the F99 extinction curve.
    }
    \label{fig:low_ion}
\end{figure*}

\begin{figure}
    \includegraphics[width=0.46\textwidth]{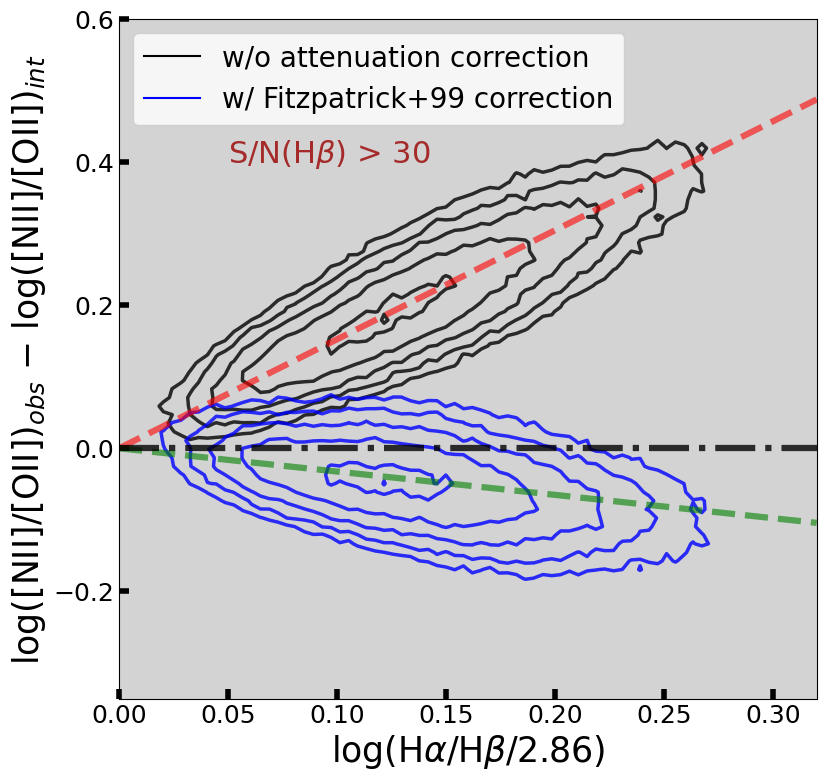}
    \caption{The reddening relation between log(\no) and log(H$\alpha$/H$\beta$/2.86). 
    Only data with S/N > 30 in H$\beta$ are included.
    We obtained the intrinsic log(\no) value at log(H$\alpha$/H$\beta$/2.86) = 0 and subtracted it from the observed log(\no) for each 3D bin. The black contours show the distribution of the data without any attenuation correction. Whereas the blue contours show the density distribution of the data with their log(\no) corrected by the F99 extinction curve.
    The contour levels trace the number density of the data points and are equally spaced on a log scale. The outermost contour and the innermost contour enclose 90\% and 10\% of the data, respectively.
    The dashed red line and the dashed green line trace the median trends in the 3D bins for the two data sets.
    The dash-dotted line is a horizontal line for reference.
    }
    \label{fig:n2o2_f99}
\end{figure}

We examined the attenuation seen by low ionization line ratios in this subsection.
Specifically, we used line ratios involving [O\,{\sc ii}] to constrain the attenuation curve blueward of H$\delta$.

{The right panel of Figure~\ref{fig:bin_eg} shows a reddening relation between log(\no) and log(H$\alpha$/H$\beta$) in one of the 3D bins. In this case, unlike the linear models for Balmer lines and high ionization lines, the best-fit model shows a clear offset from the line predicted by the F99 curve.}
We plotted the distributions of the linear parameters of all 3D bins in Figure~\ref{fig:low_ion}. Our method yielded 
$m^{\prime}_{\rm [NII],[OII]} = 1.525\pm 0.002$
, and the standard deviation of the slope distribution, $\sigma _{\rm std}$, is 0.13.
In comparison, the F99 extinction curve predicts that $m^{\rm F99}_{\rm [NII],[OII]} = 1.844$. 
{Our derived slope is smaller than the corresponding F99 value by roughly $2.4\sigma _{std}$ or $126\sigma ^{\prime}$, where $\sigma ^{\prime}$ is the uncertainty of the median.
This difference is large enough to drive a systematic bias of 13\% in the attenuation-corrected \no\ at $A_V=1$~mag, if one uses the F99 extinction curve.
}
Furthermore, the derived slope is significantly smaller than $m^{\prime}_{\rm H\alpha ,H\delta}$, which cannot be explained by {any single extinction curve}. Since $\rm \lambda _{[NII]} > \lambda _{H\alpha}$ and $\rm \lambda _{[OII]} < \lambda _{H\delta}$, for any given extinction curve that predicts the extinction to monotonically increase with decreasing wavelength in the optical, one expects $\rm A_{H\delta} - A_{H\alpha} < A_{[OII]} - A_{[NII]}$ and thus $\rm m_{H\alpha ,H\delta} < m_{[NII],[OII]}$, contrary to the observed relation.
Therefore, our result implies that the F99 extinction curve correctly describes the attenuation probed by Balmer lines and high ionization lines, but overpredicts the attenuation probed by low ionization lines.
{If we interpret this result as [N\,{\sc ii}] and [O\,{\sc ii}] both having a different magnitude of attenuation compared to that of Balmer lines, that is, $A_{V, {\rm Low}} \ne A_{V,{\rm Balmer}}$, then from the definition of the reddening relation, we have}

\begin{equation}
    A_{V,{\rm Low}} \approx \frac{m^{\prime}_{\rm [NII],[OII]}}{m^{\rm F99}_{\rm [NII],[OII]}}A_{V,{\rm Balmer}}\approx 0.83A_{V,{\rm Balmer}}.
\end{equation}
{However, as we discuss in Section~\ref{3d_att_bias}, simply using different $A_V$ cannot explain our results and one should not use this derived ``apparent $A_V$'' to do corrections. Another complicating factor is that different low ionization lines do not necessarily have the same $A_V$ either (see Section~\ref{subsec:hybrid} and Section~\ref{subsec:toy_model}).
}

Another way to see this discrepancy is to inspect log(\no) corrected by the F99 extinction curve as a function of log(H$\alpha$/H$\beta$) in each 3D bin. Figure~\ref{fig:n2o2_f99} shows such an example.
We removed the fitted intercept for the log(\no) versus log(H$\alpha$/H$\beta$/2.86) relation in each 3D bin (which effectively removed the intrinsic value of log(\no) at zero attenuation), and plotted all the intercept-removed data in this figure.
Clearly, if \no\ is corrected by the F99 extinction law before fitting, the overall slope becomes negative. This implies that the F99 extinction law overcorrects \no\ in our sample. Meanwhile, the distribution of $m^{\prime}_{\rm [NII],[OII]}$ is relatively tight and the intrinsic scatter is small, indicating that the intrinsic scatter plays a minor role in affecting our measurement of the slope.
{In Figure~\ref{fig:n2o2_f99}, we only show data with high S/N in H$\beta$. This is because the measurement uncertainty in log(H$\alpha$/H$\beta$/2.86) can make the attenuation-corrected trend to appear steeper than it should be as the measured value of log(H$\alpha$/H$\beta$/2.86) is used to derive the correction. This effect is negligible when the S/N of H$\beta$ is high. However, we note that if there is intrinsic scatter along the $x$ axis, a similar effect can occur even at high S/N. We discuss the effect associated with the potential intrinsic scatter in detail in Section~\ref{subsec:int_scatter}.
}

We made a similar measurement using another combination of low ionization lines, [S\,{\sc ii}] and [O\,{\sc ii}], and found that $m^{\prime}_{\rm [SII] ,[OII]}$ is also significantly smaller than the prediction of the F99 curve, as shown in Table~\ref{tab:derived}.
One might wonder if this discrepancy is due to some hidden bias induced by our 3D binning method. Indeed, [N\,{\sc ii}] and [S\,{\sc ii}] are among the emission lines we used to construct the 3D space. As a sanity check, we computed $m_{\rm H\gamma ,[OII]}$ using our 3D binning method. Both H$\gamma$ and [O\,{\sc ii}] are not involved in the construction of the 3D bins.
We found that
$m^{\prime}_{\rm H\gamma ,[OII]} = 0.161\pm 0.003$
(whose $\sigma _{\rm std}$ is 0.14), which is much smaller than $m^{F99}_{\rm H\gamma ,[OII]} = 0.439$. This again indicates that Balmer lines and low ionization lines (in this case the [O\,{\sc ii}] line) are not subject to the same attenuation law.

\subsection{Line ratios involving different species of ions}
\label{subsec:hybrid}

\begin{figure}
    \centering
    \includegraphics[width=0.44\textwidth]{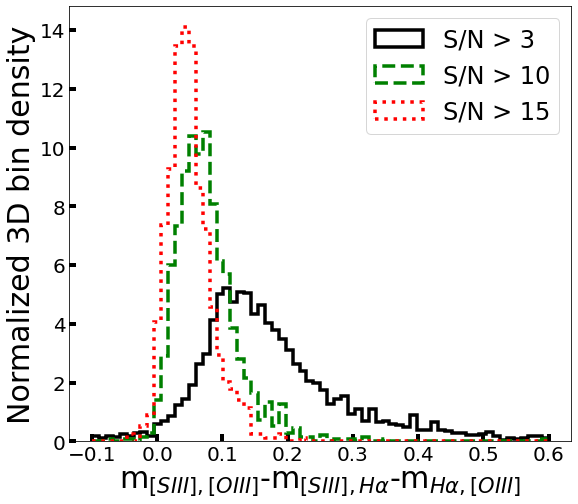}
    \caption{Distributions of the slope differences under different S/N cuts. The black solid histogram shows data with S/N of [S\,{\sc iii}], [O\,{\sc iii}], and H$\beta$ greater than 3; the dashed green histogram shows data with S/N of the aforementioned lines greater than 10; the dotted red histogram shows data with S/N of the aforementioned lines greater than 15. The slope difference peaks toward smaller values with the increasing S/N limit.
    {However, the difference is actually due to the dependence of the slope on the surface brightness of the lines.}
    }
    \label{fig:add_s3o3}
\end{figure}

In previous subsections, we found evidence suggesting different attenuation laws for different species of lines.
To further investigate this issue, we explored slopes for reddening relations involving combinations of Balmer lines, high ionization lines, and low ionization lines (which we call hybrid line ratios hereafter).

We first checked [S\,{\sc iii}]/H$\alpha$, which is a hybrid line ratio combining a high ionization line and a Balmer line. Judging from our previous measurements, one might expect that the derived slope for this line ratio should be consistent with the prediction of the F99 curve. However, we found that $m^{\prime}_{\rm [SIII] ,H\alpha } = 0.600\pm 0.006$ (the $\sigma _{\rm std}$ is 0.28), which deviates from the F99 value, $m^{\rm F99}_{\rm [SIII] ,H\alpha } = 0.811$, by $0.75\sigma _{\rm std}$ or $38\sigma ^{\prime}$.
{The difference indicates a fractional bias of 8\% for the attenuation-corrected [S\,{\sc iii}]/H$\alpha$ at $A_V = 1$~mag, if one uses the F99 extinction curve for corrections.
}
In fact, for almost all of the hybrid line ratios we checked in Table~\ref{tab:derived}, the slopes of their reddening relations appear considerably lower than the corresponding F99 values.

Another noticeable effect is related to the additivity of the slope measurements.
Ideally, one expects if line ratios A, B, and C satisfy $\log(A) = \log(B) + \log(C)$, then $m^{\prime}_{A}=m^{\prime}_{B}+m^{\prime}_{C}$. All line ratios should follow the law of additivity if they are subject to the same attenuation curve.
From Table~\ref{tab:derived}, we see that some slopes do show additivity. For example, $m^{\prime}_{\rm [SIII] ,H\beta } = m^{\prime}_{\rm [SIII] ,H\alpha } + m^{\prime}_{\rm H\alpha ,H\beta } = m^{\prime}_{\rm [SIII] ,H\alpha } + 1$, $m^{\prime}_{\rm [SIII] ,[OII] } \approx m^{\prime}_{\rm [SIII] ,[OIII] }+m^{\prime}_{\rm [OIII] ,[OII] }$, etc.
However, there are also a few cases where the additivity seems violated. For instance, $m^{\prime}_{\rm [SIII] ,[OIII] } > m^{\prime}_{\rm [SIII] ,H\alpha } + m^{\prime}_{\rm H\alpha , H\beta} + m^{\prime}_{\rm H\beta , [OIII]} \approx m^{\prime}_{\rm [SIII] ,H\alpha } + 0.89$, where the difference is roughly 0.16, 
{equivalent to $1.5\sigma _{std}$ or $87\sigma ^{\prime}$ ($\sigma _{\rm std}$ and $\sigma ^{\prime}$ are calculated from the distribution of $m_{\rm [SIII] ,[OIII] } - m_{\rm [SIII] ,H\alpha } - m_{\rm H\alpha , H\beta}$).}
Also, $m^{\prime}_{\rm [NII] ,[OII] } > m^{\prime}_{\rm [NII] ,[OIII] } + m^{\prime}_{\rm [OIII] , [OII]} \approx m^{\prime}_{\rm [OIII] ,[OII] } + 0.89$, where the difference is roughly 0.15 ($1.3\sigma _{\rm std}$ or $80\sigma ^{\prime}$).
Interestingly, the additivity problem seems to be related to the slopes that are apparently lower than the corresponding F99 values.

One related question is whether the additivity problem comes from the bias during the fitting process associated with incorrectly estimated measurement uncertainties.
In Figure~\ref{fig:add_s3o3}, we plot distributions of $m_{\rm [SIII] ,[OIII] } - m_{\rm [SIII] ,H\alpha } - m_{\rm H\alpha , H\beta}$ using samples selected based on different S/N cuts.
It is clear from this figure that the additivity for these line ratios is asymptotically recovered at high S/N.
In particular, we found $\rm m^{\prime}_{\rm [SIII] ,H\alpha }$ increases noticeably with increasing S/N, while $\rm m^{\prime}_{\rm [SIII] ,[OIII] }$ and $\rm m^{\prime}_{\rm H\alpha ,[OIII]}$ remains roughly the same.
{If the input measurement uncertainties are biased, the best-fit parameters returned by our likelihood function would also be biased (see Appendix~\ref{appendix}), which could lead to violation of the additivity if the bias is different for different lines. On the other hand, the bias becomes negligible when the S/N is very high.}
However, this explanation seems inapplicable to the case of $m^{\prime}_{\rm [NII],[OII]}$. Even after we increased the S/N cut to 20 for [N\,{\sc ii}], [O\,{\sc ii}], [O\,{\sc iii}], and H$\beta$, $m^{\prime}_{\rm [NII],[OII]}$ is still greater than $m^{\prime}_{\rm [NII],[OIII]} + m^{\prime}_{\rm [OIII],[OII]}$ by 0.12 (2.1$\sigma _{\rm std}$ or roughly $60\sigma ^{\prime}$).
{Also, the bias does not originate from the data lost from the S/N cut, as we performed tests with mock data and found this effect was negligible even when the S/N limit was set to 3.
In Section~\ref{subsec:depend_obs_pro} we show the additivity bias is actually more related to the surface brightness of the lines.
}

The slopes of the hybrid line ratios could also be affected by our construction of the 3D bins.
Since we restricted the variation of \nha, \sha, and \ohb\ in each bin, the derived slopes for certain combinations of lines are tied together.
This tying should not introduce any bias if all line ratios follow the same attenuation curve, but would become problematic if different lines are attenuated by different curves.
As an example, since log([S\,{\sc iii}]/H$\alpha$) = log([S\,{\sc iii}]/[N\,{\sc ii}]) + log(\nha), and log(\nha) varies little in each bin, we have forced $m^{\prime}_{\rm [SIII] ,H\alpha } \approx m^{\prime}_{\rm [SIII] ,[NII] }$. Therefore, the measurement of $m_{\rm [SIII] ,H\alpha }$ is affected by $m_{\rm [SIII] ,[NII] }$.
Even if [S\,{\sc iii}] indeed shares the same attenuation law with Balmer lines, the involvement of the low ionization line [N\,{\sc ii}] could contribute to the deviation from the F99 value.
In other words, if the observed [N\,{\sc ii}] and H$\alpha$ no longer share the same attenuation law, binning data using log([N\,{\sc ii}]/H$\alpha$) can modify the measured slope for certain hybrid line ratios. In addition, our assumption of negligible variations of nebular parameters might become invalid in such cases.
Mathematically, our construction of the likelihood function becomes not precise, as the single parameter $\sigma _0$ might no longer be sufficient to describe the intrinsic variation of line ratios.

Indeed, from Table~\ref{tab:derived}, one can see that if a reddening relation directly or indirectly involves a low ionization line, the resulting median slope tends to be significantly lower than the expected value from the F99 curve.
Interestingly, even the low ionization lines might not share the same attenuation law.
One example is provided by [S\,{\sc iii}]/[O\,{\sc i}], whose median slope is slightly smaller than that of [S\,{\sc iii}]/[N\,{\sc ii}]. If both [O\,{\sc i}] and [N\,{\sc ii}] are subject to the same attenuation curve, we should have $m^{\prime}_{\rm [SIII] ,[OI] } > m^{\prime}_{\rm [SIII] ,[NII] }$.
Since [O\,{\sc i}] is an even lower ionization line compared to [N\,{\sc ii}], this discrepancy indicates a further flattening of the attenuation curve at lower ionization.
There is, however, a special case where the hybrid line ratio, [N\,{\sc ii}]/H$\gamma$, gives the slope close to the F99 value. Following the same argument we use in the preceding paragraphs, we can understand this case as follows.
Even if [N\,{\sc ii}]/H$\alpha$ is biased and is not attenuation-free, H$\alpha$/H$\gamma$ would still follow the unbiased attenuation curve since the slope measurements for Balmer lines do not rely on 3D binning. Given that log([N\,{\sc ii}]/H$\gamma$) = log(H$\alpha$/H$\gamma$) + log([N\,{\sc ii}]/H$\alpha$), and the fact that we have forced $m^{\prime}_{\rm [NII],H\alpha}\approx 0$ in each 3D bin, it is no surprise $m^{\prime}_{\rm [NII],H\gamma}\approx m^{\prime}_{\rm H\alpha,H\gamma}\approx m^{F99}_{\rm H\alpha,H\gamma}$.

Finally, another explanation could be that both high ionization lines and Balmer lines follow the F99 curve, but their magnitudes of attenuation differ systematically.
Thus, when measuring the reddening of their combined line ratios, the result deviates from the F99 curve.

In summary, most of the hybrid line ratios we check show slopes deviating from the F99 values and some of them violate additivity as well.
There could be biases associated with measurement uncertainties, which, however, cannot explain all hybrid line ratios.
Since there is already some evidence that different species of lines might follow different attenuation curves, our constructions of the 3D bins could also bias the slope measurements for certain hybrid lines.
To better understand these effects, we investigate some physical models that create different attenuation laws for different lines in the next section.


\begin{table*}
	\caption{Median slopes of the reddening relations [log(Emission line ratio) versus log(H$\alpha$/H$\beta$/2.86)] derived for a sample of 3D bins. The bins were set up in a 3D line-ratio space spanned by log(\nha), log(\sha), and log(\ohb). The size of each cell is $\rm 0.0167\times 0.0167\times 0.0167~dex^3$. 3D bins with $\rm N_{spaxel} > 180$ were selected to derive the slope distribution in most cases. We used the 68\% confidence interval of the median slope as the uncertainty ($\sigma ^{\prime}$), but we also showed the standard deviation of the slope distribution for each line ratio ($\sigma _{\rm std}$).}
	\label{tab:derived}
	\begin{tabular}{lccc} 
		\hline
		\hline
		Emission line ratio & Median slope & Standard deviation & Slope predicted by the F99 curve
		\\
		 & ($m^{\prime}\pm \sigma ^{\prime}$) & ($\sigma _{\rm std}$) & with $R_V = 3.1$ ($m^{\rm F99}$) 
		 \\
		\hline
		H$\alpha$ (6563)/H$\gamma$ (4340) & $1.343\pm 0.002$ ($1.3400\pm 0.0005$)$\rm ^a$ & 0.08 & 1.395
		\\
		H$\alpha$ (6563)/H$\delta$ (4102) & {$1.545\pm 0.003$} {($1.5683\pm 0.0009$)$\rm ^{a}$} & {0.15} & 1.563
		\\
		\hline
		$[$S\,{\sc iii}]9069,9531/[O\,{\sc iii}]5007 & $1.660\pm 0.005$ & 0.27 & 1.695$\rm ^b$
		\\
		$[$O\,{\sc iii}]5007/[Ne\,{\sc iii}]3869,3967 & {$0.68\pm 0.06$} & {1.04} & 0.81$\rm ^b$
		\\
		\hline
		$[$S\,{\sc ii}]6716,6731/[O\,{\sc ii}]3726,3729 & $1.580\pm 0.002$ & 0.13 & 1.903
		\\
		$[$N\,{\sc ii}]6583/[O\,{\sc ii}]3726,3729 & $1.525\pm 0.002$ & 0.13 & 1.844
		\\
		\hline
		$[$S\,{\sc iii}]9069,9531/[S\,{\sc ii}]6716,6731 & $0.469\pm 0.005$ & 0.28 & 0.745$^{\rm b}$
		\\
		$[$S\,{\sc iii}]9069,9531/[N\,{\sc ii}]6583 & $0.536\pm 0.005$ & 0.27 & 0.801$^{\rm b}$
		\\
		$[$S\,{\sc iii}]9069,9531/[O\,{\sc i}]6300 & {$0.493\pm 0.009$} & {0.38} & 0.928$^{\rm b}$
		\\
		$[$S\,{\sc iii}]9069,9531/[O\,{\sc ii}]3726,3729 & $2.127\pm 0.006$ & 0.32 & 2.645$^{\rm b}$
		\\
		$[$O\,{\sc iii}]5007/[O\,{\sc ii}]3726,3729 & $0.459\pm 0.003$ & 0.15 & 0.950
		\\
		\hline
		$[$S\,{\sc iii}]9069,9531/H$\alpha$ (6563) & $0.600\pm 0.006$ & 0.28 & $0.811^{\rm b}$
		\\
		$[$S\,{\sc iii}]9069,9531/H$\beta$ (4861) & $1.600\pm 0.006$ & 0.28 & $1.811^{\rm b}$
		\\
		$[$S\,{\sc iii}]9069,9531/H$\gamma$ (4340) & $1.929\pm 0.005$ & 0.26 & $2.206^{\rm b}$
		\\
		\hline
		$[$N\,{\sc ii}]6583/H$\gamma$ (4340) & $1.352\pm 0.002$ & 0.08 & $1.405$
		\\
		H$\alpha$ (6563)/[O\,{\sc ii}]3726,3729 & $1.509\pm 0.002$ & 0.13 & 1.835
		\\
		H$\beta$ (4861)/[O\,{\sc ii}]3726,3729 & $0.509\pm 0.002$ & 0.13 & 0.835
		\\
		H$\gamma$ (4340)/[O\,{\sc ii}]3726,3729 & $0.161\pm 0.003$ & 0.14 & 0.439
		\\
		\hline
	\end{tabular}
	\begin{tablenotes}
        \small
        \item $\bf Notes.$
        \item $^{\rm a}$ The value inside the parenthesis is obtained using the whole sample without binning.
        \item $^{\rm b}$ The value is the average of the doublet values as the measured fluxes of the corresponding doublet were fixed by DAP.
    \end{tablenotes}
\end{table*}

\section{Physical model}
\label{sec:explain}

In this section we present some physical models to explain how different observed emission lines can have different attenuation laws.
In addition, we compare the model predictions with the slopes measured in our sample.

\subsection{Bias from low-resolution observations}
\label{subsec:bias_for_lo}

\begin{figure*}
    \includegraphics[width=0.23\textwidth]{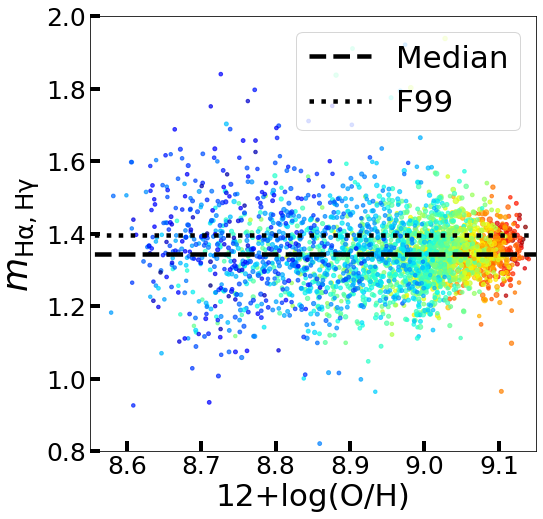}
    \includegraphics[width=0.23\textwidth]{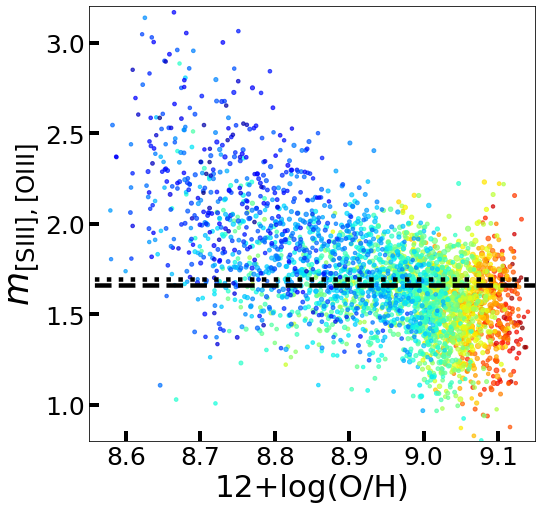}
    \includegraphics[width=0.23\textwidth]{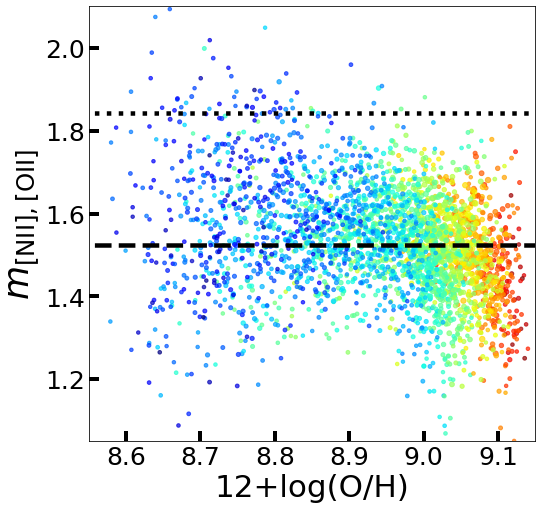}
    \includegraphics[width=0.3\textwidth]{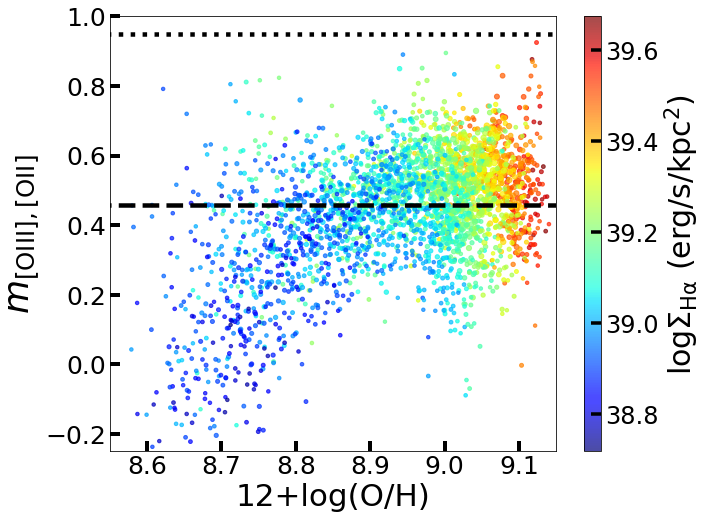}
    \caption{Derived slopes for different line ratios as a function of the gas-phase metallicity. The $y$ axes correspond to slopes of the $\log (f_{\rm line_1}/f_{\rm line_2})$ versus $\rm \log (H\alpha /H\beta /2.86)$ relations. The $x$ axes correspond to metallicities derived by using Bayesian inference with a fiducial photoionization model.
    Each data point corresponds to a 3D bin with number of spaxels greater than 180. 
    The 3D bins are color-coded according to the median H$\alpha$ surface brightness (which is corrected by the Balmer decrement assuming an F99 extinction curve) of their spaxels.
    Dashed black lines and dotted black lines correspond to the median slopes and the values predicted by an F99 extinction curve, respectively.
    }
    \label{fig:m_met}
\end{figure*}

When dealing with extragalactic observations, one usually has very limited knowledge about the geometry of the dust distribution inside the corresponding galaxy.
Still, indirect evidence concerning the dust distribution can be derived by comparing the attenuation of different sources.
The attenuation probed by nebular emissions from SF regions is usually larger than that probed by starlight \citep[e.g.,][]{fanelli1988, calzetti1994, kashino2013, price2014, pannella2015, li2021}, which is attributed to the fact that the neutral gas and dust are more concentrated and clump around individual \hii\ region \citep{charlot2000, wild2011, chevallard2013}. Therefore, one usually assumes that emission from \hii\ regions is {``extincted''} by the foreground dust rather than {``attenuated''} due to a complex mix of sources and dust, which is, however, not necessarily valid for unresolved ``\hii\ regions'' \citep{pellegrini2020b}.

Within each \hii\ region, different emission lines originate from different optical depths due to their different ionization potentials. 
The Balmer lines are produced by recombination throughout the ionized shell. In comparison, the high ionization lines and low ionization lines are mainly produced in the inner part and outer part of the \hii\ region, respectively. Since dust grains also exist inside the ionized shell, one might wonder whether the internal differential attenuation within the \hii\ region can result in different attenuation laws for different lines. However, as shown by \citet{bottorff1998}, even for a highly ionized \hii\ region with $\log U \sim -2.0$ and an Orion-type dust composition, the maximum optical depth at $\lambda \sim 0.1~{\rm \mu m}$ is still smaller than 1. For the emission lines we are concerned about, the corresponding optical depths are much smaller and the dust attenuation effect should be negligible within most \hii\ regions.

The above scenario is applicable to radiation-bounded \hii\ regions with ionized shells well confined by the background molecular clouds. In observations, however, people found that Lyman continuum photons (Lyc) can leak from dense clouds and create emission line regions in the surrounding lower density regions \citep[e.g.,][]{reynolds1990,mckee1997,ferguson1996,zurita2000,haffner2009,howard2018}, which can extend over a few kiloparsecs \citep[see e.g.,][]{zurita2002, seon2009, belfiore2022}.
These low-density regions are usually referred to as the diffuse ionized gas (DIG).
The ratios between low ionization lines and Balmer lines are generally higher in DIG, and emission lines produced in DIG are also less attenuated compared to \hii\ regions.

Apart from the leaked Lyc, there is another important contributor to the emissions in DIG. As shown by \cite{zhang2017}, the leaked Lyc from \hii\ regions cannot explain the high [O\,{\sc iii}]/H$\beta$ observed in DIG for massive galaxies. Thus, one needs extra ionizing sources apart from \hii\ regions to produce the observed [O\,{\sc iii}] fluxes. As suggested by \cite{flores-fajardo2011}, the hot low-mass evolved stars (HOLMES) are likely to play an important role in ionizing DIG. Specifically, HOLMES create high [O\,{\sc iii}]/H$\beta$ due to their harder spectra compared to \hii\ regions. It is important to note that the enhancement of [O\,{\sc iii}]/H$\beta$ in DIG depends on the stellar mass (or metallicity) of the galaxy, which has been confirmed on both kiloparsec and subkiloparsec scales \citep{zhang2017, belfiore2022}. For low-mass galaxies, [O\,{\sc iii}]/H$\beta$ barely changes from \hii\ regions to DIG. Whilst for high-mass galaxies, [O\,{\sc iii}]/H$\beta$ has a conspicuous increase toward DIG. This is because low-metallicity \hii\ regions (which are typically found in low-mass galaxies) have much higher [O\,{\sc iii}]/H$\beta$ ratios compared to high-metallicity \hii\ regions (as can be seen in the BPT diagrams), while DIG in general covers a relatively narrow range in [O\,{\sc iii}]/H$\beta$.

The above evidence suggests that both low and high ionization lines could experience different amount of attenuation compared to Balmer lines.
Since the MaNGA data we use have a typical resolution of 1\,--\,2 kpc, it is possible that a considerable amount of diffuse emission has been blended with the emission from the nearby \hii\ regions.
Due to the contribution of less-attenuated DIG, we expect the measured slopes for line ratios involving low ionization lines and high ionization lines to be different from the true slopes. This appears to be consistent with our interpretation in the previous section. In addition, this effect is likely metallicity-dependent. 

Figure~\ref{fig:m_met} shows how the measured slopes for H$\alpha$/H$\gamma$, [S\,{\sc iii}]/[O\,{\sc iii}], [N\,{\sc ii}]/[O\,{\sc ii}], and [O\,{\sc iii}]/[O\,{\sc ii}] correlate with the median gas-phase metallicity as well as the median H$\alpha$ luminosity surface density (corrected using the F99 curve) within each bin. 
{We derived the metallicities following the Bayesian method of \cite{ji2022}. In short, we first computed the likelihood of the observed \nha, \sha, and \ohb\ given the photoionization model of \cite{ji2020b}. Combining the likelihood with a flat prior in the logarithmic space, we then obtained the posterior distribution of the metallicity for each datum. Finally, the metallicity of each datum was calculated as the weighted mean using the posterior distribution.
}
We see that $\rm m_{H\alpha,H\gamma}$ has no dependency on the metallicity.
While $\rm m_{[NII],[OII]}$ appears to have a sharp decline at very high metallicities, there seems to be no obvious trend at lower metallicities\footnote{We note that the metallicity measurements at the highest metallicity might have slightly larger systematic uncertainty due to the extrapolation of the stellar spectral energy distribution in the model \citep{ji2020b}.}.
In comparison, the slopes for line ratios involving [O\,{\sc iii}] have a noticeable change with the metallicity. For high metallicity galaxies, [O\,{\sc iii}] tends to have a larger contribution from the DIG, and we see it appears less attenuated as reflected by decreasing $\rm m_{[SIII],[OIII]}$ and increasing $\rm m_{[OIII],[OII]}$.
However, it is worth noting that $\rm m_{[SIII],[OIII]}$ and $\rm m_{[OIII],[OII]}$ deviate from the F99 values at low metallicities, where the bias brought by the DIG is supposed to be lower compared to high metallicities.
The contribution of DIG to [S\,{\sc iii}] and [O\,{\sc ii}] could be the complicating factor in this explanation,
which we further explore in the next subsection.
{In addition, Figure~\ref{fig:m_met} also shows that the metallicity strongly correlates with the median H$\alpha$ surface brightness of individual 3D bins. Since the H$\alpha$ surface brightness is related the level of the DIG contamination \citep{oey2007,zhang2017}, the trends we see in Figure~\ref{fig:m_met} could also be caused by the overall change in the DIG contamination.
We further discuss the effect of selecting the sample based on the H$\alpha$ surface brightness in Section~\ref{subsec:depend_obs_pro}.
}

Besides the potential DIG contamination, there is another effect that could be important due to the limited spatial resolution. For a single \hii\ region or an association of spatially close \hii\ regions, it is possible that not all of its emission is attenuated by the same amount. One can imagine a dust screen covers part but not all of the \hii\ region. For such a ``partially covered \hii\ region,'' the effective integrated Balmer decrement depends on the relative contributions from the covered and uncovered parts to the observed emission line spectrum. If the covered fraction varies in different \hii\ regions, there would be an effective variation in the observed Balmer decrement, which can also give rise to line-specific effective attenuation or differential attenuation.

In summary, observations that are unable to resolve individual \hii\ regions could lead to differential attenuation inconsistent with a unified curve 
either due to the distribution of dust or due to the contamination from the DIG, even though there is an underlying extinction curve at work.
In the next subsection, we try setting up a few toy models to reproduce the attenuation we measured for different emission lines in the MaNGA data.

\subsection{Attenuation model}
\label{subsec:toy_model}

\begin{figure}
    \centering
    \includegraphics[width=0.42\textwidth]{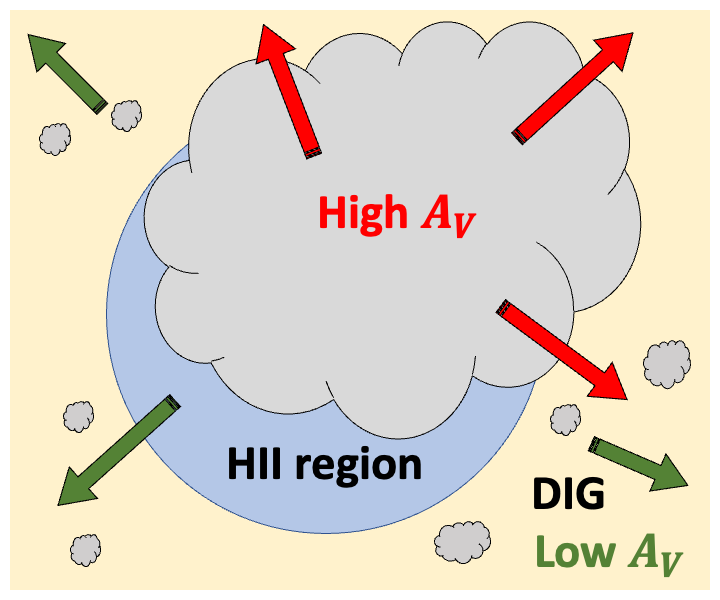}
    \caption{{Cartoon illustration of a two-component attenuation model.
    The model consists of a more attenuated component with high $A_V$ and a less attenuated component with low $A_V$. The emission lines from both components are mixed in observations.
    While the high $A_V$ component is (part of) a dusty \hii\ region, the low $A_V$ component could be a part of the \hii\ region that is less dusty, the DIG around the \hii\ region, or a combination of both.
    }}
    \label{fig:cartoon}
\end{figure}

\begin{figure*}
    \includegraphics[width=0.95\textwidth]{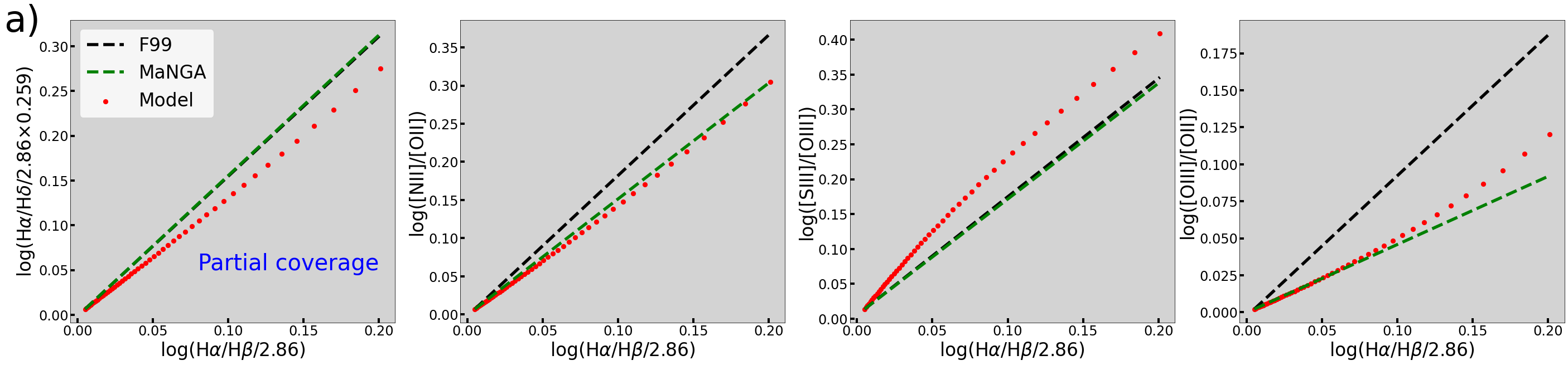}
    \includegraphics[width=0.95\textwidth]{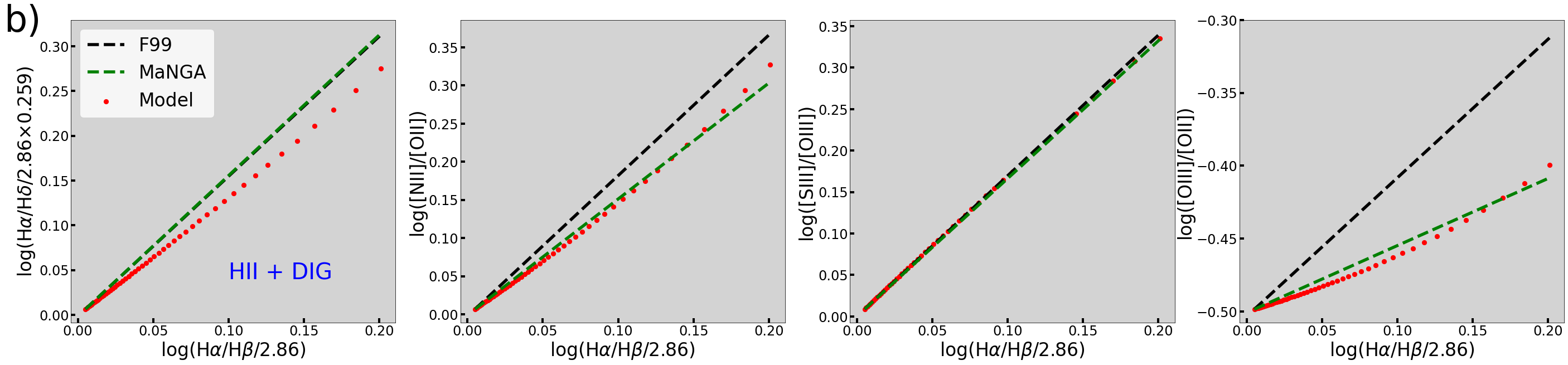}
    \includegraphics[width=0.95\textwidth]{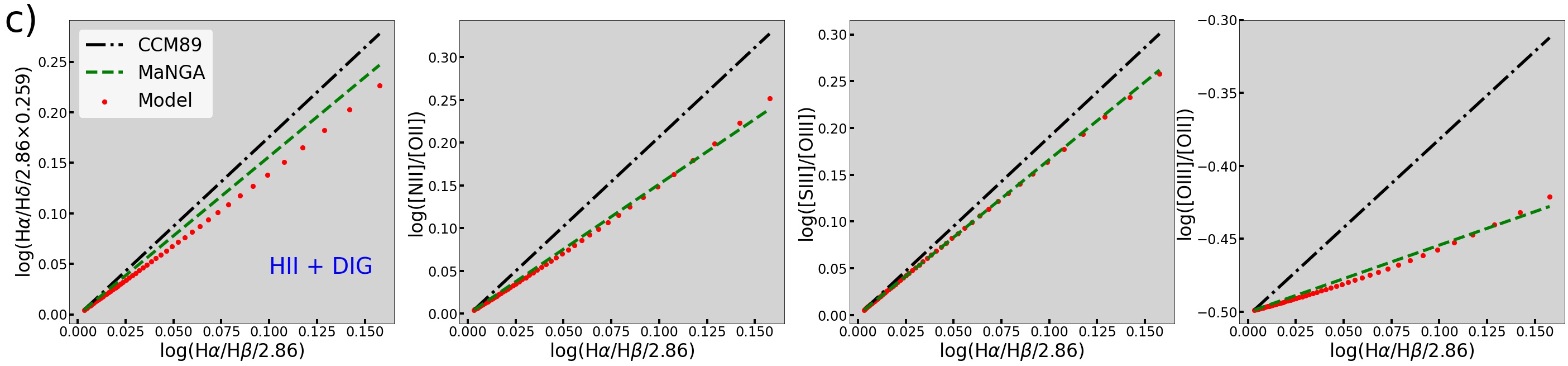}
    \caption{Effective attenuation laws generated by different models. 
    Row a) shows several logarithmic line ratio versus logarithmic Balmer derement relations predicted by a partially covered cloud model.
    For comparison, we also plotted the relation predicted by the F99 extinction curve and the median relation we found in MaNGA.
    Row b) shows the relations predicted by a two-component model with different intrinsic line ratios. We made the \hii\ component to have overall lower intrinsic line ratios (relative to H$\alpha$) compared to the DIG component.
    Row c) shows a similar model as Row b), but the underlying true attenuation curve is replaced with the CCM89 extinction curve.
    All $y$ axes have arbitrary normalization.
    }
    \label{fig:toy_model}
\end{figure*}

We started by assuming a simple cloud model, which consists of two separate components that could have different optical depths and emission line fluxes. The first component corresponds to high attenuation, which we denote as $C_{\rm High}$; the second component corresponds to low attenuation, which we denote as $C_{\rm Low}$\footnote{This model shares some similarities with the \cite{charlot2000} model, where two different attenuation components are present.}.
{Figure~\ref{fig:cartoon} shows a cartoon illustration of the two-component model. Attenuated emission lines from both $C_{\rm High}$ and $C_{\rm Low}$ are mixed in observations. In this model, $C_{\rm High}$ is a single dusty \hii\ region or a part of an \hii\ region that is more obscured, and $C_{\rm Low}$ is the DIG surrounding the \hii\ region or a part of the \hii\ region that is less obscured.
}
By definition, $A_{V, {\rm High}} > A_{V, {\rm Low}}$.
We then defined the fractional contribution to the total intrinsic H$\alpha$ flux from $C_{High}$ to be $\eta \equiv f_{\rm H\alpha ,0 (High)}/f_{\rm H\alpha ,0 (total)}=f_{\rm H\alpha ,0 (High)}/[f_{\rm H\alpha ,0 (High)}+f_{\rm H\alpha ,0 (Low)}]$ ($0 \leq \eta \leq 1$).

The first case we considered is where both $C_{\rm High}$ and $C_{\rm Low}$ have the same intrinsic line ratios, which corresponds to a partially covered \hii\ region.
In addition, we fixed $A_{V, {\rm High}}$ and $A_{V, {\rm Low}}$ for simplicity.
The only thing we varied was $\eta$, the fractional flux contributed by $C_{\rm High}$. As a result, the observed attenuation or Balmer decrement would follow the change in $\eta$. The observed H$\alpha$ flux is
\begin{equation}
    f_{\rm H\alpha, obs} = f_{\rm H\alpha, 0 (total)}[\eta 10^{-0.4 A_{V, {\rm High}}l_{\rm H\alpha}} + (1-\eta )10^{-0.4 A_{V, {\rm Low}}l_{\rm H\alpha}}].
\end{equation}
Since $f_{\rm H\beta ,0} = f_{\rm H\alpha ,0}/2.86$, we have
\begin{equation}
    (f_{H\alpha} /f_{H\beta}/2.86)_{\rm obs}=\frac{\eta 10^{-0.4 A_{V, {\rm High}}l_{\rm H\alpha}} + (1-\eta )10^{-0.4 A_{V, {\rm Low}}l_{\rm H\alpha}}}{\eta 10^{-0.4 A_{V, {\rm High}}l_{\rm H\beta}} + (1-\eta )10^{-0.4 A_{V, {\rm Low}}l_{\rm H\beta}}}.
\end{equation}
We note that both $l_{\rm H\alpha}$ and $l_{\rm H\beta}$ are given by an underlying true attenuation curve, which we assume to be the F99 curve here. It is clear from the above equation that the effective V-band attenuation derived from the observed Balmer decrement, $A_{V,{\rm eff}}$, changes from $A_{V, {\rm Low}}$ to $A_{V, {\rm High}}$ as $\eta$ changes from 0 to 1. In general, the observed flux ratio of two lines with wavelengths $\lambda _1$ and $\lambda _2$ is given by
\begin{equation}
    (f_{\lambda _1} /f_{\lambda _2})_{\rm obs}=(\frac{f_{\lambda _1}}{f_{\lambda _2}})_0\frac{\eta 10^{-0.4 A_{V, {\rm High}}l_{\rm \lambda _1}} + (1-\eta )10^{-0.4 A_{V, {\rm Low}}l_{\rm \lambda _1}}}{\eta 10^{-0.4 A_{V, {\rm High}}l_{\rm \lambda _2}} + (1-\eta )10^{-0.4 A_{V, {\rm Low}}l_{\rm \lambda _2}}},
    \label{eq:partial_cover}
\end{equation}
where we have used the assumption that the intrinsic line ratio is the same for $C_{\rm High}$ and $C_{\rm Low}$.

What we measured in observations is how $\log(f_{\lambda _1} /f_{\lambda _2})_{\rm obs}$ changes with $\log(f_{\rm H\alpha} /f_{\rm H\beta}/2.86)_{\rm obs}$, which is supposed to give us $\frac{l_{\lambda _1}-l_{\lambda _2}}{l_{\rm H\alpha}-l_{\rm H\beta}}$. However, if the change in $A_{V,{\rm eff}}$ is primarily driven by the change in the covering fraction, or $\eta$, we expect a modified attenuation curve to be observed. 
The top panel of Figure~\ref{fig:toy_model} shows such an example. Instead of specifying $A_{V, {\rm High}}$ and $A_{V, {\rm Low}}$, we set $\log(f_{\rm H\alpha} /f_{\rm H\beta}/2.86)_{\rm obs}$ to 0.4 and 0 for $C_{\rm High}$ and $C_{\rm Low}$, respectively.
We see that the final $m_{\rm [NII],[OII]}$ and $m_{\rm [OIII],[OII]}$ are lower than the expected values given by the F99 curve and are closer to the median value we measured in MaNGA.
The values of the Balmer decrement for $C_{\rm High}$ and $C_{\rm Low}$ are the only parameters that determine the final slope in this model. Surprisingly, this oversimplified model can already produce some changes in the measured attenuation curve.
However, the weakness of this model is apparent. The result depends on the wavelengths of the lines involved rather than the species of ions that produce them. As a consequence, the slopes for Balmer lines such as $\rm m_{H\alpha,H\gamma}$ and $\rm m_{H\alpha,H\delta}$ become lower than their true values, as shown in the top leftmost panel of Figure~\ref{fig:toy_model}, contrary to the observed results. 
On the other hand, for the high ionization line ratio $\rm m_{[SIII],[OIII]}$, this model yields a slope that is higher than the true value, different from the observations.

\begin{table}
	\centering
	\caption{Logarithmic difference between the intrinsic line strengths of $C_{\rm High}$ and $C_{\rm Low}$ in a DIG contamination model. Here, $r_{\lambda}$ is defined to be $f_{\lambda}/f_{\rm H\alpha}$ at zero attenuation.}
	{\fontsize{10}{10.5}\selectfont
	\label{tab:model_param}
	\begin{tabularx}{\columnwidth}{lc} 
	    \hline\hline
        Line & $\log(r_{\lambda , {\rm High}}/r_{\lambda ,{\rm Low}})^{\rm a}$\\
        \hline
        Balmer lines & 0\\
        $[$N\,{\sc ii}] & $-0.05$\\
        $[$O\,{\sc ii}] & $-0.30$\\
        $[$O\,{\sc iii}] & $-0.20$\\
        $[$S\,{\sc iii}] & $-0.20$ (F99), $-0.30$ (CCM89)\\
	    \hline
	\end{tabularx}
	}
\end{table}

To solve the discrepancy associated with $m_{\rm [SIII],[OIII]}$, we added new parameters to the model by setting different intrinsic line ratios for $C_{\rm High}$ and $C_{\rm Low}$. This model simulates the situation where two kinds of ionized regions, \hii\ regions ($C_{\rm High}$) and DIG ($C_{\rm Low}$) are mixed in observations. If we denote $r_{\rm \lambda , High}$ and $r_{\rm \lambda , Low}$ as the intrinsic values of $(f_{\lambda}/f_{\rm H\alpha})$ for $C_{\rm High}$ and $C_{\rm Low}$, Equation~\ref{eq:partial_cover} becomes
\begin{equation}
    (f_{\lambda _1} /f_{\lambda _2})_{\rm obs}=\frac{r_{\lambda _1, {\rm High}}\eta 10^{-0.4 A_{V, {\rm High}}l_{\rm \lambda _1}} + r_{\lambda _1, {\rm Low}}(1-\eta )10^{-0.4 A_{V, {\rm Low}}l_{\rm \lambda _1}}}{r_{\lambda _2, {\rm High}}\eta 10^{-0.4 A_{V, {\rm High}}l_{\rm \lambda _2}} + r_{\lambda _2, {\rm Low}}(1-\eta )10^{-0.4 A_{V, {\rm Low}}l_{\rm \lambda _2}}}.
    \label{eq:two_comp}
\end{equation}
The intrinsic flux ratio $r_{\rm \lambda, High (Low)}$ should depend on the species of the ion as well as nebular parameters such as the metallicity, but for now we chose a set of values that could roughly reproduce the slopes we measured in MaNGA, which is listed in Table~\ref{tab:model_param}. For each line, we set a value for $\log(r_{\rm \lambda , High}/r_{\rm \lambda ,Low})$. From Table~\ref{tab:model_param}, we see that in order to fit the observed slopes, we need to make $C_{\rm High}$ (\hii\ component) have overall lower intrinsic forbidden line ratios compared to $C_{\rm Low}$ (DIG component), which is qualitatively consistent with observations \citep[e.g.,][]{zhang2017,belfiore2022}\footnote{We note that the assumed line ratios for the two components will affect the resulting slopes, and some fine-tuning is required to make it work approximately.}. 

There is, however, still a discrepancy for $\rm m_{H\alpha,H\gamma}$.
One might wonder whether we can achieve better agreement with the observation by changing the underlying attenuation curve. 
Indeed, the bottom panels of Figure~\ref{fig:toy_model} shows that if we instead assume a \cite{cardelli1989} extinction curve with $R_V =3.1$ (CCM89) as the true underlying extinction curve, $m_{\rm H\alpha,H\gamma}$ gets slightly closer to the observed values. Adjusting the Balmer decrement values for $C_{\rm High}$ and $C_{\rm Low}$ can further improve the fit.
In addition, we changed $\log(r_{\rm [SIII] , High}/r_{\rm [SIII] ,Low})$ from $-0.20$ to $-0.30$ in order to fit $m_{\rm [SIII],[OIII]}$ in this case.
To determine the proper physical models and model parameters, we need evidence from high-resolution observations. One example using the MaNGA observation of the galaxy IC 342 is discussed in Appendix~\ref{subsec:high_resol_hii}.

\begin{figure}
    \includegraphics[width=0.43\textwidth]{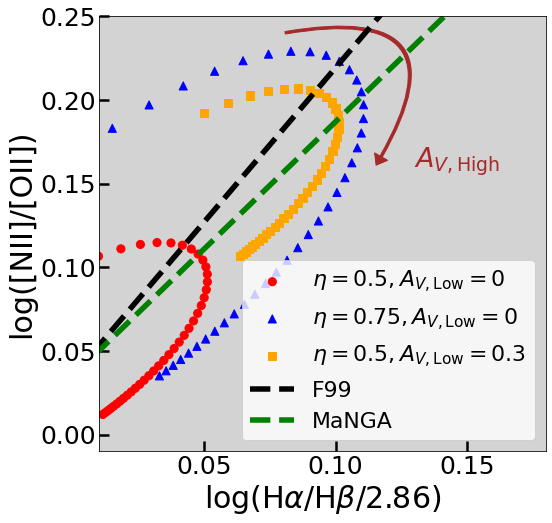}
    \caption{{Two-component attenuation model for \no, where only the attenuation of the more attenuated component is varied.
    Different sets of points correspond to models with different intrinsic flux contributions from the two components as well as different $A_{V,{\rm Low}}$.
    The arrow shows the direction of increasing $A_{V,{\rm High}}$ in the model.}
    }
    \label{fig:toy_model1}
\end{figure}

The above models assume that the two components, $C_{\rm High}$ and $C_{\rm Low}$ have fixed magnitudes of attenuation so that the attenuation differences among \hii\ regions are solely driven by the change in the weights of the two components. In reality, one expects the magnitude of the attenuation of the two (if not more) components also change.
We can try going to another extreme by assuming that $C_{\rm High}$ and $C_{\rm Low}$ have fixed weights in their contributions to H$\alpha$ and the attenuation differences among different MaNGA spaxels are solely driven by the change in the magnitudes of the attenuation for $C_{\rm High}$ and $C_{\rm Low}$.
If, for example, we keep $A_{V,{\rm Low}}$ constant but vary $A_{V,{\rm High}}$, one expects $A_{V,{\rm eff}}$ to first increase with increasing $A_{V,{\rm High}}$ but then start to decrease at some point due to the reduction of the observed fluxes from $C_{\rm High}$. This makes the Balmer decrement not a monotonic function of $A_{V,{\rm High}}$. 
Besides the Balmer decrement, other line ratios would also reach an extreme value and then turn around.
Figure~\ref{fig:toy_model1} shows such a toy model for the attenuation of \no.
The turn-around point depends on the value of $\eta$. When $\eta = 0.5$, the turn around happens when the observed Balmer decrement changes by $\sim 0.05$~dex from the value under $A_{V,{\rm Low}}$; for $\eta = 0.75$, the turn around happens when the observed Balmer decrement changes by $\sim 0.10$~dex from the value under $A_{V,{\rm Low}}$.
We did not observe this behavior clearly in our sample, indicating that this is not a dominant effect but could contribute to the scatter in the attenuation.

On the other hand, if we vary $A_{V,{\rm High}}$ and $A_{V,{\rm Low}}$ by the same amount, there is no change in the attenuation curve. This is because changing both $A_V$s by the same amount effectively changes the column density of the common foreground dust screen.

The observations likely include both changes in the intrinsic flux ratios between $C_{\rm High}$ and $C_{\rm Low}$ and variations in the magnitudes of $A_{V,{\rm High}}$ and $A_{V,{\rm Low}}$. Within our 3D bins, the latter effect might be present as scatters rather than dominating the trend for different {reddening relations}.
To check whether these toy models are applicable to realistic data, we examined the observational evidence in the next subsection.

\subsection{Comparison with observations}
\label{subsec:obs_evidence}

In Section~\ref{subsec:bias_for_lo}, we showed how the slopes of different reddening relations change with the gas-phase metallicity in Figure~\ref{fig:m_met}. Among the slope measurements we showed, $\rm m_{[SIII],[OIII]}$ and $\rm m_{[OIII],[OII]}$ exhibit the most clear dependence on the metallicity, which we suspected to be related to the DIG contamination on the observed [O\,{\sc iii}]. Previous studies show the \ohb\ difference between the \hii\ region and the DIG increases with increasing galaxy masses \citep{zhang2017,belfiore2022}, and thus increasing metallicity. The key question is whether the two-component model can explain the trends for $\rm m_{[SIII],[OIII]}$ and $\rm m_{[OIII],[OII]}$ given the observational evidence on the change of the \ohb\ difference. 

Intuitively, when the observed \ohb\ has a larger contribution from the DIG, one might expect [O\,{\sc iii}] becomes less attenuated compared to other lines as the DIG contains less dust compared to \hii\ regions. As a consequence, $m_{\rm [SIII],[OIII]}\equiv (A_{\rm [SIII]}-A_{\rm [OIII]})/(A_{\rm H\alpha}-A_{\rm H\beta})$ would decrease and $m_{\rm [OIII],[OII]}\equiv (A_{\rm [OIII]}-A_{\rm [OII]})/(A_{\rm H\alpha}-A_{\rm H\beta})$ would increase, assuming the intrinsic line ratios, $r_{\rm \lambda, High}/r_{\rm \lambda, Low}$, of other lines remain unchanged. Whereas as we have pointed out, this does not explain why $m_{\rm [SIII],[OIII]}$ and $\rm m_{\rm [OIII],[OII]}$ deviate more from the expected F99 values at lower metallicities where the DIG has more similar intrinsic line ratios as \hii\ regions.

Using our two-component model, we find the opposite of the observed trends for $m_{\rm [SIII],[OIII]}$ and $m_{\rm [OIII],[OII]}$. When we fixed $r_{\rm \lambda, High}/r_{\rm \lambda, Low}$ for other line ratios while increasing $r_{\rm \lambda, High}/r_{\rm \lambda, Low}$ for [O\,{\sc iii}] (which effectively reduces the difference in \ohb\ between \hii\ regions and the DIG), $m_{\rm [SIII],[OIII]}$ became smaller and $m_{\rm [OIII],[OII]}$ became larger.
This is because in our two-component model, the observed change in the Balmer decrement is driven by the change of the fractional contributions from the \hii\ component and the DIG component. As the Balmer decrement becomes larger, the \hii\ component gradually dominates and the observed line ratios approach the intrinsic line ratios of the \hii\ component. Therefore, once the relative strength of [O\,{\sc iii}] increases in the \hii\ component (i.e., less DIG contamination), our model actually predicts a boost in the increase of total [O\,{\sc iii}] when the observed Balmer decrement increases. As a result, if only the change in \ohb\ is considered, the two-component model would fail to describe the metallicity dependence of $m_{\rm [SIII],[OIII]}$ and $m_{\rm [OIII],[OII]}$.

However, a complicating factor here is that [O\,{\sc iii}] is not the only line whose relative strength in \hii\ regions and DIG changes with the metallicity.
In fact, the value of [O\,{\sc ii}]/H$\beta$ in the DIG is also likely a function of the stellar mass and metallicity \citep{zhang2017}. When there are other line ratios changing together with \ohb, the final slope for the reddening relation would depend on the both the intrinsic shape of the true extinction curve and the relative change in different line ratios.
In principle we can tune the model predictions to follow the observed trends of $m_{\rm [SIII],[OIII]}$ and $m_{\rm [OIII],[OII]}$, but without further evidence from resolved observations we opt for not adding extra parameters related to different line ratios so as to avoid overfitting.
Furthermore, in MaNGA SF spaxels, the metallicity is found to correlate with the H$\alpha$ surface brightness, which is negatively correlated with the DIG contamination \citep{oey2007,zhang2017}.
Therefore, the overall level of the DIG contamination could also change with the metallicity of the MaNGA SF spaxels, which is not captured by our simple toy model.

In summary, in order to explain the observed dependence of the reddening relations on the metallicity, our model requires the relative strengths of various lines to change in the DIG along with \ohb. This extra assumption needs verification from resolved observations of \hii\ regions with a range of host galaxy masses and metallicities and is beyond the scope of the current paper.

{A more direct test of the scale-dependent attenuation model is to measure the attenuation of different emission lines using IFU data with high spatial resolutions. In Appendix~\ref{subsec:high_resol_hii}, we apply our method to the high-resolution IFU data within a single galaxy, IC 342. While the effect of multicomponent attenuation appears to vanish in this case, the uncertainties of the derived slopes are large due to the small number of spaxels. It would be interesting to determine a physical scale, around which the attenuations of different emission lines start to have good agreements. We plan to further investigate this problem using a larger set of high-resolution IFU data
in future work.
}

{Thus far, we have discussed the physical picture for the multicomponent attenuation, but we have not discussed its impact on observational studies. In the next section, we detail the biases from improper attenuation corrections.
}

\section{Impact on nebular diagnositcs}
\label{sec:impact}

{There are many widely used strong-line metallicity calibrations and ionization parameter calibrations relying on attenuation-sensitive line ratios. For example, the calibration of N/O in \hii\ regions relies on the attenuation corrected \no\ \citep{thurston1996}, the R23 metallicity estimator involves the attenuation corrected [O\,{\sc ii}]/H$\beta$ \citep{pagel1979}, the ionization parameter calibration usually involves the attenuation corrected \oo\ or [S\,{\sc iii}]/[S\,{\sc ii}] \citep{diaz1991,kewley2002}, and so forth.
}

{If the attenuation correction is biased, the derived nebular parameters based on the line ratios above would be biased. Our results have shown that using a single attenuation curve for correcting emission lines from different transitions is inaccurate. However, how to quantify this impact is nontrivial.
}

{If we ignored the underlying dust model and thought different lines were simply described by different attenuation curves, the fractional bias in nebular diagnostics from using a single attenuation curve was in general around 10\%\,--\,20\% depending on the calibration method and the overall attenuation.
For example, our results show the F99 extinction curve with $R_V = 3.1$ fits the attenuation of Balmer line ratios well but does not fits the attenuation of \no. From Table~\ref{tab:derived}, one can see using the F99 curve would underestimate the intrinsic \no, creating an average systematic bias of 0.05 dex in N/O derived from \no\ using \cite{thurston1996}'s prescription at $A_V = 1$, where $A_V$ is derived from the Balmer decrement.
As another example, using the F99 curve to correct [S\,{\sc iii}]/[S\,{\sc ii}] would underestimate the ionization parameter by 0.08 dex at $A_V = 1$ \citep{diaz1991}. Compared to the systematic difference between different strong-line calibrations \citep{kewley2019}, the amount of bias estimated this way is not very significant.
}

{However, as we have shown in Section~\ref{sec:explain}, the potential existence of unresolved multicomponent attenuation cannot be simply described by different attenuation curves.
In our two-component dust model, one expect the attenuation correction bias is small when the observed fluxes are dominated by $C_{\rm High}$ or $C_{\rm Low}$, but becomes larger when $C_{\rm High}$ and $C_{\rm Low}$ have similar contributions to the observed fluxes.
Using the same model as we constructed in Section~\ref{sec:impact}, where the intrinsic log(H$\alpha$/H$\beta$/2.86) is set to 0.4 for $C_{\rm High}$ and 0 for $C_{\rm Low}$, we can calculate how the line ratios corrected based on the apparent log(H$\alpha$/H$\beta$/2.86) differ from the intrinsic line ratios.
For a partial coverage dust model, the bias due to the F99 correction in \no\ is 0.06 dex when the apparent $A_V = 1$.
For a \hii\ + DIG model assuming the parameters in Table~\ref{tab:model_param}, the bias becomes 0.25 dex at $A_V = 1$, which would significantly underestimate the intrinsic N/O by 44\%.
Meanwhile, there is also a bias in the corrected H$\alpha$ flux, which is roughly 0.23 dex at $A_V = 1$ in both cases, which would bias the estimated SFR by 0.16 dex or 0.23 dex depending on whether the less attenuated component is from DIG or \hii\ regions.
We note that the bias estimation strongly relies on the dust model and could vary in galaxies.
For example, the larger the attenuation difference between $C_{\rm High}$ and $C_{\rm Low}$ or the larger the line ratio difference between $C_{\rm High}$ and $C_{\rm Low}$, the larger the resulting bias.
Regardless, estimating the bias using dust models are more physically motivated than simply using different attenuation curves.}

{We thereby conclude that the impact from using a single attenuation curve depends on the dust model and the physical conditions in galaxies. In our two-component model, the impact would be negligible if a partial coverage scenario is assumed, but would be significant if the differently attenuated components have different intrinsic line ratios.
Again, to constrain the dust model, we need high-resolution data in future work.
}

To proceed further from the physical picture we proposed, it is important to verify the assumptions we made in our method.
In the next section, we discuss the validity of our method in detail.

\section{Discussion}
\label{sec:discussions}

{In this section we examine three important assumptions we made in this work. First, we discuss the form of the likelihood function and check the effect of the intrinsic scatter. Second, we discuss the 3D binning scheme and check its impact on the results. Third, we check the sample selection effect and its implications.}


\subsection{Impact of the intrinsic scatter in the Balmer decrement}
\label{subsec:int_scatter}

\begin{figure*}
    \centering
    \includegraphics[width=0.38\textwidth]{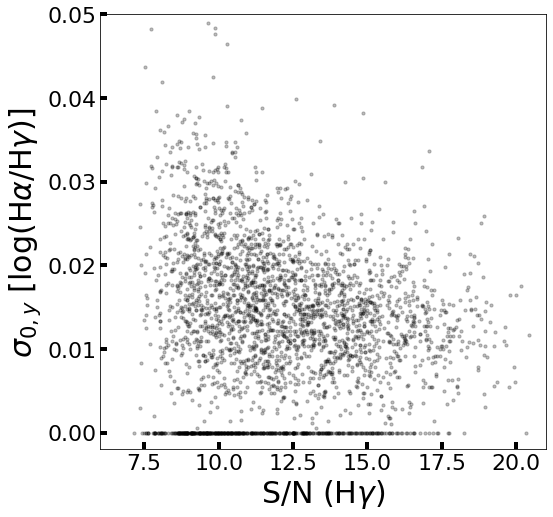}
    \includegraphics[width=0.38\textwidth]{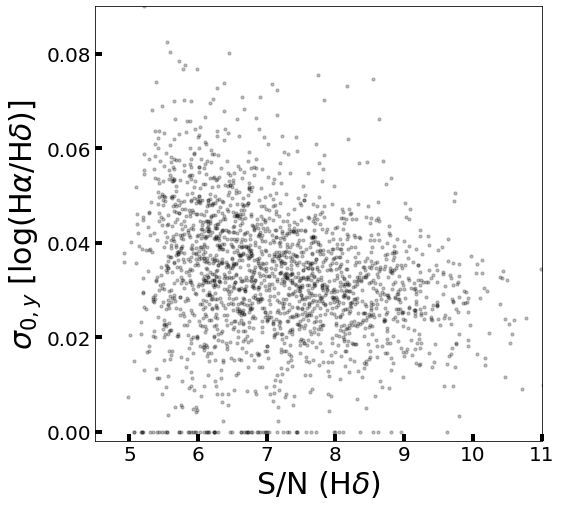}
    \caption{Intrinsic scatters in $y$ for the Balmer reddening relations as functions of the S/N of the weaker Balmer lines. Each point represents a 3D bin, and its median S/N is plotted. {\it Left}: how the derived intrinsic scatter in log(H$\alpha$/H$\gamma$) varies with the S/N of H$\gamma$. {\it Right}: how the derived intrinsic scatter in log(H$\alpha$/H$\delta$) varies with the S/N of H$\delta$.
    }
    \label{fig:int_bal}
\end{figure*}

{Our likelihood function, Equation~\ref{eq:lf_i}, considers measurement errors on both axes, but assumes the intrinsic scatter is only in $y$, the dependent variable. When there is truly no intrinsic scatter in $x$, our function can well reproduce the true slope, intercept, and the intrinsic scatter in $y$, as is shown in Appendix~\ref{appendix}.}

We also performed tests where intrinsic scatter was manually added to $x$. In such cases, ignoring the intrinsic scatter in $x$ during the fits results in an underestimation of the true slope. Also, it is noteworthy that not a great amount of the intrinsic scatter is needed to result in considerable change in the derived slope. For example, a Gaussian intrinsic scatter with $\sigma \approx 0.022$ in the Balmer decrement is enough to make one underestimate the $m_{\rm [NII],[OII]}$ by $\sim 0.3$, changing it from the F99 value to the median value we measured in MaNGA.

Despite its important effect, there is much evidence against the presence of the intrinsic scatter in $x$, the Balmer decrement. The most likely origin of the intrinsic scatter in $x$ is the variation in the intrinsic H$\alpha$/H$\beta$ due to the temperature and density variation \citep{agn3}. However, by our construction of the 3D bins, variations in the metallicity and ionization parameter are negligibly small within each bin, which makes the temperature variations to be small as well. In fact, according to our {\sc cloudy} model, even when we consider a wide range of metallicity and ionization parameter, with $\rm 7.39\leq 12+\log (O/H) \leq 9.19$ and $\rm -4.0\leq \log(U) \leq -1.5$, the {maximum variation} in H$\alpha$/H$\beta$ is only 0.022~dex from minimum to maximum. The expected variation in each bin is much smaller and cannot explain the discrepancy between the F99 values and our measurements. Furthermore, we performed another test where we added an extra dimension using the density-sensitive log([S\,{\sc ii}]$\lambda 6716$/[S\,{\sc ii}]$\lambda 6731$) and constructed 4D bins, within which the density variations are well constrained. The results based on the 4D bins are nearly identical to the ones based on the 3D bins, meaning that density variations do not introduce any noticeable intrinsic scatter.

Even if we assume the intrinsic scatter in the Balmer decrement were somehow responsible for the discrepancy between the observed slopes and F99, we would have to adopt different values of intrinsic scatter for different slope measurements.
For example, while making $m_{\rm [NII],[OII]}$ agree with F99 requires an intrinsic scatter of 0.022~dex for the Balmer decrement, $m_{\rm [OIII],[OII]}$ needs a significantly different intrinsic scatter of 0.034~dex. In addition, slopes of Balmer line ratios and high ionization line ratios are already in good agreement with F99 and only allow an intrinsic scatter $< 0.01$~dex. These results are not self-consistent, as the intrinsic scatter in $x$ should only depend on the data distribution along the $x$ axis, which is the same for all the slope measurements using the same set of 3D bins.

However, according to Figure~\ref{fig:balmer}, our likelihood function predicts nonzero intrinsic scatters for H$\alpha$/H$\gamma$ and H$\alpha$/H$\delta$, which seems to contradict our assumption. We note that the origin of this intrinsic scatter might come from the imperfect estimations on the measurement errors. As shown in Figure~\ref{fig:int_bal}, the derived intrinsic scatters for these Balmer ratios are negatively correlated with the S/N of Balmer lines.
H$\alpha$/H$\delta$ also shows larger intrinsic scatter compared to H$\alpha$/H$\gamma$, which could come from the overall poorer detection on H$\delta$.
{Specifically, the errors resulted from the stellar continuum subtraction could play a role, which is supposed to be anticorrelated with the S/N.}
It is possible that the likelihood function attempts to compensate the effects from the total errors with the intrinsic scatter. If it is true, one expects H$\alpha$/H$\beta$ to show even smaller intrinsic scatter ($< 0.01$), as H$\beta$ is has overall higher S/N than H$\gamma$
\footnote{One might also wonder about the impact from flux calibrations. Since we defined the reddening relations in logarithmic line-ratio spaces, any systematic bias in flux calibrations would only change the intercept rather than the slope.}.

We used a series of tests to further check if there could be any intrinsic scatter in H$\alpha$/H$\beta$. First, we found that when there was significant intrinsic scatter in $x$ that was ignored, the measured intrinsic scatter for $y$ would be overestimated, and the measured slope would be underestimated. Second, we measured a new reddening relation between log(\no) and log(H$\alpha$/H$\gamma$). The F99 value for this relation is 1.321, whereas the value based on our previous measurements, $m^{\prime}_{\rm [NII],[OII]}/m^{\prime}_{\rm H\alpha,H\gamma}$, is 1.134. When we ignored the intrinsic scatter in H$\alpha$/H$\gamma$, we obtained a slope of 1.08, which is indeed an underestimation. If we included the intrinsic scatter in H$\alpha$/H$\gamma$ in the likelihood function, we obtained a slope of 1.145, which is more consistent with the value based on our previous measurements. In addition, the derived intrinsic scatter for log(\no) is 0.033, in good agreement with our previous measurement, which is 0.032. If H$\alpha$/H$\beta$ has significant intrinsic scatter, one expects the intrinsic scatters in both log(\no) and log(H$\alpha$/H$\gamma$) were overestimated previously. As a result, using the overestimated intrinsic scatter in H$\alpha$/H$\gamma$, one should overestimate the slope of the log(\no) versus log(H$\alpha$/H$\gamma$) relation. In other words, if the intrinsic scatter in H$\alpha$/H$\beta$ is truly responsible for making $m^{\prime}_{\rm [NII],[OII]}$ deviate from the F99 value, our derived slope for the log(\no) versus log(H$\alpha$/H$\gamma$) relation should actually be {larger} than the F99 value.
From our tests, we see that the intrinsic scatter in H$\alpha$/H$\beta$, even if present, should be negligibly small.

Finally, we would like to comment on the general approach to linear regression problems with measurement uncertainties and intrinsic scatters. The method we adopted from \cite{tremaine2002} cannot simultaneously estimate intrinsic scatters in both axes. Only when one knows the intrinsic scatter in one axis, can one use this method to obtain an unbiased estimation of the intrinsic scatter in the other axis. The general approach by \cite{kelly2007} also does not deal with the case where both axes have intrinsic scatters. However, in principle one can further generalize \cite{kelly2007}'s approach. Another question is whether a Gaussian component is enough to describe the intrinsic scatter, especially when it is related to measurement errors. We intend to further explore the possibility of a general solution in future work.

\subsection{Impact of the binning method on the derived slopes}
\label{subsec:binning_eff}

Most of our slope measurements are based on cubic bins in the log(\nha)-log(\sha)-log(\ohb) space with number of spaxels greater than 180. We performed a series of tests to see how the size and the number cut for the 3D bins influence our results.

On the one hand, one expects the larger the bin size, the greater the intrinsic scatter within each bin, which increases the uncertainty of the measured slope. To test this effect, we made the {volume of each bin eight times as big (with a bin size of $0.034\times0.034\times0.034~{\rm dex^{3}}$)}, but keep the cut on the number of spaxels. While we found $m^{\prime}_{\rm [NII],[OII]}$ remains roughly unchanged, the $\sigma _{\rm std}$ of the slope distribution increases from 0.13 to 0.17. For reddening relations that have shallower slopes and thus are more susceptible to the intrinsic scatter, one might expect the median slopes are also affected by the bin size. For example, $m^{\prime}_{\rm [OIII],[OII]}$ has a significant change from 0.46 to 0.30 after we adopted larger bins. However, the change in the median metallicity could play an important role in the change as well. Since the high metallicity MaNGA spaxels are more densely populated in the line-ratio space, using larger bins while keeping the cut on the number of spaxels allows more low metallicity bins to be considered while reducing the total number of high metallicity bins. This tends to lower the median slope for [O\,{\sc iii}]/[O\,{\sc ii}] (see e.g., Figure~\ref{fig:m_met}).

{Our conclusions also remain valid for measurements based on smaller bins. We performed another test by making the volume of each bin 5.7 times as small (with a bin size of $0.0084\times0.0084\times0.0116~{\rm dex^{3}}$). In this case no bin has more than 180 spaxels. By lowering the number cut to 55, we obtained 2783 bins, whose $m^{\prime}_{\rm [NII],[OII]} = 1.514\pm0.004$ and $m^{\prime}_{\rm [OIII],[OII]} = 0.508\pm0.004$ are again close to our results from the original binning.
Since each small-size bin contains fewer spaxels, the statistical uncertainty of the linear fitting becomes larger in each bin.
One can see there are different uncertainties for large bins and small bins. Regardless, we have shown that our conclusions remain valid over a wide range in bin size.
}

On the other hand, changing the cut on the number of spaxels also has an effect on the median and the width of the slope distribution.
We expect that increasing (decreasing) the lower limit of the cut reduces (enlarges) the width of the slope distribution while biasing the median slope to the high (low) metallicity value.
To check this, we kept the size of each bin unchanged and performed two additional sets of fitting with bins having more than 300 data points and 100 data points, which include 527 bins and 7287 bins, respectively.
We obtained $m^{\prime}_{\rm [NII],[OII]} = 1.51$ and $\rm \sigma _{std; [NII],[OII]}=0.10$ for bins having more than 300 data points, and $m^{\prime}_{\rm [NII],[OII]} = 1.51$ and $\rm \sigma _{std; [NII],[OII]}=0.16$ for bins having more than 100 data points.
Whereas $m^{\prime}_{\rm [OIII],[OII]} = 0.53$ and $\rm \sigma _{std; [OIII],[OII]}=0.10$ for bins having more than 300 data points, and $m^{\prime}_{\rm [OIII],[OII]} = 0.40$ and $\rm \sigma _{std; [OIII],[OII]}=0.23$ for bins having more than 100 data points. These results are consistent with our expectations. Specifically, $m^{\prime}_{\rm [OIII],[OII]}$ is increasingly reflecting the low metallicities regions as we lowered the number cut.

In brief, both the size of the 3D bins and the spaxel number cut could affect the width and the median values of the slope distribution, with the latter effect being mainly a reflection of the metallicity dependence of certain slopes.
Nevertheless, our main conclusion about the inconsistent attenuation laws between different species of lines still holds, nearly irrespective of the binning method.

\subsection{Bias from an attenuation dependent line-ratio space}
\label{3d_att_bias}


The next important effect associated with our binning method is the correlation between different slopes. As we have seen in Section~\ref{sec:results}, since we constructed bins by constraining the variation in \nha, \sha, and \ohb, slopes of some line ratios are tied together. This would not be a problem if all emission lines are seeing the same attenuation curve. However, if there are different attenuation laws for different lines due to, for example, contamination from the DIG, our assumption breaks down since \nha, \sha, and \ohb\ no longer form an attenuation-insensitive space.
This could bias the slope measurements.
In addition, there are problems associated with the additivity for certain line ratios, especially the hybrid line ratios, which could also be related to the attenuation dependence of the 3D space.

To understand this effect, we performed a simple test to recompute $m^{\prime}_{\rm [NII],[OII]}$ and $m^{\prime}_{\rm [OIII],[OII]}$ with a semi-empirical data set. To generate the data set, we first extracted the observed H$\alpha$ and H$\beta$ fluxes from the spaxels that were identified as SF regions in MaNGA and set them as the attenuated fluxes with no uncertainties for the simulated data. We then treated the H$\alpha$ and H$\beta$ fluxes corrected by the F99 extinction curve as the true unattenuated fluxes. The simulated ``observed'' fluxes were computed by adding Gaussian errors to the attenuated fluxes using the uncertainty estimations of DAP.

Next, we set the observed \nha, \sha, and \ohb\ ratios as the intrinsic ratios for the simulated data set.
Combining these with the intrinsic fluxes of Balmer lines, we obtained the intrinsic fluxes of [N\,{\sc ii}], [S\,{\sc ii}], and [O\,{\sc iii}]. However, when computing the attenuated fluxes for these lines, we used different attenuation curves for different lines. Specifically, we made $l_{\rm [NII]}=0.827~l^{\rm F99}_{\rm [NII]}$ and $l_{\rm [SII]}=0.827~l^{\rm F99}_{\rm [SII]}$, and $l_{\rm [OIII]} = 0.948~l^{\rm F99}_{\rm [OIII]}$. Here we assumed all low ionization lines share the same attenuation curve, while all high ionization lines share a different one. This assumption is oversimplified, but it can make the expected slopes including $m^{\prime}_{\rm [NII],[OII]}$ and $m^{\prime}_{\rm [OIII],[OII]}$ to be close to our measured values in Table~\ref{tab:derived}.
Measurement errors were subsequently added to these lines.

The intrinsic fluxes for [O\,{\sc ii}], on the other hand, were set by a theoretical photoionization model. With the SF-ionized model computed by \cite{ji2020b}, we estimated the metallicity and ionization parameter for every SF spaxel using the assumed true values of \nha, \sha, and \ohb. Then for each spaxel, we find the model prediction for \no\ using the estimated metallicity and ionization parameter. Combining \no\ with the intrinsic fluxes of [N\,{\sc ii}], we obtained the intrinsic fluxes for [O\,{\sc ii}]. The attenuated [O\,{\sc ii}] was then computed by setting $l_{\rm [OII]}=0.827l^{\rm F99}_{\rm [OII]}$.
After we obtained the simulated observed line ratios, we added small intrinsic scatters to these line ratios. Next, we corrected the manually attenuated \nha, \sha, and \ohb\ using the F99 curve and created 3D bins.
As the final step, we remeasured $m^{\prime}_{\rm [NII],[OII]}$ and $m^{\prime}_{\rm [OIII],[OII]}$ based on the semi-empirical data set.

As expected, the resulting slopes deviate from the F99 values due to our modification of the attenuation law. Interestingly, they are not entirely consistent with the expected values from the modified curves either. We found $m^{\prime}_{\rm [NII],[OII]} \approx 1.59$ and $m^{\prime}_{\rm [OIII],[OII]} \approx 0.61$, whereas the modified curves predict $m^{\prime}_{[\rm NII],[OII]} \approx 1.53$ and $m^{\prime}_{\rm [OIII],[OII]} \approx 0.45$. In addition, the additivity of slopes is violated and $m^{\prime}_{\rm [NII],[OII]}-m^{\prime}_{\rm [NII],[OIII]}-m^{\prime}_{\rm [OIII],[OII]}\approx 0.09$.
We note that $m^{\prime}_{\rm [NII],[OIII]}\approx0.89\approx m^{\rm F99}_{\rm [NII],[OIII]}$ even after we modified the attenuation law. In comparison, the expected modified value for $m^{\prime}_{\rm [NII],[OIII]}$ is 1.07. This indicates $m^{\prime}_{\rm [NII],[OIII]}$ is likely fixed by our construction of the 3D bins and the preapplied extinction correction based on the F99 curve, making it almost always follow the F99 extinction curve.

When generating the semi-empirical data, we noticed another effect associated with the measurement errors. If we made the propagated measurement errors for the logarithmic line ratios Gaussians rather than made the flux errors Gaussians, the additivity seemed to be retrieved, although the slopes were still different from the expected values.
In such a case, $m^{\prime}_{\rm [NII],[OII]} \approx 1.53$, $m^{\prime}_{\rm [OIII],[OII]} \approx 0.68$, and $|m^{\prime}_{\rm [NII],[OII]}-m^{\prime}_{\rm [NII],[OII]}-m^{\prime}_{\rm [NII],[OIII]}|< 0.01$.
This effect suggests the loss of the additivity is associated with the error distributions. If the flux errors are Gaussians and the propagated errors for logarithmic line ratios deviate from Gaussians significantly, the likelihood function seems to bias the results and violate the additivity.
Once we manually set the measurement errors to be 1/2 of the current values, the additivity was satisfied to a good approximation even if we assumed Gaussian errors for line fluxes.
However, as we have seen in Section~\ref{subsec:hybrid}, increasing the S/N does not improve the additivity for \no,\oo, and [N\,{\sc ii}]/[O\,{\sc iii}] in the MaNGA sample. In addition, we did not find such a bias associated with the error distribution in another simulated data set in Appendix~\ref{appendix} even after we manually lowered the S/N of lines to be close to 3.

In fact, by checking the dependence of the derived slopes on various physical parameters, we find the surface brightness of H$\alpha$ seems more relevant to the additivity issue compared to the S/N, which we discuss in Section~\ref{subsec:depend_obs_pro}. All the evidence implies that although our semi-empirical data set can provide hints on the effects of binning biases, it is not realistic enough to reproduce our measurements with the MaNGA main sample.

In summary, when there exists different attenuation laws for different lines, the final slopes derived from 3D bins are affected but could still be different from the expected values from the modified curves. Meanwhile, the assumption of the error distribution could lead to the loss of the additivity of slopes, but it cannot explain the results we found in our sample.

\subsection{Reliability of the BPT diagrams}

A fundamental assumption of our 3D binning method is that the line ratio variations in the 3D-BPT space are reflecting the intrinsic variations in the metallicity, ionization parameter, and other physical parameters (except the attenuation) in SF regions. This assumption is widely adopted since the theoretical photoionization models with varying nebular parameters can well reproduce the shapes of the SF loci in the BPT diagrams \citep[e.g.,][]{kewley2001,dopita2013,dagostino2019,kewley2019}.
In addition, although photoionization models show significant degeneracy between the metallicity and ionization parameter in the 2D-BPT diagrams, the degeneracy disappears in the 3D-BPT space \citep{vogt2014,ji2020b}. However, there are still several caveats we would like to discuss.

First, the SF loci in the BPT diagrams could be affected by the DIG contamination. As shown by \cite{sanders2020}, the resolved \hii\ regions in the CHemical Abundances Of Spirals survey \citep[CHAOS,][]{berg2015,croxall2015,croxall2016,berg2020} show a systematic offset from the MaNGA SF spaxels in the [S\,{\sc ii}]-based BPT diagram. If the DIG contamination is strong enough to shift observed SF regions in the BPT diagrams, there would be important systematic uncertainties when people use these line ratios to constrain metallicities and ionization parameters. However, as noted by \cite{mannucci2021}, the aperture effect for \hii\ regions become important for high-resolution studies. When a single \hii\ region
is not fully sampled in space, the line ratios would depend on the size of the aperture and the results cannot be directly compared with photoionization models that calculate the integrated emission line spectra for the whole \hii\ region.
After taking the aperture effect into account, \cite{mannucci2021} argue the DIG contamination only has a secondary effect on the BPT diagnostics of MaNGA SF spaxels. Therefore, the DIG contamination is likely not strong enough to shift the whole SF locus in the line-ratio space. Whereas the DIG contamination on the observed attenuation within small regions of the line-ratio space cannot be ruled out.

Second, there could be degeneracy between the evolutionary stage and the metallicity of \hii\ regions. By modeling the time evolution of an \hii\ region, \cite{pellegrini2020} were able to cover a large area in the [N\,{\sc ii}]-based BPT diagrams with a photoionization model grid at fixed metallicity. If the observed \hii\ regions really show a significant variation in their evolutionary stages, they would show scatter similar to the metallicity variation in BPT diagrams \citep[see also][]{byler2017}.
Still, the level of potential degeneracy introduced by this effect is still unclear.
In surveys like MaNGA, the unresolved \hii\ regions are effectively light-weighted ionized regions. Since \hii\ regions at different evolutionary phases are characterized by different brightnesses, it is likely that the observations are biased toward some average phases, which might alleviate this degeneracy.

Finally, this work also raises a question about whether the BPT space is truly attenuation-free at low spatial resolution.
Although we started our calculation with the attenuation-free assumption, the results we obtained suggest differential attenuation that depends on the species of lines, undermining the initial assumption we made.
Under different attenuation curves, variations in metallicities and ionization parameters are no longer well constrained in individual bins, which is reflected by the results we show in the previous subsection.
Still the amount of bias associated with this process depends on the mechanism that results in different apparent attenuation laws, which could be much more complicated than the DIG contamination model we proposed.

To conclude, despite being powerful diagnostic tools, BPT diagrams are still not fully understood in terms of their applicability and their implication.
A large spectroscopic sample of well-resolved \hii\ regions and more sophisticated photoionization models are both needed to solve the aforementioned problems.

\subsection{Dependence of the derived attenuation law on galaxy properties}
\label{subsec:depend_obs_pro}

\begin{table*}
	\centering
	\caption{Median slopes of reddening relations of different line ratios derived from samples selected by $\rm \Sigma _{H\alpha}$ and $b/a^{\rm a}$.}
	{\fontsize{9}{11}\selectfont
	\label{tab:hainav_com}
	\begin{tabularx}{2\columnwidth}{lcccccc} 
	    \hline\hline
        Sample selection & $m^{\prime}_{\rm [SIII],[OIII]}$ & $m^{\prime}_{\rm [SIII],H\alpha}$ & $m^{\prime}_{\rm H\alpha ,[OIII]}$ & $m^{\prime}_{\rm [NII],[OII]}$ & $m^{\prime}_{\rm [NII],[OIII]}$ & $m^{\prime}_{\rm [OIII],[OII]}$\\
        \hline
        $\rm \Sigma _{H\alpha} < 10^{39}~erg/s/kpc^{2}$ & $3.16\pm 0.05$ & $1.13\pm 0.07$ & $0.8918\pm 0.0006$ & $2.05\pm 0.03$ & $0.9055\pm 0.0008$ & $-0.25\pm 0.02$\\
        \hline
        $\rm \Sigma _{H\alpha} > 2\times10^{39}~erg/s/kpc^{2}$ & $1.705\pm 0.008$ & $0.770\pm 0.008$ & $0.8890\pm 0.0005$ & $1.456\pm 0.005$ & $0.9031\pm 0.0007$ & $0.502\pm 0.005$\\
        \hline
        $b/a < 0.6$ & $1.767\pm 0.009$ & $0.74\pm 0.01$ & $0.8888\pm 0.0004$ & $1.525\pm 0.004$ & $0.9020\pm 0.0006$ & $0.490\pm 0.005$\\
        \hline
        $b/a > 0.7$ & $1.68\pm 0.02$ & $0.33\pm 0.02$ & $0.8915\pm 0.0006$ & $1.495\pm 0.007$ & $0.9068\pm 0.0008$ & $0.187\pm 0.010$\\
        \hline
        \hline
        F99 curve with $R_V = 3.1$ & 1.695 & 0.811 & 0.8845 & 1.844 & 0.8935 & 0.950\\
        \hline
    \end{tabularx}
    \begin{tablenotes}
        \small
        \item $\bf Notes.$
        \item $^{\rm a}$ 
        When splitting the sample according to each of the parameters above, we only selected 3D bins with the number of spaxels satisfying each selection criterion greater than $60\sim 90$ so that the final number of bins is $\sim 1000$.
    \end{tablenotes}
	}
\end{table*}

As summarized by \cite{salim2020}, observations have shown the shape of the attenuation curve depends on a number of galaxy properties including the metallicity, stellar mass, inclination, dust column density, etc. Thus far, we have seen the relative attenuation we derived shows dependencies on the metallicity and the S/N of the lines. Some of the dependencies might come from the true variation of the attenuation law, while others might be related to the observational bias induced by differently attenuated components.
In this subsection we examine these potential dependencies by checking how the sample selections based on the H$\alpha$ surface brightness ($\rm \Sigma _{H\alpha}$), inclination ($b/a$), and $A_V$ influence our results. The H$\alpha$ surface brightness is anticorrelated with the contamination from the DIG \citep{oey2007,zhang2017}, and it is also partly related to the S/N of lines and the metallicity of the galaxy.
Therefore, the observational bias associated with the unresolved DIG could be reflected by the dependence of the attenuation on $\rm \Sigma _{H\alpha}$.
The inclination and $A_V$ are closely related to the column density of the dust and could reflect the intrinsic variations of the attenuation curve due to dust geometries and compositions.

Table~\ref{tab:hainav_com} summarizes the resulting slopes for several line ratios with different sample selection criteria.
We inspected two sets of line ratios: 1) [S\,{\sc iii}]/[O\,{\sc iii}], [S\,{\sc iii}]/H$\alpha$, and H$\alpha$/[O\,{\sc iii}], 2) \no, [N\,{\sc ii}]/[O\,{\sc iii}], \oo. Ideally, slopes for each set of line ratios should satisfy additivity.
When doing the sample selection, we made sure the number of spaxels selected by each selection criterion is larger than a minimum number (roughly $60\sim 90$). Meanwhile, we used the same set of 3D bins when making cut on the same physical parameter to control the metallicity effect on the resulting slopes, though the spaxels in each bin could be somewhat different. {We chose not to divide the sample according to $A_V$, of which the reason will be discussed later in this subsection.}

For the sample selection based on $\rm \Sigma _{H\alpha}$, we made two samples with $\rm \Sigma _{H\alpha} < 10^{39}~erg/s/kpc^{2}$ and $\rm \Sigma _{H\alpha} > 2\times10^{39}~erg/s/kpc^{2}$, where $\rm \Sigma _{H\alpha}$ is corrected for extinction using the F99 extinction curve.
One can see for the low $\rm \Sigma _{H\alpha}$ sample, $m^{\prime}_{\rm [SIII],[OIII]}$ is much larger than $m^{\rm F99}_{\rm [SIII],[OIII]}$ and $m^{\prime}_{\rm [SIII],[OIII]} > m^{\prime}_{\rm [SIII],H\alpha} + m^{\prime}_{\rm H\alpha ,[OIII]}$.
On the other hand, $m^{\prime}_{\rm [NII],[OII]}$ is slightly closer to $m^{\rm F99}_{\rm [NII],[OII]}$ at low $\rm \Sigma _{H\alpha}$, but has much larger uncertainty. Meanwhile, at low $\rm \Sigma _{H\alpha}$, $m^{\prime}_{\rm [OIII],[OII]}$ even becomes negative.
In fact, all measured slopes except $m_{\rm H\alpha,[OIII]}$ and $m_{\rm [NII],[OIII]}$ show wide distributions at low $\rm \Sigma _{H\alpha}$, which are reflected by the uncertainties of the medians.
For $m^{\prime}_{\rm H\alpha,[OIII]}$ and $m^{\prime}_{\rm [NII],[OIII]}$, their measured values are fixed by our construction of the 3D bins (see Section~\ref{3d_att_bias}).

At high $\rm \Sigma _{H\alpha}$, both sets of line ratios follow the additivity better, although $m^{\prime}_{\rm [NII],[OII]}$ deviates more from the F99 value.
The behavior of the first set of line ratios involving [S\,{\sc iii}] and [O\,{\sc iii}] seems to support the scenario of the DIG contamination, which should be reduced in brighter regions dominated by \hii\ regions.
As $\rm \Sigma _{H\alpha}$ increases, both $m^{\prime}_{\rm [SIII],[OIII]}$ and $m^{\prime}_{\rm [SIII],H\alpha}$ become much closer to the expected F99 values.
While for the second set of line ratios, despite the retrieval of its additivity at high $\rm \Sigma _{H\alpha}$, $m^{\prime}_{\rm [NII],[OII]}$ and $m^{\prime}_{\rm [OIII],[OII]}$ remain biased at high $\rm \Sigma _{H\alpha}$, although $m^{\prime}_{\rm [OIII],[OII]}$ does get closer to the corresponding F99 value. One might wonder whether the slopes of Balmer lines also change with $\rm \Sigma _{H\alpha}$. We found both $m^{\prime}_{\rm H\alpha,H\gamma}$ and $m^{\prime}_{\rm H\alpha,H\delta}$ to slightly decrease at high $\rm \Sigma _{H\alpha}$, but $m^{\prime}_{\rm H\alpha,H\delta}$ is still greater than $m^{\prime}_{\rm [NII],[OII]}$ by roughly 0.05, meaning they still cannot be fitted by the same attenuation curve.

In Section~\ref{subsec:hybrid}, we found the S/N cut was able to improve the additivity for the first set of line ratios, but failed for the second set of line ratios.
Compared to the cut on S/N, the cut on $\rm \Sigma _{H\alpha}$ provides an overall better recovery of the additivity, but the physical picture behind is still unclear. In addition, the DIG contamination scenario is not able to explain the deviations of $m^{\prime}_{\rm [NII],[OII]}$ and $m^{\prime}_{\rm [OIII],[OII]}$ from the F99 values. 
{We also ran another test that divided the sample according to the equivalent width of H$\alpha$, which can also be used to remove the DIG \citep{cidfernandes2011}. However, we obtained similar results where $m^{\prime}_{\rm [NII],[OII]}$ and $m^{\prime}_{\rm [OIII],[OII]}$ deviate from the F99 value.}
Still, geometric effects caused by partially covered \hii\ regions could create a modified attenuation curve at low resolution without the DIG (see Figure~\ref{fig:cartoon} and Figure~\ref{fig:toy_model}). Then the question becomes how the dust geometry would be different for different lines in bright \hii\ regions. Again high-resolution data are the key to fully understanding this problem, which we will explore in future work.

For the sample selection based on $b/a$, we made two samples with $b/a < 0.6$ and $b/a > 0.7$, respectively.
This time the differences between the slopes derived from the two samples are smaller. The high inclination sample ($b/a < 0.6$) shows higher $m^{\prime}_{\rm [SIII],[OIII]}$ and $m^{\prime}_{\rm [SIII],H\alpha }$, corresponding to a larger $R_V$ under the parameterization of \cite{fitzpatrick1999}. If these differences are truly driven by the change of the attenuation curve rather than the observational bias, it means the more inclined regions with larger dust column density exhibits a grayer attenuation curve, which is qualitatively consistent with the conclusion of \cite{salim2020}.
However, $m^{\prime}_{\rm [NII],[OII]}$ and $m^{\prime}_{\rm [OIII],[OII] }$ also become larger at high inclination, which requires a lower $R_V$ instead.
This suggests either the intrinsic variation in the shape of the attenuation curve due to the change of the inclination is overwhelmed by the observation bias, or the intrinsic variation cannot be completely characterized by the variation in $R_V$.


\begin{figure}
    \centering
    \includegraphics[width=0.4\textwidth]{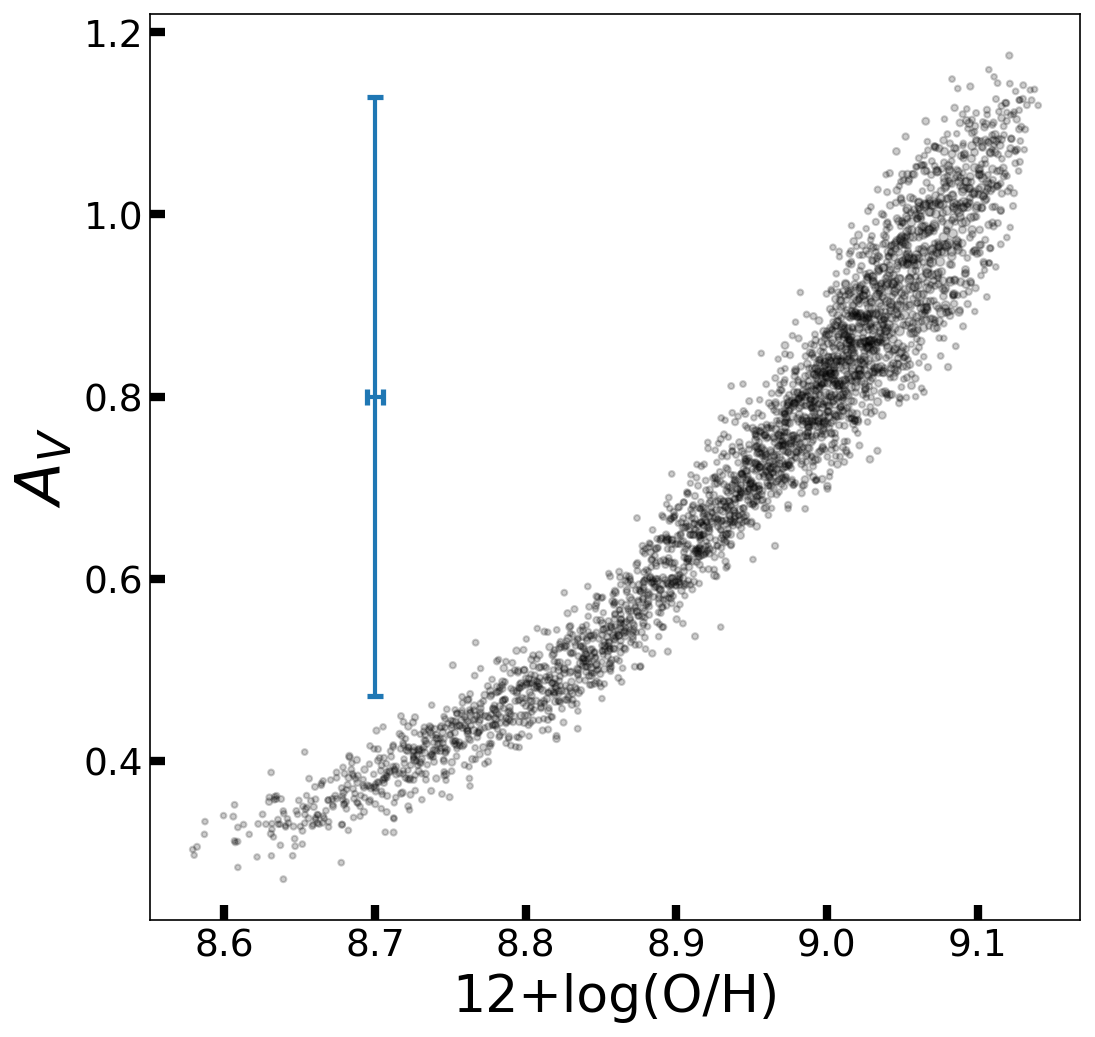}
    \caption{Correlation between the metallicity and the V-band attenuation $A_V$ for the 3D bins that contain more than 180 spaxels. The V-band attenuation is derived from the Balmer decrement. For each 3D bin, the median values of 12+log(O/H) and $A_V$ are plotted. 
    The vertical and horizontal errorbars show the typical (median) standard deviations of $A_V$ and 12+log(O/H), respectively, in each 3D bin.
    }
    \label{fig:avoh}
\end{figure}

Finally, we checked the dependence of our results on $A_V$. According to \cite{salim2020}, $A_V$ is a more fundamental parameter that correlates with the variation of the attenuation curve, {which has been extensively investigated in works using galaxy spectral energy distribution (SED) fitting. 
Still, the observational determination of the $A_V$ dependence based on galaxy SED fitting might be subject to systematic biases from correlated uncertainties \citep{qin2022}.}
We calculated $A_V$ using the Balmer decrement assuming the F99 extinction curve.
{However, we cannot directly construct subsamples with $A_V$ like we did with the previous two parameters. This is because $A_V$ is exactly the $x$ axis in the reddening relation we defined. Making cuts on the observed independent variable with measurement errors can easily bias the linear regression results, which is similar to the ``missing data problem'' \citep[see e.g.,][]{gelman1995,little2002,kelly2007}.
Indeed, we found in many cases that within a 3D bin, the slopes given by our maximum-likelihood method using the high $A_V$ spaxels and low $A_V$ spaxels are both larger than when using all spaxels.}
Therefore, instead of making subsamples, we checked how the derived slopes depend on the median $A_V$ within each bin. We found the slope versus $A_V$ relations are very similar to the slope versus metallicity relations.
In fact, as shown by Figure~\ref{fig:avoh}, the median $A_V$ shows a strong positive correlation with the metallicity, indicating more metal-rich clouds are also dustier on average.
If the changes in the slopes are mainly driven by the change in $R_V$ due to increasing $A_V$, one expects in Figure~\ref{fig:m_met}, $m^{\prime}_{\rm [SIII],[OIII]}$ should increase with increasing metallicity (and $A_V$) and $m^{\prime}_{\rm [OIII],[OII]}$ should decrease with increasing metallicity (and $A_V$) as the attenuation curve become grayer.
Given the opposite trends we obtained, the metallicity effect is more likely to be responsible for the changes of slopes.
{Due to the tight correlation between the median $A_V$ and metallicity, we cannot find any clear dependence of the slopes on the median $A_V$ at fixed metallicities.}



In summary, due to the observational bias possibly related to the contamination of the DIG, the relative attenuation of forbidden lines show complex dependencies on various galaxy properties. Specifically, increasing $\rm \Sigma _{H\alpha}$ seems to improve the additivity of the slopes of the reddening relations, but the deviation from the expected F99 curve remains for $m^{\prime}_{\rm [NII],[OII]}$ and $m^{\prime}_{\rm [OIII],[OII]}$. Changing $b/a$ and $A_V$ also affects the derived slopes, but the results are likely dominated by the observational bias rather than the intrinsic variation in $R_V$.

\section{Summary and conclusions}
\label{sec:conclusions}


In this work we study the nebular attenuation probed by different optical-to-NIR emission lines in MaNGA via a novel method. This method bins observed SF spaxels in a 3D line-ratio space spanned by \nha, \sha, and \ohb. In theory, within a small region in this attenuation-insensitive space, variations in nebular parameters such as the metallicity and ionization parameter are well constrained, whereas the magnitude of nebular attenuation is free to vary. Therefore, variations in the attenuation-sensitive line ratios are dominated by nebular attenuation in each small 3D bin. By fitting linear models to reddening relations between logarithmic attenuation-sensitive line ratios and logarithmic Balmer decrements, we measured the nebular attenuation curve using different species of lines. We summarize our results as follows.

First, reddening relations of high ionization line ratios, including \so\ and [O\,{\sc iii}]/[Ne\,{\sc iii}], shows median slopes consistent with the predictions of a \cite{fitzpatrick1999} extinction with $R_V = 3.1$. The same attenuation curve can well describe the attenuation of Balmer lines as well. In contrast, line ratios involving low ionization lines appear to follow a grayer attenuation curve inconsistent with high ionization line ratios and Balmer line ratios. As a result, a single attenuation curve cannot simultaneously explain the reddening of high ionization lines, Balmer lines, and low ionization lines.

Second, we suspect the low spatial resolution of MaNGA causes this discrepancy and propose a toy model that can produce different observed attenuation laws for different species of lines. In the toy model, two distinct line-emitting components with different levels of attenuation and different intrinsic line ratios are mixed and their contributions to the integrated line fluxes vary in different observations. The DIG contamination could fit into this scenario and explain the variations of the attenuation curves as a result of the different intrinsic line ratios in the DIG and \hii\ regions, but the model parameters are not well constrained and high-resolution data are needed for verification. We found evidence in the MaNGA ancillary observations of the nearby galaxy IC 342 that at a higher spatial resolution, all emission lines follow an attenuation curve more consistent with the \cite{fitzpatrick1999} one. However, the uncertainties of these measurements are too large and we need a larger volume of data to confirm these results.

{Third, the impact of the multicomponent dust attenuation we discovered on nebular diagnostics has a nontrivial dependence on the dust model and the physical conditions in galaxies. If one uses a single attenuation curve determined by Balmer lines to correct observed forbidden line ratios, the corrected line ratios and nebular parameters inferred from these line ratios could be biased by 0.06\,--\,0.25 dex when the apparent $A_V = 1$, depending on the assumptions in the attenuation model.}

Finally, we checked the assumptions we made in our measurements including {the intrinsic scatter}, the binning method, and the sample selection.
{Although the intrinsic scatter in the Balmer decrement, if present, could change our slope measurements, our tests rule out any significant amount of intrinsic scatter in H$\alpha$/H$\beta$ in our sample.}
We noticed the reddening relations of some line ratios are correlated due to our construction of the 3D bins. If our 3D space is no longer attenuation-free as indicated by the different attenuation laws, there would be biases associated with the measurements. Still, the main conclusion about the differential attenuation remains qualitatively unchanged, unless the 3D BPT diagrams we used are not reliable tools to reveal the overall variations in the metallicity and ionization parameter.
In addition, we found strong dependence of the measured slopes on the H$\alpha$ surface brightness. At high H$\alpha$ surface brightness, the reddening of line ratios related to [S\,{\sc iii}] converges to the predictions of the \cite{fitzpatrick1999} extinction curve, whereas the reddening of line ratios involving [O\,{\sc ii}] remain different from the predictions of this curve. This result cannot be fully explained by the DIG contamination, implying the different dust geometries inside \hii\ regions might also play a role.

Our results challenge the widely adopted assumption that different emission lines should follow the same attenuation law on all spatial scales, which is the basis for a large number of nebular diagnostics including the BPT diagnostics and metallicity calibrations.
However, since the physical picture behind this phenomenon is still unclear, currently we do not have a reliable and well-tested way to make corrections to the observed emission lines, and we will try to address this issue in future work.
The key to solving this issue is to perform similar analyses on large spectroscopic data sets with high spatial resolution and wide spectral coverage. Upcoming surveys such as AMASE \citep{yan2020} and LVM \citep{kollmeier2017} will provide valuable data for further investigations.
Meanwhile, the PHANGS-MUSE \citep{emsellem2022} data, despite covering a narrower spectral range, could provide hints on the spatial distribution of dust. In addition, PHANGS-MUSE covers a few low ionization lines such as [O\,{\sc i}]$\lambda 6300$, which could be used to compare with the attenuation measurements at MaNGA's resolution. We will compare the two data sets in a future paper.

\begin{acknowledgements}

The authors thank Kathryn Kreckel for providing the
observational data of IC 342 in the data release of SDSS-IV MaNGA.

The work described in this paper was partially supported by a grant from the Research Grants Council of the Hong Kong Special Administrative Region, China [Project No: CUHK 14302522].

RY also acknowledges support by the Hong Kong Global STEM Scholar scheme, by the Hong Kong Jockey Club Charities Trust through the JC STEM Lab of Astronomical Instrumentation, and by the Direct Grant of CUHK Faculty of Science.

MB acknowledges support from FONDECYT regular grant 1211000 and by the ANID BASAL project FB210003.

{RR thanks to Conselho Nacional de Desenvolvimento Cient\'{i}fico e Tecnol\'ogico (CNPq, Proj. 311223/2020-6, 304927/2017-1 and  400352/2016-8), Funda\c{c}\~ao de amparo \`{a} pesquisa do Rio Grande do Sul (FAPERGS, Proj. 16/2551-0000251-7 and 19/1750-2), and Coordena\c{c}\~ao de Aperfei\c{c}oamento de Pessoal de N\'{i}vel Superior (CAPES, Proj. 0001).}

R.A.R. acknowledges the support from Conselho Nacional de Desenvolvimento Cient\'ifico e Tecnol\'ogico and Funda\c c\~ao de Amparo \`a pesquisa do Estado do Rio Grande do Sul.

Funding for the Sloan Digital Sky Survey IV has been provided by the Alfred P. Sloan Foundation, the U.S. Department of Energy Office of Science, and the Participating Institutions. SDSS acknowledges support and resources from the Center for High-Performance Computing at the University of Utah. The SDSS web site is www.sdss.org.

SDSS is managed by the Astrophysical Research Consortium for the Participating Institutions of the SDSS Collaboration including the Brazilian Participation Group, the Carnegie Institution for Science, Carnegie Mellon University, the Chilean Participation Group, the French Participation Group, Harvard-Smithsonian Center for Astrophysics, Instituto de Astrof\'isica de Canarias, The Johns Hopkins University, Kavli Institute for the Physics and Mathematics of the Universe (IPMU) / University of Tokyo, the Korean Participation Group, Lawrence Berkeley National Laboratory, Leibniz Institut f\"ur Astrophysik Potsdam (AIP), Max-Planck-Institut f\"ur Astronomie (MPIA Heidelberg), Max-Planck-Institut f\"ur Astrophysik (MPA Garching), Max-Planck-Institut f\"ur Extraterrestrische Physik (MPE), National Astronomical Observatories of China, New Mexico State University, New York University, University of Notre Dame, Observat\'orio Nacional / MCTI, The Ohio State University, Pennsylvania State University, Shanghai Astronomical Observatory, United Kingdom Participation Group, Universidad Nacional Aut\'onoma de M\'exico, University of Arizona, University of Colorado Boulder, University of Oxford, University of Portsmouth, University of Utah, University of Virginia, University of Washington, University of Wisconsin, Vanderbilt University, and Yale University.

\end{acknowledgements}

%
   \bibliographystyle{aa} 
   \bibliography{ref} 

\begin{thebibliography}{102}
\expandafter\ifx\csname natexlab\endcsname\relax\def\natexlab#1{#1}\fi

\bibitem[{{Abdurro'uf} {et~al.}(2022){Abdurro'uf}, {Accetta}, {Aerts}, {Silva
  Aguirre}, {Ahumada}, {Ajgaonkar}, {Filiz Ak}, {Alam}, {Allende Prieto},
  {Almeida}, {Anders}, {Anderson}, {Andrews}, {Anguiano}, {Aquino-Ort{\'\i}z},
  {Arag{\'o}n-Salamanca}, {Argudo-Fern{\'a}ndez}, {Ata}, {Aubert},
  {Avila-Reese}, {Badenes}, {Barb{\'a}}, {Barger}, {Barrera-Ballesteros},
  {Beaton}, {Beers}, {Belfiore}, {Bender}, {Bernardi}, {Bershady}, {Beutler},
  {Bidin}, {Bird}, {Bizyaev}, {Blanc}, {Blanton}, {Boardman}, {Bolton},
  {Boquien}, {Borissova}, {Bovy}, {Brandt}, {Brown}, {Brownstein}, {Brusa},
  {Buchner}, {Bundy}, {Burchett}, {Bureau}, {Burgasser}, {Cabang}, {Campbell},
  {Cappellari}, {Carlberg}, {Wanderley}, {Carrera}, {Cash}, {Chen}, {Chen},
  {Cherinka}, {Chiappini}, {Choi}, {Chojnowski}, {Chung}, {Clerc}, {Cohen},
  {Comerford}, {Comparat}, {da Costa}, {Covey}, {Crane}, {Cruz-Gonzalez},
  {Culhane}, {Cunha}, {Dai}, {Damke}, {Darling}, {Davidson}, {Davies},
  {Dawson}, {De Lee}, {Diamond-Stanic}, {Cano-D{\'\i}az}, {S{\'a}nchez},
  {Donor}, {Duckworth}, {Dwelly}, {Eisenstein}, {Elsworth}, {Emsellem},
  {Eracleous}, {Escoffier}, {Fan}, {Farr}, {Feng}, {Fern{\'a}ndez-Trincado},
  {Feuillet}, {Filipp}, {Fillingham}, {Frinchaboy}, {Fromenteau}, {Galbany},
  {Garc{\'\i}a}, {Garc{\'\i}a-Hern{\'a}ndez}, {Ge}, {Geisler}, {Gelfand},
  {G{\'e}ron}, {Gibson}, {Goddy}, {Godoy-Rivera}, {Grabowski}, {Green},
  {Greener}, {Grier}, {Griffith}, {Guo}, {Guy}, {Hadjara}, {Harding},
  {Hasselquist}, {Hayes}, {Hearty}, {Hern{\'a}ndez}, {Hill}, {Hogg},
  {Holtzman}, {Horta}, {Hsieh}, {Hsu}, {Hsu}, {Huber}, {Huertas-Company},
  {Hutchinson}, {Hwang}, {Ibarra-Medel}, {Chitham}, {Ilha}, {Imig}, {Jaekle},
  {Jayasinghe}, {Ji}, {Johnson}, {Jones}, {J{\"o}nsson}, {Katkov}, {Khalatyan},
  {Kinemuchi}, {Kisku}, {Knapen}, {Kneib}, {Kollmeier}, {Kong}, {Kounkel},
  {Kreckel}, {Krishnarao}, {Lacerna}, {Lane}, {Langgin}, {Lavender}, {Law},
  {Lazarz}, {Leung}, {Leung}, {Lewis}, {Li}, {Li}, {Lian}, {Liang}, {Lin},
  {Lin}, {Lin}, {Lintott}, {Long}, {Longa-Pe{\~n}a}, {L{\'o}pez-Cob{\'a}},
  {Lu}, {Lundgren}, {Luo}, {Mackereth}, {de la Macorra}, {Mahadevan},
  {Majewski}, {Manchado}, {Mandeville}, {Maraston}, {Margalef-Bentabol},
  {Masseron}, {Masters}, {Mathur}, {McDermid}, {Mckay}, {Merloni},
  {Merrifield}, {Meszaros}, {Miglio}, {Di Mille}, {Minniti}, {Minsley},
  {Monachesi}, {Moon}, {Mosser}, {Mulchaey}, {Muna}, {Mu{\~n}oz}, {Myers},
  {Myers}, {Nadathur}, {Nair}, {Nandra}, {Neumann}, {Newman}, {Nidever},
  {Nikakhtar}, {Nitschelm}, {O'Connell}, {Garma-Oehmichen}, {Luan Souza de
  Oliveira}, {Olney}, {Oravetz}, {Ortigoza-Urdaneta}, {Osorio}, {Otter},
  {Pace}, {Padilla}, {Pan}, {Pan}, {Parikh}, {Parker}, {Peirani}, {Pe{\~n}a
  Ram{\'\i}rez}, {Penny}, {Percival}, {Perez-Fournon}, {Pinsonneault},
  {Poidevin}, {Poovelil}, {Price-Whelan}, {B{\'a}rbara de Andrade Queiroz},
  {Raddick}, {Ray}, {Rembold}, {Riddle}, {Riffel}, {Riffel}, {Rix}, {Robin},
  {Rodr{\'\i}guez-Puebla}, {Roman-Lopes}, {Rom{\'a}n-Z{\'u}{\~n}iga}, {Rose},
  {Ross}, {Rossi}, {Rubin}, {Salvato}, {S{\'a}nchez}, {S{\'a}nchez-Gallego},
  {Sanderson}, {Santana Rojas}, {Sarceno}, {Sarmiento}, {Sayres}, {Sazonova},
  {Schaefer}, {Schiavon}, {Schlegel}, {Schneider}, {Schultheis}, {Schwope},
  {Serenelli}, {Serna}, {Shao}, {Shapiro}, {Sharma}, {Shen}, {Shetrone}, {Shu},
  {Simon}, {Skrutskie}, {Smethurst}, {Smith}, {Sobeck}, {Spoo}, {Sprague},
  {Stark}, {Stassun}, {Steinmetz}, {Stello}, {Stone-Martinez},
  {Storchi-Bergmann}, {Stringfellow}, {Stutz}, {Su}, {Taghizadeh-Popp},
  {Talbot}, {Tayar}, {Telles}, {Teske}, {Thakar}, {Theissen}, {Tkachenko},
  {Thomas}, {Tojeiro}, {Hernandez Toledo}, {Troup}, {Trump}, {Trussler},
  {Turner}, {Tuttle}, {Unda-Sanzana}, {V{\'a}zquez-Mata}, {Valentini},
  {Valenzuela}, {Vargas-Gonz{\'a}lez}, {Vargas-Maga{\~n}a}, {Alfaro},
  {Villanova}, {Vincenzo}, {Wake}, {Warfield}, {Washington}, {Weaver},
  {Weijmans}, {Weinberg}, {Weiss}, {Westfall}, {Wild}, {Wilde}, {Wilson},
  {Wilson}, {Wilson}, {Wolf}, {Wood-Vasey}, {Yan}, {Zamora}, {Zasowski},
  {Zhang}, {Zhao}, {Zheng}, {Zheng}, \& {Zhu}}]{dr17}
{Abdurro'uf}, {Accetta}, K., {Aerts}, C., {et~al.} 2022, \apjs, 259, 35

\bibitem[{{Alloin} {et~al.}(1979){Alloin}, {Collin-Souffrin}, {Joly}, \&
  {Vigroux}}]{alloin1979}
{Alloin}, D., {Collin-Souffrin}, S., {Joly}, M., \& {Vigroux}, L. 1979, \aap,
  78, 200

\bibitem[{{Baldwin} {et~al.}(1981){Baldwin}, {Phillips}, \&
  {Terlevich}}]{baldwin1981}
{Baldwin}, J.~A., {Phillips}, M.~M., \& {Terlevich}, R. 1981, \pasp, 93, 5

\bibitem[{{Beers} {et~al.}(1990){Beers}, {Flynn}, \& {Gebhardt}}]{beers1990}
{Beers}, T.~C., {Flynn}, K., \& {Gebhardt}, K. 1990, \aj, 100, 32

\bibitem[{{Belfiore} {et~al.}(2017){Belfiore}, {Maiolino}, {Tremonti},
  {S{\'a}nchez}, {Bundy}, {Bershady}, {Westfall}, {Lin}, {Drory}, {Boquien},
  {Thomas}, \& {Brinkmann}}]{belfiore2017}
{Belfiore}, F., {Maiolino}, R., {Tremonti}, C., {et~al.} 2017, \mnras, 469, 151

\bibitem[{{Belfiore} {et~al.}(2022){Belfiore}, {Santoro}, {Groves},
  {Schinnerer}, {Kreckel}, {Glover}, {Klessen}, {Emsellem}, {Blanc}, {Congiu},
  {Barnes}, {Boquien}, {Chevance}, {Dale}, {Diederik Kruijssen}, {Leroy},
  {Pan}, {Pessa}, {Schruba}, \& {Williams}}]{belfiore2022}
{Belfiore}, F., {Santoro}, F., {Groves}, B., {et~al.} 2022, \aap, 659, A26

\bibitem[{{Belfiore} {et~al.}(2019){Belfiore}, {Westfall}, {Schaefer},
  {Cappellari}, {Ji}, {Bershady}, {Tremonti}, {Law}, {Yan}, {Bundy}, {Shetty},
  {Drory}, {Thomas}, {Emsellem}, \& {S{\'a}nchez}}]{belfiore2019}
{Belfiore}, F., {Westfall}, K.~B., {Schaefer}, A., {et~al.} 2019, \aj, 158, 160

\bibitem[{{Berg} {et~al.}(2020){Berg}, {Pogge}, {Skillman}, {Croxall},
  {Moustakas}, {Rogers}, \& {Sun}}]{berg2020}
{Berg}, D.~A., {Pogge}, R.~W., {Skillman}, E.~D., {et~al.} 2020, \apj, 893, 96

\bibitem[{{Berg} {et~al.}(2015){Berg}, {Skillman}, {Croxall}, {Pogge},
  {Moustakas}, \& {Johnson-Groh}}]{berg2015}
{Berg}, D.~A., {Skillman}, E.~D., {Croxall}, K.~V., {et~al.} 2015, \apj, 806,
  16

\bibitem[{{Blanton} {et~al.}(2017){Blanton}, {Bershady}, {Abolfathi},
  {Albareti}, {Allende Prieto}, {Almeida}, {Alonso-Garc{\'\i}a}, {Anders},
  {Anderson}, {Andrews}, {Aquino-Ort{\'\i}z}, {Arag{\'o}n-Salamanca},
  {Argudo-Fern{\'a}ndez}, {Armengaud}, {Aubourg}, {Avila-Reese}, {Badenes},
  {Bailey}, {Barger}, {Barrera-Ballesteros}, {Bartosz}, {Bates}, {Baumgarten},
  {Bautista}, {Beaton}, {Beers}, {Belfiore}, {Bender}, {Berlind}, {Bernardi},
  {Beutler}, {Bird}, {Bizyaev}, {Blanc}, {Blomqvist}, {Bolton}, {Boquien},
  {Borissova}, {van den Bosch}, {Bovy}, {Brandt}, {Brinkmann}, {Brownstein},
  {Bundy}, {Burgasser}, {Burtin}, {Busca}, {Cappellari}, {Delgado Carigi},
  {Carlberg}, {Carnero Rosell}, {Carrera}, {Chanover}, {Cherinka}, {Cheung},
  {G{\'o}mez Maqueo Chew}, {Chiappini}, {Choi}, {Chojnowski}, {Chuang},
  {Chung}, {Cirolini}, {Clerc}, {Cohen}, {Comparat}, {da Costa}, {Cousinou},
  {Covey}, {Crane}, {Croft}, {Cruz-Gonzalez}, {Garrido Cuadra}, {Cunha},
  {Damke}, {Darling}, {Davies}, {Dawson}, {de la Macorra}, {Dell'Agli}, {De
  Lee}, {Delubac}, {Di Mille}, {Diamond-Stanic}, {Cano-D{\'\i}az}, {Donor},
  {Downes}, {Drory}, {du Mas des Bourboux}, {Duckworth}, {Dwelly}, {Dyer},
  {Ebelke}, {Eigenbrot}, {Eisenstein}, {Emsellem}, {Eracleous}, {Escoffier},
  {Evans}, {Fan}, {Fern{\'a}ndez-Alvar}, {Fernandez-Trincado}, {Feuillet},
  {Finoguenov}, {Fleming}, {Font-Ribera}, {Fredrickson}, {Freischlad},
  {Frinchaboy}, {Fuentes}, {Galbany}, {Garcia-Dias},
  {Garc{\'\i}a-Hern{\'a}ndez}, {Gaulme}, {Geisler}, {Gelfand},
  {Gil-Mar{\'\i}n}, {Gillespie}, {Goddard}, {Gonzalez-Perez}, {Grabowski},
  {Green}, {Grier}, {Gunn}, {Guo}, {Guy}, {Hagen}, {Hahn}, {Hall}, {Harding},
  {Hasselquist}, {Hawley}, {Hearty}, {Gonzalez Hern{\'a}ndez}, {Ho}, {Hogg},
  {Holley-Bockelmann}, {Holtzman}, {Holzer}, {Huehnerhoff}, {Hutchinson},
  {Hwang}, {Ibarra-Medel}, {da Silva Ilha}, {Ivans}, {Ivory}, {Jackson},
  {Jensen}, {Johnson}, {Jones}, {J{\"o}nsson}, {Jullo}, {Kamble}, {Kinemuchi},
  {Kirkby}, {Kitaura}, {Klaene}, {Knapp}, {Kneib}, {Kollmeier}, {Lacerna},
  {Lane}, {Lang}, {Law}, {Lazarz}, {Lee}, {Le Goff}, {Liang}, {Li}, {Li},
  {Lian}, {Lima}, {Lin}, {Lin}, {Bertran de Lis}, {Liu}, {de Icaza Lizaola},
  {Long}, {Lucatello}, {Lundgren}, {MacDonald}, {Deconto Machado}, {MacLeod},
  {Mahadevan}, {Geimba Maia}, {Maiolino}, {Majewski}, {Malanushenko},
  {Malanushenko}, {Manchado}, {Mao}, {Maraston}, {Marques-Chaves}, {Masseron},
  {Masters}, {McBride}, {McDermid}, {McGrath}, {McGreer}, {Medina Pe{\~n}a},
  {Melendez}, {Merloni}, {Merrifield}, {Meszaros}, {Meza}, {Minchev},
  {Minniti}, {Miyaji}, {More}, {Mulchaey}, {M{\"u}ller-S{\'a}nchez}, {Muna},
  {Munoz}, {Myers}, {Nair}, {Nandra}, {Correa do Nascimento}, {Negrete},
  {Ness}, {Newman}, {Nichol}, {Nidever}, {Nitschelm}, {Ntelis}, {O'Connell},
  {Oelkers}, {Oravetz}, {Oravetz}, {Pace}, {Padilla}, {Palanque-Delabrouille},
  {Alonso Palicio}, {Pan}, {Parejko}, {Parikh}, {P{\^a}ris}, {Park}, {Patten},
  {Peirani}, {Pellejero-Ibanez}, {Penny}, {Percival}, {Perez-Fournon},
  {Petitjean}, {Pieri}, {Pinsonneault}, {Pisani}, {Poleski}, {Prada},
  {Prakash}, {Queiroz}, {Raddick}, {Raichoor}, {Barboza Rembold}, {Richstein},
  {Riffel}, {Riffel}, {Rix}, {Robin}, {Rockosi}, {Rodr{\'\i}guez-Torres},
  {Roman-Lopes}, {Rom{\'a}n-Z{\'u}{\~n}iga}, {Rosado}, {Ross}, {Rossi}, {Ruan},
  {Ruggeri}, {Rykoff}, {Salazar-Albornoz}, {Salvato}, {S{\'a}nchez}, {Aguado},
  {S{\'a}nchez-Gallego}, {Santana}, {Santiago}, {Sayres}, {Schiavon}, {da Silva
  Schimoia}, {Schlafly}, {Schlegel}, {Schneider}, {Schultheis}, {Schuster},
  {Schwope}, {Seo}, {Shao}, {Shen}, {Shetrone}, {Shull}, {Simon}, {Skinner},
  {Skrutskie}, {Slosar}, {Smith}, {Sobeck}, {Sobreira}, {Somers}, {Souto},
  {Stark}, {Stassun}, {Stauffer}, {Steinmetz}, {Storchi-Bergmann},
  {Streblyanska}, {Stringfellow}, {Su{\'a}rez}, {Sun}, {Suzuki}, {Szigeti},
  {Taghizadeh-Popp}, {Tang}, {Tao}, {Tayar}, {Tembe}, {Teske}, {Thakar},
  {Thomas}, {Thompson}, {Tinker}, {Tissera}, {Tojeiro}, {Hernandez Toledo}, {de
  la Torre}, {Tremonti}, {Troup}, {Valenzuela}, {Martinez Valpuesta},
  {Vargas-Gonz{\'a}lez}, {Vargas-Maga{\~n}a}, {Vazquez}, {Villanova}, {Vivek},
  {Vogt}, {Wake}, {Walterbos}, {Wang}, {Weaver}, {Weijmans}, {Weinberg},
  {Westfall}, {Whelan}, {Wild}, {Wilson}, {Wood-Vasey}, {Wylezalek}, {Xiao},
  {Yan}, {Yang}, {Ybarra}, {Y{\`e}che}, {Zakamska}, {Zamora}, {Zarrouk},
  {Zasowski}, {Zhang}, {Zhao}, {Zheng}, {Zheng}, {Zhou}, {Zhou}, {Zhu},
  {Zoccali}, \& {Zou}}]{blanton2017}
{Blanton}, M.~R., {Bershady}, M.~A., {Abolfathi}, B., {et~al.} 2017, \aj, 154,
  28

\bibitem[{{Bottorff} {et~al.}(1998){Bottorff}, {Lamothe}, {Momjian}, {Verner},
  {Vinkovi{\'c}}, \& {Ferland}}]{bottorff1998}
{Bottorff}, M., {Lamothe}, J., {Momjian}, E., {et~al.} 1998, \pasp, 110, 1040

\bibitem[{{Bundy} {et~al.}(2015){Bundy}, {Bershady}, {Law}, {Yan}, {Drory},
  {MacDonald}, {Wake}, {Cherinka}, {S{\'a}nchez-Gallego}, {Weijmans}, {Thomas},
  {Tremonti}, {Masters}, {Coccato}, {Diamond-Stanic}, {Arag{\'o}n-Salamanca},
  {Avila-Reese}, {Badenes}, {Falc{\'o}n-Barroso}, {Belfiore}, {Bizyaev},
  {Blanc}, {Bland-Hawthorn}, {Blanton}, {Brownstein}, {Byler}, {Cappellari},
  {Conroy}, {Dutton}, {Emsellem}, {Etherington}, {Frinchaboy}, {Fu}, {Gunn},
  {Harding}, {Johnston}, {Kauffmann}, {Kinemuchi}, {Klaene}, {Knapen},
  {Leauthaud}, {Li}, {Lin}, {Maiolino}, {Malanushenko}, {Malanushenko}, {Mao},
  {Maraston}, {McDermid}, {Merrifield}, {Nichol}, {Oravetz}, {Pan}, {Parejko},
  {Sanchez}, {Schlegel}, {Simmons}, {Steele}, {Steinmetz}, {Thanjavur},
  {Thompson}, {Tinker}, {van den Bosch}, {Westfall}, {Wilkinson}, {Wright},
  {Xiao}, \& {Zhang}}]{bundy2015}
{Bundy}, K., {Bershady}, M.~A., {Law}, D.~R., {et~al.} 2015, \apj, 798, 7

\bibitem[{{Byler} {et~al.}(2017){Byler}, {Dalcanton}, {Conroy}, \&
  {Johnson}}]{byler2017}
{Byler}, N., {Dalcanton}, J.~J., {Conroy}, C., \& {Johnson}, B.~D. 2017, \apj,
  840, 44

\bibitem[{{Calzetti} {et~al.}(1994){Calzetti}, {Kinney}, \&
  {Storchi-Bergmann}}]{calzetti1994}
{Calzetti}, D., {Kinney}, A.~L., \& {Storchi-Bergmann}, T. 1994, \apj, 429, 582

\bibitem[{{Cardelli} {et~al.}(1989){Cardelli}, {Clayton}, \&
  {Mathis}}]{cardelli1989}
{Cardelli}, J.~A., {Clayton}, G.~C., \& {Mathis}, J.~S. 1989, \apj, 345, 245

\bibitem[{{Charlot} \& {Fall}(2000)}]{charlot2000}
{Charlot}, S. \& {Fall}, S.~M. 2000, \apj, 539, 718

\bibitem[{{Chevallard} {et~al.}(2013){Chevallard}, {Charlot}, {Wandelt}, \&
  {Wild}}]{chevallard2013}
{Chevallard}, J., {Charlot}, S., {Wandelt}, B., \& {Wild}, V. 2013, \mnras,
  432, 2061

\bibitem[{{Cid Fernandes} {et~al.}(2011){Cid Fernandes}, {Stasi{\'n}ska},
  {Mateus}, \& {Vale Asari}}]{cidfernandes2011}
{Cid Fernandes}, R., {Stasi{\'n}ska}, G., {Mateus}, A., \& {Vale Asari}, N.
  2011, \mnras, 413, 1687

\bibitem[{{Croxall} {et~al.}(2015){Croxall}, {Pogge}, {Berg}, {Skillman}, \&
  {Moustakas}}]{croxall2015}
{Croxall}, K.~V., {Pogge}, R.~W., {Berg}, D.~A., {Skillman}, E.~D., \&
  {Moustakas}, J. 2015, \apj, 808, 42

\bibitem[{{Croxall} {et~al.}(2016){Croxall}, {Pogge}, {Berg}, {Skillman}, \&
  {Moustakas}}]{croxall2016}
{Croxall}, K.~V., {Pogge}, R.~W., {Berg}, D.~A., {Skillman}, E.~D., \&
  {Moustakas}, J. 2016, \apj, 830, 4

\bibitem[{{D'Agostino} {et~al.}(2019){D'Agostino}, {Kewley}, {Groves}, {Byler},
  {Sutherland}, {Nicholls}, {Leitherer}, \& {Stanway}}]{dagostino2019}
{D'Agostino}, J.~J., {Kewley}, L.~J., {Groves}, B., {et~al.} 2019, \apj, 878, 2

\bibitem[{{Diaz} {et~al.}(1991){Diaz}, {Terlevich}, {Vilchez}, {Pagel}, \&
  {Edmunds}}]{diaz1991}
{Diaz}, A.~I., {Terlevich}, E., {Vilchez}, J.~M., {Pagel}, B. E.~J., \&
  {Edmunds}, M.~G. 1991, \mnras, 253, 245

\bibitem[{{Dopita} {et~al.}(2000){Dopita}, {Kewley}, {Heisler}, \&
  {Sutherland}}]{dopita2000}
{Dopita}, M.~A., {Kewley}, L.~J., {Heisler}, C.~A., \& {Sutherland}, R.~S.
  2000, \apj, 542, 224

\bibitem[{{Dopita} {et~al.}(2013){Dopita}, {Sutherland}, {Nicholls}, {Kewley},
  \& {Vogt}}]{dopita2013}
{Dopita}, M.~A., {Sutherland}, R.~S., {Nicholls}, D.~C., {Kewley}, L.~J., \&
  {Vogt}, F. P.~A. 2013, \apjs, 208, 10

\bibitem[{{Draine}(2003)}]{draine2003}
{Draine}, B.~T. 2003, \apj, 598, 1017

\bibitem[{{Drory} {et~al.}(2015){Drory}, {MacDonald}, {Bershady}, {Bundy},
  {Gunn}, {Law}, {Smith}, {Stoll}, {Tremonti}, {Wake}, {Yan}, {Weijmans},
  {Byler}, {Cherinka}, {Cope}, {Eigenbrot}, {Harding}, {Holder}, {Huehnerhoff},
  {Jaehnig}, {Jansen}, {Klaene}, {Paat}, {Percival}, \& {Sayres}}]{drory2015}
{Drory}, N., {MacDonald}, N., {Bershady}, M.~A., {et~al.} 2015, \aj, 149, 77

\bibitem[{{Emsellem} {et~al.}(2022){Emsellem}, {Schinnerer}, {Santoro},
  {Belfiore}, {Pessa}, {McElroy}, {Blanc}, {Congiu}, {Groves}, {Ho}, {Kreckel},
  {Razza}, {Sanchez-Blazquez}, {Egorov}, {Faesi}, {Klessen}, {Leroy}, {Meidt},
  {Querejeta}, {Rosolowsky}, {Scheuermann}, {Anand}, {Barnes},
  {Be{\v{s}}li{\'c}}, {Bigiel}, {Boquien}, {Cao}, {Chevance}, {Dale},
  {Eibensteiner}, {Glover}, {Grasha}, {Henshaw}, {Hughes}, {Koch}, {Kruijssen},
  {Lee}, {Liu}, {Pan}, {Pety}, {Saito}, {Sandstrom}, {Schruba}, {Sun},
  {Thilker}, {Usero}, {Watkins}, \& {Williams}}]{emsellem2022}
{Emsellem}, E., {Schinnerer}, E., {Santoro}, F., {et~al.} 2022, \aap, 659, A191

\bibitem[{{Fanelli} {et~al.}(1988){Fanelli}, {O'Connell}, \&
  {Thuan}}]{fanelli1988}
{Fanelli}, M.~N., {O'Connell}, R.~W., \& {Thuan}, T.~X. 1988, \apj, 334, 665

\bibitem[{{Ferguson} {et~al.}(1996){Ferguson}, {Wyse}, {Gallagher}, \&
  {Hunter}}]{ferguson1996}
{Ferguson}, A. M.~N., {Wyse}, R. F.~G., {Gallagher}, J.~S., I., \& {Hunter},
  D.~A. 1996, \aj, 111, 2265

\bibitem[{{Ferland} {et~al.}(2017){Ferland}, {Chatzikos}, {Guzm{\'a}n},
  {Lykins}, {van Hoof}, {Williams}, {Abel}, {Badnell}, {Keenan}, {Porter}, \&
  {Stancil}}]{ferland2017}
{Ferland}, G.~J., {Chatzikos}, M., {Guzm{\'a}n}, F., {et~al.} 2017, \rmxaa, 53,
  385

\bibitem[{{Fitzpatrick}(1999)}]{fitzpatrick1999}
{Fitzpatrick}, E.~L. 1999, \pasp, 111, 63

\bibitem[{{Fitzpatrick} \& {Massa}(2007)}]{fitzpatrick2007}
{Fitzpatrick}, E.~L. \& {Massa}, D. 2007, \apj, 663, 320

\bibitem[{{Flaherty} {et~al.}(2007){Flaherty}, {Pipher}, {Megeath}, {Winston},
  {Gutermuth}, {Muzerolle}, {Allen}, \& {Fazio}}]{flaherty2007}
{Flaherty}, K.~M., {Pipher}, J.~L., {Megeath}, S.~T., {et~al.} 2007, \apj, 663,
  1069

\bibitem[{{Flores-Fajardo} {et~al.}(2011){Flores-Fajardo}, {Morisset},
  {Stasi{\'n}ska}, \& {Binette}}]{flores-fajardo2011}
{Flores-Fajardo}, N., {Morisset}, C., {Stasi{\'n}ska}, G., \& {Binette}, L.
  2011, \mnras, 415, 2182

\bibitem[{{Foreman-Mackey} {et~al.}(2013){Foreman-Mackey}, {Hogg}, {Lang}, \&
  {Goodman}}]{emcee}
{Foreman-Mackey}, D., {Hogg}, D.~W., {Lang}, D., \& {Goodman}, J. 2013, \pasp,
  125, 306

\bibitem[{Gelman {et~al.}(1995)Gelman, Carlin, Stern, \& Rubin}]{gelman1995}
Gelman, A., Carlin, J.~B., Stern, H.~S., \& Rubin, D.~B. 1995, Bayesian data
  analysis (Chapman and Hall/CRC)

\bibitem[{{Gunn} {et~al.}(2006){Gunn}, {Siegmund}, {Mannery}, {Owen}, {Hull},
  {Leger}, {Carey}, {Knapp}, {York}, {Boroski}, {Kent}, {Lupton}, {Rockosi},
  {Evans}, {Waddell}, {Anderson}, {Annis}, {Barentine}, {Bartoszek}, {Bastian},
  {Bracker}, {Brewington}, {Briegel}, {Brinkmann}, {Brown}, {Carr},
  {Czarapata}, {Drennan}, {Dombeck}, {Federwitz}, {Gillespie}, {Gonzales},
  {Hansen}, {Harvanek}, {Hayes}, {Jordan}, {Kinney}, {Klaene}, {Kleinman},
  {Kron}, {Kresinski}, {Lee}, {Limmongkol}, {Lindenmeyer}, {Long}, {Loomis},
  {McGehee}, {Mantsch}, {Neilsen}, {Neswold}, {Newman}, {Nitta}, {Peoples},
  {Pier}, {Prieto}, {Prosapio}, {Rivetta}, {Schneider}, {Snedden}, \&
  {Wang}}]{gunn2006}
{Gunn}, J.~E., {Siegmund}, W.~A., {Mannery}, E.~J., {et~al.} 2006, \aj, 131,
  2332

\bibitem[{{Haffner} {et~al.}(2009){Haffner}, {Dettmar}, {Beckman}, {Wood},
  {Slavin}, {Giammanco}, {Madsen}, {Zurita}, \& {Reynolds}}]{haffner2009}
{Haffner}, L.~M., {Dettmar}, R.~J., {Beckman}, J.~E., {et~al.} 2009, Reviews of
  Modern Physics, 81, 969

\bibitem[{{Hogg} {et~al.}(2010){Hogg}, {Bovy}, \& {Lang}}]{hogg2010}
{Hogg}, D.~W., {Bovy}, J., \& {Lang}, D. 2010, arXiv e-prints, arXiv:1008.4686

\bibitem[{{Howard} {et~al.}(2018){Howard}, {Pudritz}, {Harris}, \&
  {Klessen}}]{howard2018}
{Howard}, C.~S., {Pudritz}, R.~E., {Harris}, W.~E., \& {Klessen}, R.~S. 2018,
  \mnras, 475, 3121

\bibitem[{{Inoue}(2005)}]{inoue2005}
{Inoue}, A.~K. 2005, \mnras, 359, 171

\bibitem[{{Ji} \& {Yan}(2020)}]{ji2020b}
{Ji}, X. \& {Yan}, R. 2020, \mnras, 499, 5749

\bibitem[{{Ji} \& {Yan}(2022)}]{ji2022}
{Ji}, X. \& {Yan}, R. 2022, \aap, 659, A112

\bibitem[{{Kashino} {et~al.}(2013){Kashino}, {Silverman}, {Rodighiero},
  {Renzini}, {Arimoto}, {Daddi}, {Lilly}, {Sanders}, {Kartaltepe}, {Zahid},
  {Nagao}, {Sugiyama}, {Capak}, {Carollo}, {Chu}, {Hasinger}, {Ilbert},
  {Kajisawa}, {Kewley}, {Koekemoer}, {Kova{\v{c}}}, {Le F{\`e}vre}, {Masters},
  {McCracken}, {Onodera}, {Scoville}, {Strazzullo}, {Symeonidis}, \&
  {Taniguchi}}]{kashino2013}
{Kashino}, D., {Silverman}, J.~D., {Rodighiero}, G., {et~al.} 2013, \apjl, 777,
  L8

\bibitem[{{Kauffmann} {et~al.}(2003){Kauffmann}, {Heckman}, {Tremonti},
  {Brinchmann}, {Charlot}, {White}, {Ridgway}, {Brinkmann}, {Fukugita}, {Hall},
  {Ivezi{\'c}}, {Richards}, \& {Schneider}}]{kauffmann2003}
{Kauffmann}, G., {Heckman}, T.~M., {Tremonti}, C., {et~al.} 2003, \mnras, 346,
  1055

\bibitem[{{Kelly}(2007)}]{kelly2007}
{Kelly}, B.~C. 2007, \apj, 665, 1489

\bibitem[{{Kelly}(2012)}]{kelly2012}
{Kelly}, B.~C. 2012, in Statistical Challenges in Modern Astronomy V, Vol. 902,
  147--162

\bibitem[{{Kennicutt} \& {Evans}(2012)}]{kennicutt2012}
{Kennicutt}, R.~C. \& {Evans}, N.~J. 2012, \araa, 50, 531

\bibitem[{{Kewley} \& {Dopita}(2002)}]{kewley2002}
{Kewley}, L.~J. \& {Dopita}, M.~A. 2002, \apjs, 142, 35

\bibitem[{{Kewley} {et~al.}(2001){Kewley}, {Dopita}, {Sutherland}, {Heisler},
  \& {Trevena}}]{kewley2001}
{Kewley}, L.~J., {Dopita}, M.~A., {Sutherland}, R.~S., {Heisler}, C.~A., \&
  {Trevena}, J. 2001, \apj, 556, 121

\bibitem[{{Kewley} \& {Ellison}(2008)}]{kewley2008}
{Kewley}, L.~J. \& {Ellison}, S.~L. 2008, \apj, 681, 1183

\bibitem[{{Kewley} {et~al.}(2019){Kewley}, {Nicholls}, \&
  {Sutherland}}]{kewley2019}
{Kewley}, L.~J., {Nicholls}, D.~C., \& {Sutherland}, R.~S. 2019, \araa, 57, 511

\bibitem[{{Kollmeier} {et~al.}(2017){Kollmeier}, {Zasowski}, {Rix}, {Johns},
  {Anderson}, {Drory}, {Johnson}, {Pogge}, {Bird}, {Blanc}, {Brownstein},
  {Crane}, {De Lee}, {Klaene}, {Kreckel}, {MacDonald}, {Merloni}, {Ness},
  {O'Brien}, {Sanchez-Gallego}, {Sayres}, {Shen}, {Thakar}, {Tkachenko},
  {Aerts}, {Blanton}, {Eisenstein}, {Holtzman}, {Maoz}, {Nandra}, {Rockosi},
  {Weinberg}, {Bovy}, {Casey}, {Chaname}, {Clerc}, {Conroy}, {Eracleous},
  {G{\"a}nsicke}, {Hekker}, {Horne}, {Kauffmann}, {McQuinn}, {Pellegrini},
  {Schinnerer}, {Schlafly}, {Schwope}, {Seibert}, {Teske}, \& {van
  Saders}}]{kollmeier2017}
{Kollmeier}, J.~A., {Zasowski}, G., {Rix}, H.-W., {et~al.} 2017, arXiv
  e-prints, arXiv:1711.03234

\bibitem[{{Law} {et~al.}(2016){Law}, {Cherinka}, {Yan}, {Andrews}, {Bershady},
  {Bizyaev}, {Blanc}, {Blanton}, {Bolton}, {Brownstein}, {Bundy}, {Chen},
  {Drory}, {D'Souza}, {Fu}, {Jones}, {Kauffmann}, {MacDonald}, {Masters},
  {Newman}, {Parejko}, {S{\'a}nchez-Gallego}, {S{\'a}nchez}, {Schlegel},
  {Thomas}, {Wake}, {Weijmans}, {Westfall}, \& {Zhang}}]{law2016}
{Law}, D.~R., {Cherinka}, B., {Yan}, R., {et~al.} 2016, \aj, 152, 83

\bibitem[{{Law} {et~al.}(2021{\natexlab{a}}){Law}, {Ji}, {Belfiore},
  {Bershady}, {Cappellari}, {Westfall}, {Yan}, {Bizyaev}, {Brownstein},
  {Drory}, \& {Andrews}}]{law2021b}
{Law}, D.~R., {Ji}, X., {Belfiore}, F., {et~al.} 2021{\natexlab{a}}, \apj, 915,
  35

\bibitem[{{Law} {et~al.}(2021{\natexlab{b}}){Law}, {Westfall}, {Bershady},
  {Cappellari}, {Yan}, {Belfiore}, {Bizyaev}, {Brownstein}, {Chen}, {Cherinka},
  {Drory}, {Lazarz}, \& {Shetty}}]{law2021}
{Law}, D.~R., {Westfall}, K.~B., {Bershady}, M.~A., {et~al.}
  2021{\natexlab{b}}, \aj, 161, 52

\bibitem[{{Law} {et~al.}(2015){Law}, {Yan}, {Bershady}, {Bundy}, {Cherinka},
  {Drory}, {MacDonald}, {S{\'a}nchez-Gallego}, {Wake}, {Weijmans}, {Blanton},
  {Klaene}, {Moran}, {Sanchez}, \& {Zhang}}]{law2015}
{Law}, D.~R., {Yan}, R., {Bershady}, M.~A., {et~al.} 2015, \aj, 150, 19

\bibitem[{{Li} {et~al.}(2021){Li}, {Li}, {Mo}, {Zhou}, {Liang}, {Boquien},
  {Drory}, {Fern{\'a}ndez-Trincado}, {Greener}, \& {Riffel}}]{li2021}
{Li}, N., {Li}, C., {Mo}, H., {et~al.} 2021, \apj, 917, 72

\bibitem[{{Little} \& {Rubin}(2002)}]{little2002}
{Little}, R. J.~A. \& {Rubin}, D.~B. 2002, {Statistical analysis with missing
  data}

\bibitem[{{Mannucci} {et~al.}(2021){Mannucci}, {Belfiore}, {Curti}, {Cresci},
  {Maiolino}, {Marasco}, {Marconi}, {Mingozzi}, {Tozzi}, \&
  {Amiri}}]{mannucci2021}
{Mannucci}, F., {Belfiore}, F., {Curti}, M., {et~al.} 2021, \mnras, 508, 1582

\bibitem[{{McKee} \& {Williams}(1997)}]{mckee1997}
{McKee}, C.~F. \& {Williams}, J.~P. 1997, \apj, 476, 144

\bibitem[{{Mingozzi} {et~al.}(2020){Mingozzi}, {Belfiore}, {Cresci}, {Bundy},
  {Bershady}, {Bizyaev}, {Blanc}, {Boquien}, {Drory}, {Fu}, {Maiolino},
  {Riffel}, {Schaefer}, {Storchi-Bergmann}, {Telles}, {Tremonti}, {Zakamska},
  \& {Zhang}}]{mingozzi2020}
{Mingozzi}, M., {Belfiore}, F., {Cresci}, G., {et~al.} 2020, \aap, 636, A42

\bibitem[{Nelder \& Mead(1965)}]{nelder1965}
Nelder, J.~A. \& Mead, R. 1965, The Computer Journal, 7, 308

\bibitem[{{Oey} {et~al.}(2007){Oey}, {Meurer}, {Yelda}, {Furst},
  {Caballero-Nieves}, {Hanish}, {Levesque}, {Thilker}, {Walth},
  {Bland-Hawthorn}, {Dopita}, {Ferguson}, {Heckman}, {Doyle}, {Drinkwater},
  {Freeman}, {Kennicutt}, {Kilborn}, {Knezek}, {Koribalski}, {Meyer}, {Putman},
  {Ryan-Weber}, {Smith}, {Staveley-Smith}, {Webster}, {Werk}, \&
  {Zwaan}}]{oey2007}
{Oey}, M.~S., {Meurer}, G.~R., {Yelda}, S., {et~al.} 2007, \apj, 661, 801

\bibitem[{{Osterbrock} \& {Ferland}(2006)}]{agn3}
{Osterbrock}, D.~E. \& {Ferland}, G.~J. 2006, {Astrophysics of gaseous nebulae
  and active galactic nuclei}

\bibitem[{{Pagel} {et~al.}(1979){Pagel}, {Edmunds}, {Blackwell}, {Chun}, \&
  {Smith}}]{pagel1979}
{Pagel}, B.~E.~J., {Edmunds}, M.~G., {Blackwell}, D.~E., {Chun}, M.~S., \&
  {Smith}, G. 1979, \mnras, 189, 95

\bibitem[{{Pannella} {et~al.}(2015){Pannella}, {Elbaz}, {Daddi}, {Dickinson},
  {Hwang}, {Schreiber}, {Strazzullo}, {Aussel}, {Bethermin}, {Buat},
  {Charmandaris}, {Cibinel}, {Juneau}, {Ivison}, {Le Borgne}, {Le Floc'h},
  {Leiton}, {Lin}, {Magdis}, {Morrison}, {Mullaney}, {Onodera}, {Renzini},
  {Salim}, {Sargent}, {Scott}, {Shu}, \& {Wang}}]{pannella2015}
{Pannella}, M., {Elbaz}, D., {Daddi}, E., {et~al.} 2015, \apj, 807, 141

\bibitem[{{Pellegrini} {et~al.}(2020{\natexlab{a}}){Pellegrini}, {Rahner},
  {Reissl}, {Glover}, {Klessen}, {Rousseau-Nepton}, \&
  {Herrera-Camus}}]{pellegrini2020}
{Pellegrini}, E.~W., {Rahner}, D., {Reissl}, S., {et~al.} 2020{\natexlab{a}},
  \mnras, 496, 339

\bibitem[{{Pellegrini} {et~al.}(2020{\natexlab{b}}){Pellegrini}, {Reissl},
  {Rahner}, {Klessen}, {Glover}, {Pakmor}, {Herrera-Camus}, \&
  {Grand}}]{pellegrini2020b}
{Pellegrini}, E.~W., {Reissl}, S., {Rahner}, D., {et~al.} 2020{\natexlab{b}},
  \mnras, 498, 3193

\bibitem[{{Press} {et~al.}(1986){Press}, {Flannery}, \&
  {Teukolsky}}]{numerical_recipes}
{Press}, W.~H., {Flannery}, B.~P., \& {Teukolsky}, S.~A. 1986, {Numerical
  recipes. The art of scientific computing}

\bibitem[{{Price} {et~al.}(2014){Price}, {Kriek}, {Brammer}, {Conroy},
  {F{\"o}rster Schreiber}, {Franx}, {Fumagalli}, {Lundgren}, {Momcheva},
  {Nelson}, {Skelton}, {van Dokkum}, {Whitaker}, \& {Wuyts}}]{price2014}
{Price}, S.~H., {Kriek}, M., {Brammer}, G.~B., {et~al.} 2014, \apj, 788, 86

\bibitem[{{Qin} {et~al.}(2022){Qin}, {Zheng}, {Fang}, {Pan}, {Wuyts}, {Shi},
  {Peng}, {Gonzalez}, {Bian}, {Huang}, {Gu}, {Liu}, {Tan}, {Shi}, {Ren},
  {Zhang}, {Qiao}, {Wen}, \& {Liu}}]{qin2022}
{Qin}, J., {Zheng}, X.~Z., {Fang}, M., {et~al.} 2022, \mnras, 511, 765

\bibitem[{{Reddy} {et~al.}(2020){Reddy}, {Shapley}, {Kriek}, {Steidel},
  {Shivaei}, {Sanders}, {Mobasher}, {Coil}, {Siana}, {Freeman}, {Azadi},
  {Fetherolf}, {Leung}, {Price}, \& {Zick}}]{reddy2020}
{Reddy}, N.~A., {Shapley}, A.~E., {Kriek}, M., {et~al.} 2020, \apj, 902, 123

\bibitem[{{Reddy} \& {Steidel}(2004)}]{reddy2004}
{Reddy}, N.~A. \& {Steidel}, C.~C. 2004, \apjl, 603, L13

\bibitem[{{Reynolds}(1990)}]{reynolds1990}
{Reynolds}, R.~J. 1990, in The Galactic and Extragalactic Background Radiation,
  ed. S.~{Bowyer} \& C.~{Leinert}, Vol. 139, 157

\bibitem[{{Rezaee} {et~al.}(2021){Rezaee}, {Reddy}, {Shivaei}, {Fetherolf},
  {Emami}, \& {Khostovan}}]{rezaee2021}
{Rezaee}, S., {Reddy}, N., {Shivaei}, I., {et~al.} 2021, \mnras, 506, 3588

\bibitem[{{Saha} {et~al.}(2002){Saha}, {Claver}, \& {Hoessel}}]{saha2002}
{Saha}, A., {Claver}, J., \& {Hoessel}, J.~G. 2002, \aj, 124, 839

\bibitem[{{Salim} \& {Narayanan}(2020)}]{salim2020}
{Salim}, S. \& {Narayanan}, D. 2020, \araa, 58, 529

\bibitem[{{S{\'a}nchez} {et~al.}(2012){S{\'a}nchez}, {Kennicutt}, {Gil de Paz},
  {van de Ven}, {V{\'\i}lchez}, {Wisotzki}, {Walcher}, {Mast}, {Aguerri},
  {Albiol-P{\'e}rez}, {Alonso-Herrero}, {Alves}, {Bakos}, {Bart{\'a}kov{\'a}},
  {Bland-Hawthorn}, {Boselli}, {Bomans}, {Castillo-Morales}, {Cortijo-Ferrero},
  {de Lorenzo-C{\'a}ceres}, {Del Olmo}, {Dettmar}, {D{\'\i}az}, {Ellis},
  {Falc{\'o}n-Barroso}, {Flores}, {Gallazzi}, {Garc{\'\i}a-Lorenzo},
  {Gonz{\'a}lez Delgado}, {Gruel}, {Haines}, {Hao}, {Husemann},
  {Igl{\'e}sias-P{\'a}ramo}, {Jahnke}, {Johnson}, {Jungwiert}, {Kalinova},
  {Kehrig}, {Kupko}, {L{\'o}pez-S{\'a}nchez}, {Lyubenova}, {Marino},
  {M{\'a}rmol-Queralt{\'o}}, {M{\'a}rquez}, {Masegosa}, {Meidt},
  {Mendez-Abreu}, {Monreal-Ibero}, {Montijo}, {Mour{\~a}o}, {Palacios-Navarro},
  {Papaderos}, {Pasquali}, {Peletier}, {P{\'e}rez}, {P{\'e}rez}, {Quirrenbach},
  {Rela{\~n}o}, {Rosales-Ortega}, {Roth}, {Ruiz-Lara},
  {S{\'a}nchez-Bl{\'a}zquez}, {Sengupta}, {Singh}, {Stanishev}, {Trager},
  {Vazdekis}, {Viironen}, {Wild}, {Zibetti}, \& {Ziegler}}]{sanchez2012}
{S{\'a}nchez}, S.~F., {Kennicutt}, R.~C., {Gil de Paz}, A., {et~al.} 2012,
  \aap, 538, A8

\bibitem[{{Sanders} {et~al.}(2020){Sanders}, {Jones}, {Shapley}, {Reddy},
  {Kriek}, {Coil}, {Siana}, {Mobasher}, {Shivaei}, {Price}, {Freeman}, {Azadi},
  {Leung}, {Fetherolf}, {Zick}, {de Groot}, {Barro}, \&
  {Fornasini}}]{sanders2020}
{Sanders}, R.~L., {Jones}, T., {Shapley}, A.~E., {et~al.} 2020, \apjl, 888, L11

\bibitem[{{Schaefer} {et~al.}(2022){Schaefer}, {Tremonti}, {Kauffmann},
  {Andrews}, {Bershady}, {Boardman}, {Bundy}, {Drory},
  {Fern{\'a}ndez-Trincado}, {Preece}, {Riffel}, {Riffel}, \&
  {S{\'a}nchez}}]{schaefer2022}
{Schaefer}, A.~L., {Tremonti}, C., {Kauffmann}, G., {et~al.} 2022, \apj, 930,
  160

\bibitem[{{Seon}(2009)}]{seon2009}
{Seon}, K.-I. 2009, \apj, 703, 1159

\bibitem[{{Smee} {et~al.}(2013){Smee}, {Gunn}, {Uomoto}, {Roe}, {Schlegel},
  {Rockosi}, {Carr}, {Leger}, {Dawson}, {Olmstead}, {Brinkmann}, {Owen},
  {Barkhouser}, {Honscheid}, {Harding}, {Long}, {Lupton}, {Loomis}, {Anderson},
  {Annis}, {Bernardi}, {Bhardwaj}, {Bizyaev}, {Bolton}, {Brewington}, {Briggs},
  {Burles}, {Burns}, {Castander}, {Connolly}, {Davenport}, {Ebelke}, {Epps},
  {Feldman}, {Friedman}, {Frieman}, {Heckman}, {Hull}, {Knapp}, {Lawrence},
  {Loveday}, {Mannery}, {Malanushenko}, {Malanushenko}, {Merrelli}, {Muna},
  {Newman}, {Nichol}, {Oravetz}, {Pan}, {Pope}, {Ricketts}, {Shelden},
  {Sandford}, {Siegmund}, {Simmons}, {Smith}, {Snedden}, {Schneider},
  {SubbaRao}, {Tremonti}, {Waddell}, \& {York}}]{smee2013}
{Smee}, S.~A., {Gunn}, J.~E., {Uomoto}, A., {et~al.} 2013, \aj, 146, 32

\bibitem[{{Spitzer}(1978)}]{spitzer1978}
{Spitzer}, L. 1978, {Physical processes in the interstellar medium}

\bibitem[{{Thurston} {et~al.}(1996){Thurston}, {Edmunds}, \&
  {Henry}}]{thurston1996}
{Thurston}, T.~R., {Edmunds}, M.~G., \& {Henry}, R.~B.~C. 1996, \mnras, 283,
  990

\bibitem[{{Tremaine} {et~al.}(2002){Tremaine}, {Gebhardt}, {Bender}, {Bower},
  {Dressler}, {Faber}, {Filippenko}, {Green}, {Grillmair}, {Ho}, {Kormendy},
  {Lauer}, {Magorrian}, {Pinkney}, \& {Richstone}}]{tremaine2002}
{Tremaine}, S., {Gebhardt}, K., {Bender}, R., {et~al.} 2002, \apj, 574, 740

\bibitem[{{Vale Asari} {et~al.}(2020){Vale Asari}, {Wild}, {de Amorim},
  {Werle}, {Zheng}, {Kennicutt}, {Johnson}, {Galametz}, {Pellegrini},
  {Klessen}, {Reissl}, {Glover}, \& {Rahner}}]{vale2020}
{Vale Asari}, N., {Wild}, V., {de Amorim}, A.~L., {et~al.} 2020, \mnras, 498,
  4205

\bibitem[{{Veilleux} \& {Osterbrock}(1987)}]{veilleux1987}
{Veilleux}, S. \& {Osterbrock}, D.~E. 1987, \apjs, 63, 295

\bibitem[{{Vogt} {et~al.}(2014){Vogt}, {Dopita}, {Kewley}, {Sutherland},
  {Scharw{\"a}chter}, {Basurah}, {Ali}, \& {Amer}}]{vogt2014}
{Vogt}, F. P.~A., {Dopita}, M.~A., {Kewley}, L.~J., {et~al.} 2014, \apj, 793,
  127

\bibitem[{{Wake} {et~al.}(2017){Wake}, {Bundy}, {Diamond-Stanic}, {Yan},
  {Blanton}, {Bershady}, {S{\'a}nchez-Gallego}, {Drory}, {Jones}, {Kauffmann},
  {Law}, {Li}, {MacDonald}, {Masters}, {Thomas}, {Tinker}, {Weijmans}, \&
  {Brownstein}}]{wake2017}
{Wake}, D.~A., {Bundy}, K., {Diamond-Stanic}, A.~M., {et~al.} 2017, \aj, 154,
  86

\bibitem[{{Westfall} {et~al.}(2019){Westfall}, {Cappellari}, {Bershady},
  {Bundy}, {Belfiore}, {Ji}, {Law}, {Schaefer}, {Shetty}, {Tremonti}, {Yan},
  {Andrews}, {Brownstein}, {Cherinka}, {Coccato}, {Drory}, {Maraston},
  {Parikh}, {S{\'a}nchez-Gallego}, {Thomas}, {Weijmans}, {Barrera-Ballesteros},
  {Du}, {Goddard}, {Li}, {Masters}, {Ibarra Medel}, {S{\'a}nchez}, {Yang},
  {Zheng}, \& {Zhou}}]{westfall2019}
{Westfall}, K.~B., {Cappellari}, M., {Bershady}, M.~A., {et~al.} 2019, \aj,
  158, 231

\bibitem[{{Wild} {et~al.}(2011{\natexlab{a}}){Wild}, {Charlot}, {Brinchmann},
  {Heckman}, {Vince}, {Pacifici}, \& {Chevallard}}]{wild2011b}
{Wild}, V., {Charlot}, S., {Brinchmann}, J., {et~al.} 2011{\natexlab{a}},
  \mnras, 417, 1760

\bibitem[{{Wild} {et~al.}(2011{\natexlab{b}}){Wild}, {Groves}, {Heckman},
  {Sonnentrucker}, {Armus}, {Schiminovich}, {Johnson}, {Martins}, \&
  {Lamassa}}]{wild2011}
{Wild}, V., {Groves}, B., {Heckman}, T., {et~al.} 2011{\natexlab{b}}, \mnras,
  410, 1593

\bibitem[{{Yan} {et~al.}(2020){Yan}, {Bershady}, {Smith}, {MacDonald},
  {Bizyaev}, {Bundy}, {Chattopadhyay}, {Gunn}, {Westfall}, \& {Wolf}}]{yan2020}
{Yan}, R., {Bershady}, M.~A., {Smith}, M.~P., {et~al.} 2020, in Society of
  Photo-Optical Instrumentation Engineers (SPIE) Conference Series, Vol. 11447,
  Society of Photo-Optical Instrumentation Engineers (SPIE) Conference Series,
  114478Y

\bibitem[{{Yan} {et~al.}(2016{\natexlab{a}}){Yan}, {Bundy}, {Law}, {Bershady},
  {Andrews}, {Cherinka}, {Diamond-Stanic}, {Drory}, {MacDonald},
  {S{\'a}nchez-Gallego}, {Thomas}, {Wake}, {Weijmans}, {Westfall}, {Zhang},
  {Arag{\'o}n-Salamanca}, {Belfiore}, {Bizyaev}, {Blanc}, {Blanton},
  {Brownstein}, {Cappellari}, {D'Souza}, {Emsellem}, {Fu}, {Gaulme}, {Graham},
  {Goddard}, {Gunn}, {Harding}, {Jones}, {Kinemuchi}, {Li}, {Li}, {Maiolino},
  {Mao}, {Maraston}, {Masters}, {Merrifield}, {Oravetz}, {Pan}, {Parejko},
  {Sanchez}, {Schlegel}, {Simmons}, {Thanjavur}, {Tinker}, {Tremonti}, {van den
  Bosch}, \& {Zheng}}]{yan2016}
{Yan}, R., {Bundy}, K., {Law}, D.~R., {et~al.} 2016{\natexlab{a}}, \aj, 152,
  197

\bibitem[{{Yan} {et~al.}(2006){Yan}, {Newman}, {Faber}, {Konidaris}, {Koo}, \&
  {Davis}}]{yan2006}
{Yan}, R., {Newman}, J.~A., {Faber}, S.~M., {et~al.} 2006, \apj, 648, 281

\bibitem[{{Yan} {et~al.}(2016{\natexlab{b}}){Yan}, {Tremonti}, {Bershady},
  {Law}, {Schlegel}, {Bundy}, {Drory}, {MacDonald}, {Bizyaev}, {Blanc},
  {Blanton}, {Cherinka}, {Eigenbrot}, {Gunn}, {Harding}, {Hogg},
  {S{\'a}nchez-Gallego}, {S{\'a}nchez}, {Wake}, {Weijmans}, {Xiao}, \&
  {Zhang}}]{yan2016a}
{Yan}, R., {Tremonti}, C., {Bershady}, M.~A., {et~al.} 2016{\natexlab{b}}, \aj,
  151, 8

\bibitem[{{Zafar} {et~al.}(2015){Zafar}, {M{\o}ller}, {Watson}, {Fynbo},
  {Krogager}, {Zafar}, {Saturni}, {Geier}, \& {Venemans}}]{zafar2015}
{Zafar}, T., {M{\o}ller}, P., {Watson}, D., {et~al.} 2015, \aap, 584, A100

\bibitem[{{Zhang} {et~al.}(2017){Zhang}, {Yan}, {Bundy}, {Bershady}, {Haffner},
  {Walterbos}, {Maiolino}, {Tremonti}, {Thomas}, {Drory}, {Jones}, {Belfiore},
  {S{\'a}nchez}, {Diamond-Stanic}, {Bizyaev}, {Nitschelm}, {Andrews},
  {Brinkmann}, {Brownstein}, {Cheung}, {Li}, {Law}, {Roman Lopes}, {Oravetz},
  {Pan}, {Storchi Bergmann}, \& {Simmons}}]{zhang2017}
{Zhang}, K., {Yan}, R., {Bundy}, K., {et~al.} 2017, \mnras, 466, 3217

\bibitem[{{Zibetti} {et~al.}(2009){Zibetti}, {Charlot}, \& {Rix}}]{zibetti2009}
{Zibetti}, S., {Charlot}, S., \& {Rix}, H.-W. 2009, \mnras, 400, 1181

\bibitem[{{Zurita} {et~al.}(2002){Zurita}, {Beckman}, {Rozas}, \&
  {Ryder}}]{zurita2002}
{Zurita}, A., {Beckman}, J.~E., {Rozas}, M., \& {Ryder}, S. 2002, \aap, 386,
  801

\bibitem[{{Zurita} {et~al.}(2000){Zurita}, {Rozas}, \& {Beckman}}]{zurita2000}
{Zurita}, A., {Rozas}, M., \& {Beckman}, J.~E. 2000, \aap, 363, 9

\end{thebibliography}
%

\appendix
\section{Linear regression with measurement errors and intrinsic scatter}
\label{appendix}

\begin{figure*}
    \includegraphics[width=0.33\textwidth]{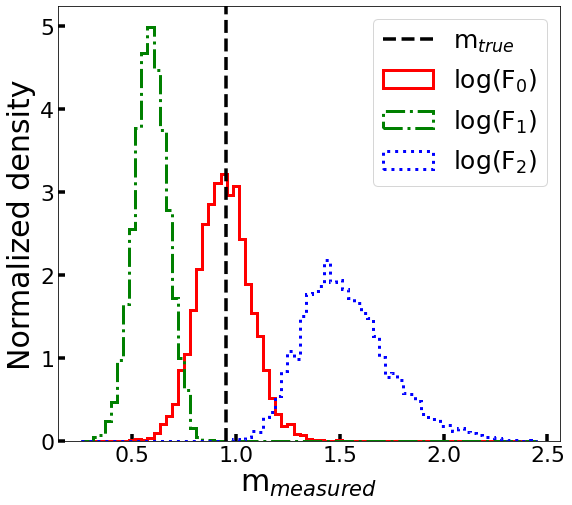}
    \includegraphics[width=0.33\textwidth]{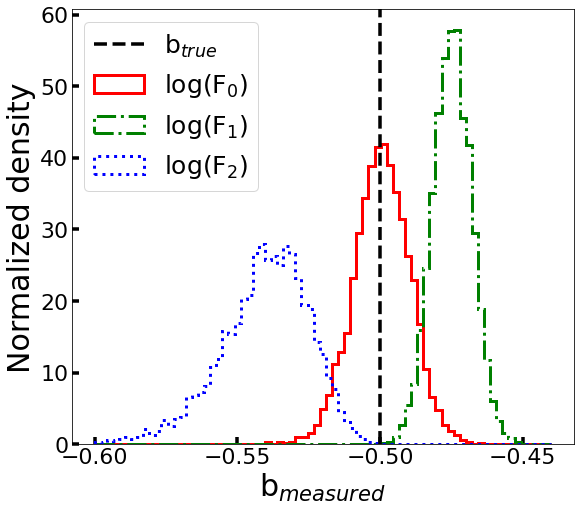}
    \includegraphics[width=0.33\textwidth]{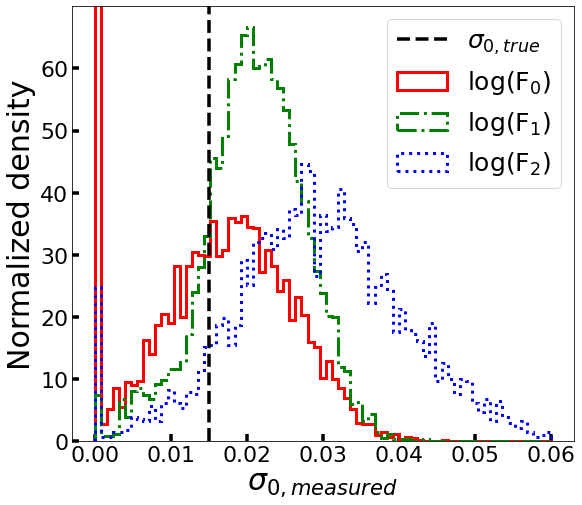}
    \caption{Distributions of slopes, intercepts, and intrinsic scatters estimated by different likelihood functions for the same simulated data set. In each panel, the vertical dashed line indicates the true value of the corresponding parameter for the simulated data set. Clearly $F_0$ best reproduces the true parameters of the linear relation. It is also the default likelihood function we used in the main paper.
    }
    \label{fig:mle_com0}
\end{figure*}

\begin{figure*}
    \includegraphics[width=0.33\textwidth]{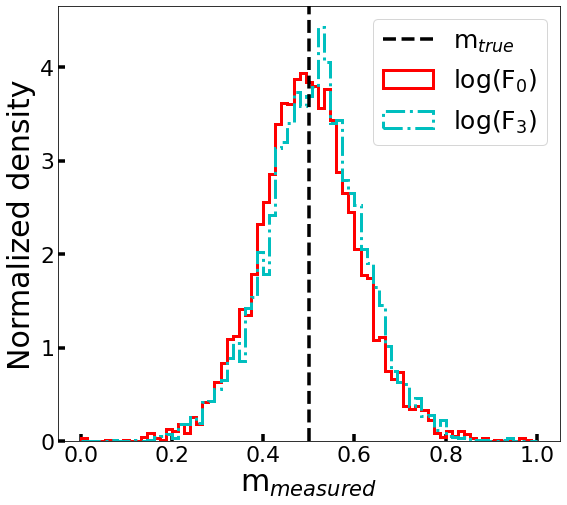}
    \includegraphics[width=0.33\textwidth]{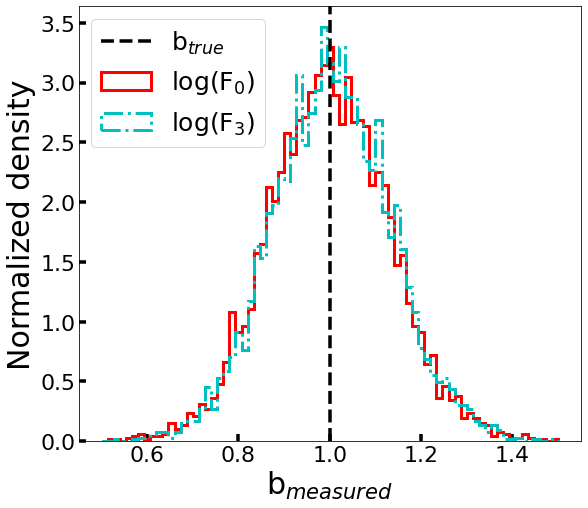}
    \includegraphics[width=0.33\textwidth]{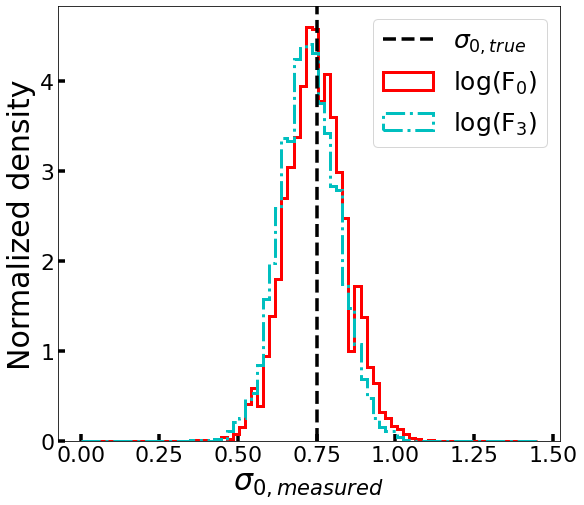}
    \caption{Distributions of slopes, intercepts, and intrinsic scatters estimated by our default likelihood function $F_0$ and \protect\cite{kelly2007}'s likelihood function $F_3$ for the same simulated data set. In each panel, the vertical dashed line indicates the true value of the corresponding parameter for the simulated data set.
    }
    \label{fig:mle_comk}
\end{figure*}

\begin{table*}
	\centering
	\caption{Median slopes given by different likelihood functions when measurement uncertainties are correctly or incorrectly estimated. {($\bf m_{true} = 0.95$)}}
	{\fontsize{10}{11}\selectfont
	\label{tab:mle_com}
	\begin{tabularx}{1.5\columnwidth}{ccccc} 
	    \hline\hline
        $\sigma _{x,{\rm estimated}}/\sigma _{x,{\rm true}}$ & $\sigma _{y,{\rm estimated}}/\sigma _{y,{\rm true}}$ & Median $m$ ($F_0$) & Median $m$ ($F_1$) & Median $m$ ($F_2$) \\
        \hline
        1 & 1 & $0.94\pm 0.13^{\rm a}$ & $0.59\pm 0.08$ & $1.51\pm 0.23$\\
        \hline
        1.25 & 1 & $1.21\pm 0.16$ & $0.58\pm 0.08$ & $1.51\pm 0.21$\\
        \hline
        1 & 1.25 & $0.83\pm 0.11$ & $0.56\pm 0.07$ & $1.60\pm 0.21$\\
        \hline
        1.25 & 1.25 & $0.98\pm 0.13$ & $0.53\pm 0.07$ & $1.68\pm 0.23$\\
        \hline
        0.75 & 1 & $0.75\pm 0.10$ & $0.59\pm 0.08$ & $1.49\pm 0.22$\\
        \hline
        1 & 0.75 & $0.97\pm 0.14$ & $0.60\pm 0.09$ & $1.54\pm 0.24$\\
        \hline
        0.75 & 0.75 & $0.75\pm 0.11$ & $0.58\pm 0.09$ & $1.54\pm 0.25$\\
        \hline
    \end{tabularx}
    \begin{tablenotes}
        \small
        \item $\bf Notes.$
        \item $^{\rm a}$ The value shown here (as well as in other similar places in the table) represents the standard deviation of the slope distribution rather than the uncertainty of the median.
    \end{tablenotes}
	}
\end{table*}

Fitting a straight line to a data set that has errors on both dependent and independent variables have been a problem discussed by many authors with different solutions \citep[e.g.,][]{numerical_recipes,tremaine2002,kelly2007,hogg2010}. Since this process is of great significance for our analyses, we examine the robustness of our adopted method through some ideal simulations in this appendix.

For our situation, the problem can be described as follows. There is a data set with two observed parameters, $x$ and $y$ (i.e., logarithmic line ratios and logarithmic Balmer decrement), which we believe to be linearly correlated according to the theory. However, there are measurement errors for both parameters, of which the variances are estimated to be $\sigma _x ^2$ and $\sigma _y ^2$, respectively. Besides known observational uncertainties, there are also intrinsic variations in $y$ (caused by variations in some nebular parameters) that have an unknown variance $\sigma _0 ^2$. Meanwhile, the intrinsic variations in $x$ is negligible (since the Balmer line ratios are roughly constants in \hii\ regions). The goal is to find a linear model, $y=mx +b$ that best represents the underlying true relation.

The simple least squares method does not work because of the extra uncertainty in $x$. A widely adopted solution is to construct and numerically maximize a likelihood function of the model parameters.
Our discussion in this appendix will mainly focus on methods based on likelihood functions. For a good review of other methods, see e.g., \cite{kelly2007,kelly2012}.

Once we have the proper likelihood function, we can numerically maximize it and find the best-fit model. However, the exact form of the likelihood function as well as the maximization method varies in the literature.
In the main paper, we followed the method of \cite{tremaine2002}, whose logarithmic likelihood function is given by
\begin{equation}
    \ln F_{0} = -0.5(y-mx-b)^2/(m^2\sigma _x^2 + \sigma _y^2 + \sigma _0^2) = -0.5 \chi ^2.
\end{equation}
One immediately notices that the above equation is monotonically increasing with increasing $\sigma _0^2$. Thus, if it is directly maximized, the solution always diverges with $\sigma _0^2 \rightarrow \infty$. \cite{tremaine2002}'s maximization method is to start by setting $\sigma _0^2 = 0$. After obtaining the first set of solution, $(m,b)$, one checks the corresponding $\chi ^2 /(N-2)$, where $N$ is the number of data points. If $\chi ^2/(N-2) > 1$, increase $\sigma _0^2$ by a small amount and recompute $(m,b)$. This step is repeated until $\chi ^2/(N-2) \leq 1$.

There are, however, other potential forms of the likelihood function. For example,
\begin{equation}
    \begin{aligned}
        \ln F_{1} = -0.5(y-mx-b)^2/(m^2\sigma _x^2 + \sigma _y^2 + \sigma _0^2) \\
        -0.5 \ln(m^2\sigma _x^2 + \sigma _y^2 + \sigma _0^2)
    \end{aligned}
    \label{eq:lf_f1}
\end{equation}
and
\begin{equation}
    \begin{aligned}
        \ln F_{2} = -0.5(y-mx-b)^2/(m^2\sigma _x^2 + \sigma _y^2 + \sigma _0^2) \\
        -0.5 \ln[(m^2\sigma _x^2 + \sigma _y^2 + \sigma _0^2)/(1+m^2)].
    \end{aligned}
    \label{eq:lf_f2}
\end{equation}
Equation~\ref{eq:lf_f1} is equivalent to the logarithm of Equation~24 in \cite{kelly2007}\footnote{According to \cite{kelly2007}, from the perspective of Bayesian statistics, this likelihood function corresponds to the special case where the prior for the intrinsic distribution of $x$ is a uniform distribution.}.
Besides the $\chi^2$ term, $F_1$ includes an extra term related to the total variance, which is not a constant as it involves $m$ and $\sigma _0^2$. With the variance term, one can now directly maximize the likelihood function as it is no longer a monotonic function of $\sigma _0^2$. Equation~\ref{eq:lf_f2} is similar to Equation~\ref{eq:lf_f1}, but it evaluates the variance perpendicular to the linear model rather than along the $y$ axis, thus introducing the $1+m^2$ term \citep[this term cancels out in the $\chi^2$ term but not in the variance term, see e.g., Equation~32 of][]{hogg2010}.
Similar to Equation~\ref{eq:lf_f1}, Equation~\ref{eq:lf_f2} can also be directly maximized.

To compare the above three likelihood methods, we set up a test based on the observed data in MaNGA. First, we generated a subsample by choosing a random spaxel classified as an SF region and selecting spaxels that lie within 0.05 dex to this spaxel in the 3D line-ratio space. By the end of this step, we obtained $\sim 7000$ spaxels that are close to each other in the 3D line-ratio space. Second, we randomly selected 200 spaxels from this subsample. The logarithmic Balmer decrements of these spaxels were used as the {\it intrinsic values} for the independent variable, $x_{0}$. Their measurement uncertainties provided by DAP were also used to calculate $\sigma _x$. The final simulated values for $x$ were obtained by using random variables drawn from the normal distribution $N(x_0, \sigma _x^2)$\footnote{Although here we assumed the error distributions of the logarithmic line ratios were Gaussian, assuming Gaussian errors for line fluxes lead to the same conclusion.}. The {\it intrinsic values} for the dependent variable was given by $y_0 = m_{\rm true}x_0 + b_{\rm true}$, where $m_{\rm true}$ was set to $m^{\rm F99}_{\rm [OIII],[OII]}=0.95$ and $b_{\rm true}$ was set to $-0.5$. We then used the measurement uncertainties of log([O\,{\sc iii}]/[O\,{\sc ii}]) to represent $\sigma _y$, while setting the intrinsic scatter in $y$ to $\sigma _0 = 0.015$. The final simulated values for $y$ were drawn from the random distribution $y_0 + N(0,\sigma _y^2) + N(0,\sigma _0^2)$. We then used the aforementioned three likelihood functions to derive $m$, $b$, and $\sigma _0$. The second step was repeated 5,000 times and distributions of these model parameters were compared.

Figure~\ref{fig:mle_com0} shows the comparisons between model parameters estimated by different likelihood functions. Our default function, $F_0$, best reproduces the true parameters for $m$, $b$, and $\sigma _0$. Although it underestimated $\sigma _0$ by yielding $\sigma _{\rm 0,measured} = 0$ for $\sim 30$\% of the data, this does not have any noticeable impact on estimations of $m$ and $b$. In comparison, $F_1$ ($F_2$) tends to underestimate (overestimate) $m$ and overestimate (underestimate) $b$. In addition, both $F_1$ and $F_2$ overestimate $\sigma _0$. 

The above maximum-likelihood methods all require a good understanding of the measurement uncertainty. To check the impact of measurement uncertainties, we performed additional tests where these uncertainties were incorrectly estimated but were still fed into the likelihood functions. 
Table~\ref{tab:mle_com} summarizes the impact of the biased measurement uncertainties on the derived median slopes.
We found that when $\sigma _x$ was overestimated or $\sigma _y$ was underestimated, the resulting median $m$ had an increase, making $F_0$ overestimate $m_{\rm true}$. On the other hand, when $\sigma _y$ was overestimated or $\sigma _x$ was underestimated, the resulting median $m$ had a decrease, making $F_0$ underestimate $m_{\rm true}$.
Interestingly, when both $\sigma _x$ and $\sigma _y$ were overestimated by the same fractional amount, the median $m$ remained roughly the same.
For example, in one of our tests where $\sigma _x$ and $\sigma _y$ were both overestimated by $25$\%, the median $m$ returned by $F_0$ only differed from $m_{\rm true}$ by $\sim 3$\%.
In contrast, when both $\sigma _x$ and $\sigma _y$ were underestimated by 25\%, the median $m$ estimated by $F_0$ differed from $m_{\rm true}$ by $\sim 20$\%.
One can see that the fractional changes in the median $m$ given by $F_1$ and $F_2$ when the measurement uncertainties are incorrectly estimated are actually smaller than those given by $F_0$ in some cases. Still, $F_0$ is the only likelihood function that recovers $m_{\rm true}$ when there is no bias in the measurement uncertainty.
Overall $F_0$ appears slightly more robust against any bias in $\sigma _y$ rather $\sigma _x$, probably because $\sigma _0$ can be adjusted to partly compensate the biased $\sigma _y$.

{Apart from the above maximum-likelihood methods, there is a Bayesian method based on Gaussian mixture models introduced by \cite{kelly2007}, which outperforms other methods according to \cite{kelly2007}.
Basically, \cite{kelly2007} approximates the {\it intrinsic distribution} of $x$, denoted as $\psi$, as Gaussians and incorporate them into the likelihood function.
The posterior distributions of $m$, $b$, and $\psi$ can be jointly inferred from the data.}

We performed another test to see if our likelihood function produced similar results to those given by \cite{kelly2007}'s estimator.
We set up a new simulated data set similar to the one produced by \cite{kelly2007} by drawing the intrinsic values of $x$ from the probability distribution function (PDF)
\begin{equation}
    p(\xi) \propto e^{\xi}(1+e^{2.75\xi})^{-1}.
\end{equation}
This PDF can be approximated by a Gaussian distribution, $N(\mu = -0.493, \sigma ^2 = 1.1^2)$. The true slope and intercept are set to $m_{\rm true} = 0.5$ and $b_{\rm true} = 1$, which gives the intrinsic distribution of $y$.
The intrinsic scatter $\sigma _0$ is described by another Gaussian distribution, $N(0, \sigma _0^2 = 0.75^2)$.
Measurement uncertainties of $x$ and $y$ are independently drawn from a scaled inverse-$\chi ^2$ distribution whose degrees of freedom $\nu = 5$ and scale parameters $s_x = 0.5\sigma$ for $x$ and $s_y = 0.5\sigma _0$ for $y$.
We drew 50 data points each trial and performed a total of 5,000 trials to derived distributions of slopes, intercepts, and intrinsic scatter using $F_0$ as well as \cite{kelly2007}'s estimator.

Similar to \cite{kelly2007}'s test, when applying his estimator, we assumed the intrinsic distribution for $x$ was known and used the Gaussian distribution $N(\mu=-0.493,\sigma ^2=1.1^2)$ as an approximation in \cite{kelly2007}'s Equation~16 to obtain the likelihood function
\begin{equation}
        \ln F_{3} = -0.5[({\bf z}-\zeta)^TV^{-1}({\bf z}-\zeta)]
        -0.5 |V|,
    \label{eq:lf_f3}
\end{equation}
where ${\bf z} = (y,x)$, $\zeta = (m_{true}\mu + b_{true}, \mu)$, and the covariance matrix
$V = \begin{pmatrix}
m_{\rm true}^2 \sigma ^2 + \sigma _y^2 + \sigma _0^2 & m_{\rm true} \sigma ^2\\
m_{\rm true} \sigma ^2 & \sigma ^2 + \sigma _x^2
\end{pmatrix}$.
Following \cite{kelly2007}, we simply performed maximum-likelihood estimations in this test rather than derived posteriors.
Similar to $F_1$ and $F_2$, this likelihood function can be directly maximized to estimate $m$ and $b$.

Figure~\ref{fig:mle_comk} shows the linear regression results we obtained with $F_0$ and $F_3$. It is clear that both methods not only recover the true value for each parameter equally well, but also yield similar dispersions for the distributions of the parameters.
We also tried different values for the input parameters to simulate other data sets as \cite{kelly2007} did, and still found the two methods gave nearly identical distributions.
In addition to the simulated data, we applied \cite{kelly2007}'s Bayesian method to several observed reddening relations in MaNGA, and again found results almost identical to what we have obtained\footnote{Here we used the {\sc python} port of \cite{kelly2007}'s {\sc idl} package {\sc linmix\_err} shared by: https://github.com/jmeyers314/linmix.}.
We thereby concluded our adopted method is equally good as the method recommended by \cite{kelly2007} in terms of the output.

Finally, in the aforementioned tests the measurement errors are not correlated, which apply to most of our results in the main paper.
In cases when $x$ and $y$ contain a common line ratio, one needs to replace the variance with a covariance matrix in the likelihood function. For example, for some of the slope measurements in the main paper, both $x$ and $y$ have the $\rm \log(H\alpha)$ term. When measuring, for example, $m_{\rm H\alpha/H\gamma}$, we simply replaced the variance term with $m^2\sigma _x^2 + \sigma _y^2 - 2mVar[\log({\rm H}\alpha)] + \sigma _0^2$, where $Var[\log({\rm H}\alpha)]$ is the variance of $\rm \log(H\alpha)$.
However, since H$\alpha$ generally has much higher S/N compared to other lines, adding $Var[\log({\rm H}\alpha)]$ has little impact on the overall results.

\section{Results based on nonparametric emission-line measurements}
\label{appendix:b}

\begin{table}
	\centering
	\caption{Median slopes measured from different data sets}
	{\fontsize{7}{9}\selectfont
	\label{tab:npp_com}
	\begin{tabularx}{1\columnwidth}{lcccc} 
	    \hline\hline
        Flux & Uncertainty & $m^{\prime}_{\rm [NII],[OII]}$ & $m^{\prime}_{\rm [SIII],[OIII]}$ & $m^{\prime}_{\rm [OIII],[OII]}$ \\
        \hline
        Nonparametric & Nonparametric & $1.616\pm 0.005$ & $1.66\pm 0.04$ & $0.49\pm 0.03$\\
        \hline
        Nonparametric & Gaussian & $1.417\pm 0.005$ & $1.468\pm 0.007$ & $0.430\pm 0.005$\\
        \hline
        Gaussian & Gaussian & $1.430\pm 0.004$ & $1.677\pm 0.006$ & $0.458\pm 0.005$\\
        \hline
	\end{tabularx}
	}
\end{table}

This work heavily relies on emission line measurements of the observed spectra. The MaNGA data were analyzed by DAP, which produced emission line measurements by using Gaussian models as templates for emission lines and fit them simultaneously with the stellar continuum models \citep{westfall2019,belfiore2019}. We note that there are other pipelines and codes that use different fitting schemes and might produce subtle differences for the emission line measurements \citep[see e.g., Appendix of][]{belfiore2019}. In addition, a single Gaussian component might not be optimal for emission line fitting in cases where multiple kinematic components are blended in observations.
\cite{belfiore2019} carefully assessed the quality of the emission line measurements of DAP, confirming the S/N reported by DAP is an appropriate metric and the emission lines are statistically well fitted by Gaussian models for S/N < 30. When S/N > 30, line profiles start to deviate from Gaussians, but the reported Gaussian fluxes are still consistent with the non-parametric fluxes obtained by direct summing. Therefore, significant bias in flux measurements and uncertainty estimations seem unlikely in our sample.

Regardless, we performed a test based on the nonparametric emission line measurements provided by DAP.
These emission lines fluxes were obtained by first subtracting the best-fit stellar continuum models from the spectra, and then subtracting a baseline local to each emission line, and finally directly integrating the residual fluxes with predefined passbands \citep{yan2006,westfall2019}. The corresponding errors were estimated via propagation of pixelwise errors but was not fully tested or calibrated. Thus, \cite{westfall2019} warned the usage of these error measurements\footnote{According to the code descriptions of DAP, there is poor agreement between the reported nonparametric errors and those from a Monte Carlo simulation.}.
In our test, we used the nonparametric fluxes and measurement errors to replace the Gaussian values and remeasure slopes for \no, \so, and \oo, and we summarized the results in Table~\ref{tab:npp_com}.
Similar to what we did in the main paper, we applied an S/N cut of 3 for every emission line involved. This selection criterion left us $\sim 8.1\times 10^5$ spaxels, which are considerably fewer compared to the Gaussian case ($\sim 2.4\times 10^6$). This is because the estimated uncertainties for nonparametric fluxes are systematically higher than those for Gaussian fluxes due to the significant noise contribution from pixels far away from line centers.
We included all 3D bins with number of spaxels greater than 80 and ended up with 1,589 bins. Slightly adjusting this bin-selection criterion does not change the overall conclusion.

We notice $m^{\prime}_{\rm [NII],[OII]}$ given by nonparametric measurements is considerably larger than that given by Gaussian measurements (still it is significantly different from the values predicted by the F99 curve). 
For comparison, in Table~\ref{tab:npp_com} we also list the results given by the Gaussian measurements using the same sample that was selected by the S/N cuts on nonparametric measurements.
{These Gaussian results also show some differences from those in the main paper, which could be due to the fact that the sample here was not selected using the S/N cuts on Gaussian measurements.}
While the nonparametric fluxes are in good agreement with Gaussian fluxes, the nonparametric flux uncertainties are overall higher. Recalling the test on measurement uncertainties in Appendix~\ref{appendix}, we suspected the difference is mainly driven by the systematically higher non-parametric flux uncertainties. 
Indeed, once we replaced the nonparametric uncertainties with the Gaussian uncertainties, but still using the nonparametric fluxes, the resulting slopes were lowered.
Since the Gaussian uncertainties reported by MaNGA were well tested and calibrated while the non-parametric uncertainties were not \citep{westfall2019,belfiore2019}, we consider the Gaussian results to be more trustworthy.

Finally, how the different emission-line fitting methods will affect our results statistically is beyond the scope of the current paper. Still we think it is important to fully test this point in future works and apply our slope-measuring method to other large astronomical data sets.

\section{{Results derived from high-resolution observations on IC 342}}
\label{subsec:high_resol_hii}

\begin{figure}
    \includegraphics[width=0.24\textwidth]{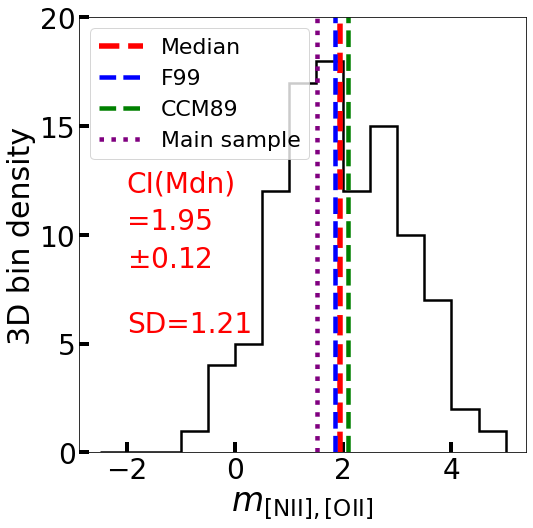}
    \includegraphics[width=0.24\textwidth]{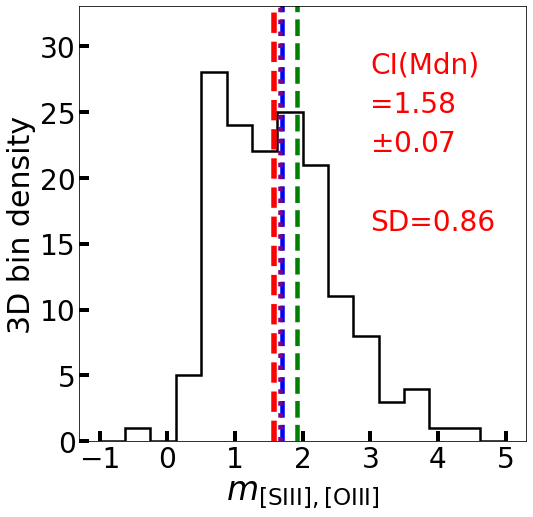}
    \includegraphics[width=0.24\textwidth]{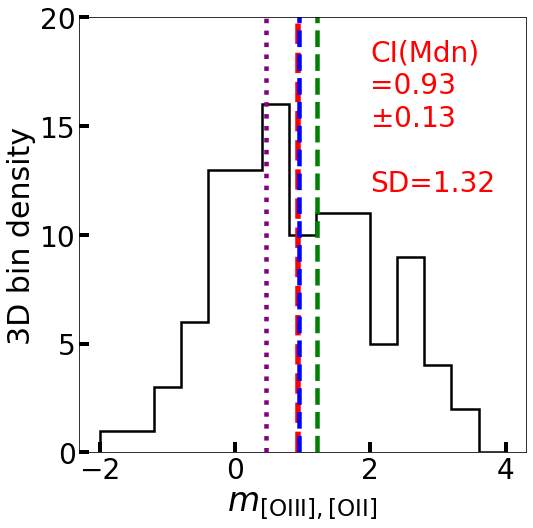}
    \includegraphics[width=0.24\textwidth]{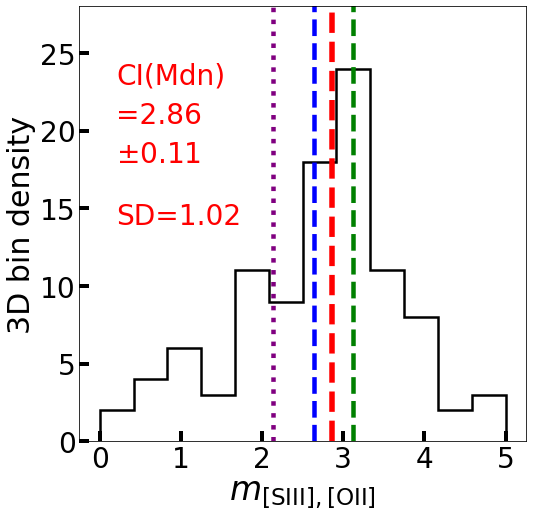}
    \caption{Slope distributions for different reddening relations in IC 342.
    {For each distribution, we show the 68\% confidence interval of the median [CI(Mdn)] and the standard deviation (SD).}
    Dashed lines mark the medians of the distributions (red), values predicted by the F99 extinction curve (blue), and values predicted by the CCM89 extinction curve (green).
    The median values found in the MaNGA main sample are indicated by the dotted purple lines.
    }
    \label{fig:m_ic342}
\end{figure}

\begin{table*}
	\centering
	\caption{Median slopes for different reddening relations in IC342 and the MaNGA main sample}
	{\fontsize{10}{11}\selectfont
	\label{tab:m_ic342}
	\begin{tabularx}{2\columnwidth}{lccccc} 
	    \hline\hline
        Line ratio & Median slope (IC342) & $\sigma _{\rm std}$ (IC342) & Median slope (main sample) & F99 ($R_V=3.1$) & CCM89 ($R_V=3.1$)\\
        \hline
        $[$N\,{\sc ii}]/[O\,{\sc ii}] & $1.95\pm 0.12$ & $1.21$ & $1.525\pm 0.002$ & $1.84$ & $2.09$\\
	    \hline
	    $[$S\,{\sc iii}]/[O\,{\sc iii}] & $1.58\pm 0.07$ & $0.86$ & $1.660\pm 0.005$ & $1.70$ & $1.91$\\
	    \hline
	    $[$O\,{\sc iii}]/[O\,{\sc ii}] & $0.93\pm 0.13$ & $1.32$ & $0.451\pm 0.005$ & $0.95$ & $1.21$\\
	    \hline
	    $[$S\,{\sc iii}]/[O\,{\sc ii}] & $2.86\pm 0.11$ & $1.02$ & $2.127\pm 0.006$ & $2.65$ & $3.12$\\
	    \hline
	\end{tabularx}
	}
\end{table*}

\begin{figure*}
    \centering
    \includegraphics[width=0.85\textwidth]{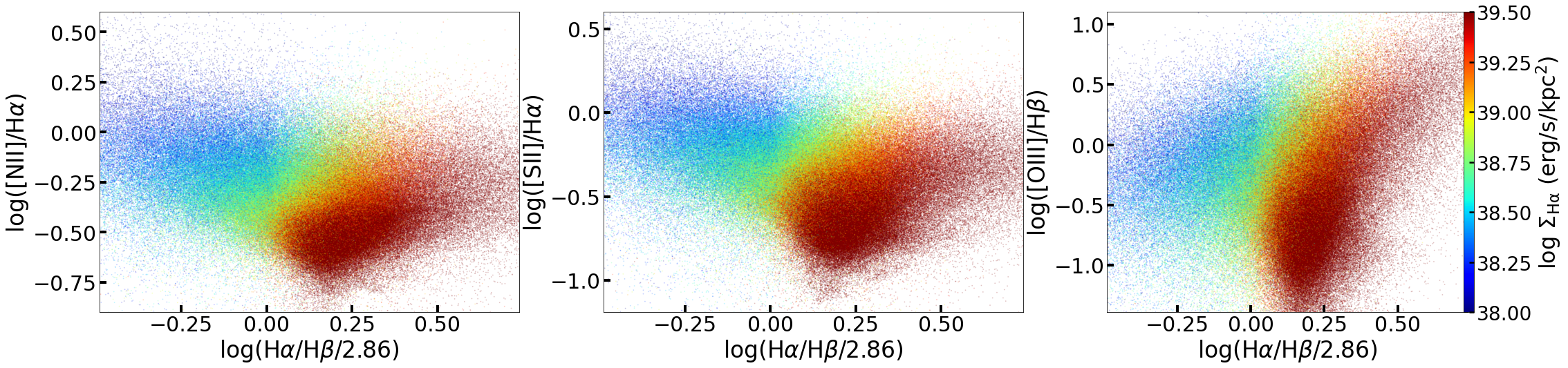}
    \caption{{Relations between forbidden line-to-Balmer line ratios and the Balmer decrement in IC 342. Data points are color-coded according to the surface brightness of H$\alpha$ line. No S/N cut was applied to the data.
    }}
    \label{fig:lrbalr_sc}
\end{figure*}

The main MaNGA sample has a typical spatial resolution of 1\,--\,2 kpc\footnote{{
One might wonder whether the variation in the physical scales of the spaxels might impact our results.
The redshift distribution of our sample is relatively narrow, with 68\% of our sample spaxels lie in the range of 0.022\,--\,0.045, corresponding to physical scales ranging from 1.2 kpc to 2.4 kpc. Meanwhile, the median standard deviation of the physical scale in 3D bins is roughly 0.5 kpc. If we narrow the redshift range by removing spaxels with $z < 0.017$ or $z > 0.050$, our conclusion still stays the same.
}}. 
However, MaNGA also observes the nearby galaxy IC 342 with a much higher resolution of $\sim 32$ pc, which is enough to resolve many large \hii\ regions and separate them from the surrounding DIG. If our speculation of a resolution-dependent attenuation law is correct, we expect to measure reddening relations that are compatible with an average MW extinction curve either described by F99 or CCM89, no matter the lines involved are low ionization lines or not. Since there are considerably fewer spaxels for IC 342 in MaNGA, we increased the sizes of the 3D bins and adjusted them based on the S/N of lines. We first selected data points with more than $3\sigma$ detections on H$\alpha$, H$\beta$, [N\,{\sc ii}], [S\,{\sc ii}], and [O\,{\sc iii}], and then removed non-\hii\ regions using the 3D diagnostic diagram.
This step results in a total of 56,425 spaxels.
When measuring reddening relations involving other lines, we applied extra cuts to remove data points with S/N < 3 for these extra lines.

Figure~\ref{fig:m_ic342} shows the distributions of $\rm m_{[NII],[OII]}$, $\rm m_{[SIII],[OIII]}$, $\rm m_{[OIII],[OII]}$, and $\rm m_{[SIII],[OII]}$, and Table~\ref{tab:m_ic342} summarizes their median values and uncertainties. The numbers of available data points for these measurements are 10,063, 50,545, 10,063, and 10,058, respectively. For reddening relations of \no, \oo, and [S\,{\sc iii}]/[O\,{\sc ii}], we selected 3D bins with the number of spaxels greater than 25. While for the reddening relation of \so, we selected 3D bins with the number of spaxels greater than 100.
The final number of 3D bins we used for \no, \so, \oo, and [S\,{\sc iii}]/[O\,{\sc ii}] is 107, 155, 107, and 107, respectively.
Clearly the S/N cut in [O\,{\sc ii}] removes a large number of data points. If we remeasured $m^{\prime}_{\rm [SIII],[OIII]}$ using a sample with S/N cuts in both [S\,{\sc iii}] and [O\,{\sc ii}] (which has 10,058 data points), the median value changes from $1.58\pm 0.07$ to $1.80\pm 0.13$.

Despite having much wider distributions compared to the MaNGA main sample, the median slopes derived from the aforementioned line ratios seem to be more consistent with the predictions of the F99 extinction curve.
Due to the limited number of spaxels in IC 342, we made larger 3D bins compared to the MaNGA main sample, which inevitably increased the uncertainties of our measurements. We made 20 bins along each of the three axes of the 3D space, covering the whole volume occupied by all the spaxels. This results in a bin size of $\sim 0.028\times 0.045\times 0.100~{\rm dex^3}$ when measuring $m^{\prime}_{\rm [NII],[OII]}$, $m^{\prime}_{\rm [OIII],[OII]}$, and $m^{\prime}_{\rm [SIII],[OII]}$, and a bin size of $\sim 0.051\times 0.063\times 0.103~{\rm dex^3}$ when measuring $m^{\prime}_{\rm [SIII],[OIII]}$.
We tried remeasuring the slopes of the MaNGA main sample using the bin size adopted for IC 342. The resulting $m^{\prime}_{\rm [NII],[OII]}$ is still significantly different from the F99 value (in fact, the deviation becomes slightly larger with these larger bins).
Therefore, it seems the different bin sizes are not the main reason for the different measurements we obtained.

{In Figure~\ref{fig:lrbalr_sc}, we show how the three BPT line ratios change with the Balmer decrement as well as the H$\alpha$ surface brightness in IC 342. Here we did no apply any S/N cut to the data. It is clear that [N\,{\sc ii}] and [S\,{\sc ii}] are relatively strong compared to H$\alpha$ when the Balmer decrement or the H$\alpha$ surface brightness is low and the emission is dominated by the DIG. In comparison, [O\,{\sc iii}] is also stronger when the Balmer decrement or the H$\alpha$ surface brightness is low, but the scatter in [O\,{\sc iii}]/H$\beta$ is large.
One can imagine that in low-resolution observations, differently attenuated regions with different line ratios are thus mixed, which could lead to the differential attenuation as we discussed.
Still, IC 342 is only one example and we need more high-resolution observations to quantify this effect.}

Overall, limited by the large standard deviations of the slope distributions, we cannot draw a strong conclusion for the attenuation law in IC 342. 
Currently there are not enough data from major IFU surveys to allow us to repeat the same statistical analysis we performed on the MaNGA data.
The Calar Alto Legacy Integral Field Area survey \citep[CALIFA,][]{sanchez2012}, for example, includes more nearby galaxies with higher spatial resolution compared to MaNGA, but its data volume is much smaller and its wavelength range does not cover [S\,{\sc iii}].
As another example, the Physics at High Angular resolution in Nearby GalaxieS - Multi Unit Spectroscopic Explorer survey \citep[PHANGS-MUSE,][]{emsellem2022} is able to provide resolved observations of \hii\ regions, but the sample size is even more limited and its wavelength range does not cover [O\,{\sc ii}] and [S\,{\sc iii}]$\lambda 9531$, the stronger line in the [S\,{\sc iii}] doublet.
{Regardless, it would be interesting to check the attenuation of low ionization lines including [N\,{\sc ii}], [S\,{\sc ii}], and [O\,{\sc i}] using these data at the cost of larger statistical uncertainties, which we plan to investigate in future work.
Upcoming surveys such as SDSS-V/Local Volume Mapper \citep[LVM,][]{kollmeier2017} and the Affordable Multi-Aperture Spectroscopy Explorer \citep[AMASE,][]{yan2020} can provide much stronger constraints on the multicomponent dust attenuation on even smaller physical scales, which we also plan to use in the future.
}

\end{document}